\documentclass[12pt,a4paper,oneside]{report}

\usepackage{aims}
\usepackage{graphicx, psfrag,hyperref}
\usepackage{xcolor} 
\usepackage[all]{xy}
\usepackage{multicol}
\usepackage{wrapfig}
\usepackage[T1]{fontenc} 
\usepackage{enumerate}
\usepackage{mflogo}
\usepackage{amssymb,amsmath, amsfonts, amsthm, latexsym}
\usepackage{bbm}
\usepackage{subfigure}

\makeatletter
\def\thickhrulefill{\leavevmode \leaders \hrule height 1ex \hfill \kern \z@}
\def\@makechapterhead#1{%
  \vspace*{5\p@}%
  {\parindent \z@ \centering \reset@font
        \thickhrulefill
        \par\nobreak
        \scshape \@chapapp{} \strut\thechapter
        \par\nobreak
        \interlinepenalty\@M
        \hrule
        \vspace*{5\p@}%
        {\Huge \bfseries #1}\par\nobreak
        \thickhrulefill
        \vspace*{5\p@}%
    \vskip 10\p@
  }}
\def\@makeschapterhead#1{%
  \vspace*{5\p@}%
  {\parindent \z@ \centering \reset@font
        \thickhrulefill
        \par\nobreak
        {\Huge \bfseries \strut #1}\par\nobreak
        \interlinepenalty\@M
        \hrule
        \vspace*{5\p@}%
    \vskip 10\p@
  }}

\newtheorem{theorem}{Theorem}[chapter]
\newenvironment{remark}[1][Remark]{\begin{trivlist}
\item[\hskip \labelsep {\bfseries #1}]}{\end{trivlist}}
\newtheorem{definition}[theorem]{Definition}
\newtheorem{property}[theorem]{Property}
\newtheorem{proposition}[theorem]{Proposition}
\newcommand{\sgn}{\mathop{\mathrm{sgn}}}
\def\babs{\begin{abstract}}             \def\eabs{\end{abstract}}
\def\barr{\begin{array}}                \def\earr{\end{array}}
\def\bcc{\begin{center}}                \def\ecc{\end{center}}

\def\bdes{\begin{description}}          \def\edes{\end{description}}
\def\bdoc{\begin{document}}             \def\edoc{\end{document}}
\def\ben{\begin{enumerate}}             \def\een{\end{enumerate}}
\def\beqn{\begin{eqnarray}}             \def\eeqn{\end{eqnarray}}
\def\beqnl#1{\beqn\label{#1}}           \def\eeqnl#1{\label{#1}\eeqn}
\def\beqnx{\begin{eqnarray*}}           \def\eeqnx{\end{eqnarray*}}
\def\bseqn{\begin{subeqnarray}}         \def\eseqn{\end{subeqnarray}}
\def\beq#1\eeq{\begin{equation}#1\end{equation}}
\def\bal#1\eal{\begin{align}#1\end{align}}
\def\balx#1\ealx{\begin{align*}#1\end{align*}}
\def\beqx{$$}                           \def\eeqx{$$}
\def\bfig{\protect\begin{figure}}       \def\efig{\protect\end{figure}}
\def\bfigx{\protect\begin{figure*}}     \def\efigx{\protect\end{figure*}}
\def\bfigt{\protect\begin{figurette}}   \def\efigt{\protect\end{figurette}}
\def\bfl{\begin{flushleft}}             \def\efl{\end{flushleft}}
\def\bfr{\begin{flushright}}            \def\efr{\end{flushright}}
\def\bit{\begin{itemize}}               \def\eit{\end{itemize}}
\def\bmi{\begin{minipage}}              \def\emi{\end{minipage}}
\def\bfmi{\begin{fminipage}}            \def\efmi{\end{fminipage}}
\def\bpic{\begin{picture}}              \def\epic{\end{picture}}
\def\bqu{\begin{quote}}                 \def\equ{\end{quote}}
\def\bqun{\begin{quotation}}            \def\equn{\end{quotation}}
\def\bsl{\begin{slide}}                 \def\esl{\end{slide}}
\def\btabb{\begin{tabbing}}             \def\etabb{\end{tabbing}}
\def\btabl{\begin{table}}               \def\etabl{\end{table}}
\def\btablx{\begin{table*}}             \def\etablx{\end{table*}}
\def\btab{\begin{tabular}} %\def\etab{\end{tabular}} Conflit avec eta gras... Elle est pas grasse, ma soeur !
\def\btabu{\begin{tabular}}             \def\etabu{\end{tabular}}
\def\btabx{\begin{tabular*}}            \def\etabx{\end{tabular*}}
\def\bbib{\begin{thebibliography}{999}} \def\ebib{\end{thebibliography}}
\def\bver{\begin{verbatim}}             \def\ever{\end{verbatim}}
\def\bca{\begin{cases}}                  \def\eca{\end{cases}}

\input{alphabet}
\input{abrege}
\def\pth#1{\left(#1\right)}                \def\stdpth#1{(#1)}
\def\acc#1{\left\{#1\right\}}              \def\stdacc#1{\{#1\}}
\def\cro#1{\left[#1\right]}                \def\stdcro#1{[#1]}
\def\bars#1{\left|#1\right|}               \def\stdbars#1{|#1|}
\def\norm#1{\left\|#1\right\|}             \def\stdnorm#1{\|#1\|}
\def\scal#1{\left\langle#1\right\rangle}   \def\stdscal#1{\langle#1\rangle}
 
\def\bigpth#1{\bigl(#1\bigr)}              \def\biggpth#1{\biggl(#1\biggr)}
\def\bigacc#1{\bigl\{#1\bigr\}}            \def\biggacc#1{\biggl\{#1\biggr\}}
\def\bigcro#1{\bigl[#1\bigr]}              \def\biggcro#1{\biggl[#1\biggr]}
\def\bigbars#1{\bigl|#1\bigr|}             \def\biggbars#1{\biggl|#1\biggr|}
\def\bignorm#1{\bigl\|#1\bigr\|}           \def\biggnorm#1{\biggl\|#1\biggr\|}
\def\bigscal#1{\bigl\langle#1\bigr\rangle} \def\biggscal#1{\biggl\langle#1\biggr\rangle}

\def\Bigpth#1{\Bigl(#1\Bigr)}              \def\Biggpth#1{\Biggl(#1\Biggr)}
\def\Bigacc#1{\Bigl\{#1\Bigr\}}            \def\Biggacc#1{\Biggl\{#1\Biggr\}}
\def\Bigcro#1{\Bigl[#1\Bigr]}              \def\Biggcro#1{\Biggl[#1\Biggr]}
\def\Bigbars#1{\Bigl|#1\Bigr|}             \def\Biggbars#1{\Biggl|#1\Biggr|}
\def\Bignorm#1{\Bigl\|#1\Bigr\|}           \def\Biggnorm#1{\Biggl\|#1\Biggr\|}
\def\Bigscal#1{\Bigl\langle#1\Bigr\rangle} \def\Biggscal#1{\Biggl\langle#1\Biggr\rangle}

\def\diag{{\mathrm{diag}}}              \def\Diag#1{{\mathrm{diag}}\bigcro{#1}} \def\Diagold#1{{\mathrm{diag}}\cro{#1}}
\def\tr{{\mathrm{tr}}\,}                \def\Tr#1{{\mathrm{tr}}\bigcro{#1}}                     \def\Trold#1{{\mathrm{tr}}\cro{#1}}
\def\rg{{\mathrm{rg}}\,}                \def\Rg#1{{\mathrm{rg}}\bigcro{#1}}                     \def\Rgold#1{{\mathrm{rg}}\cro{#1}}
\def\esp{{\mathrm{E}}\,}              \def\Esp#1{{\mathrm{E}}\bigcro{#1}}  \def\Esph#1#2{\underset{{#1}}{\mathrm{E}}\bigcro{#2}}               \def\Espold#1{{\mathrm{E}}\cro{#1}} 
\def\var{{\mathrm{var}}\,}              \def\Var#1{{\mathrm{var}}\bigcro{#1}}           \def\Varold#1{{\mathrm{var}}\cro{#1}}
\def\cov{{\mathrm{Cov}}\,}              \def\Cov#1{{\mathrm{Cov}}\bigcro{#1}}           \def\Covold#1{{\mathrm{Cov}}\cro{#1}}

\def\Cos#1{\cos\cro{#1}}
\def\Sin#1{\sin\cro{#1}}
\def\Exp#1{\exp\cro{#1}}
\def\Log#1{\log\cro{#1}}
\def\Ln#1{\ln\cro{#1}}                  
\def\Det#1{\det\bigcro{#1}}
\def\Erf#1{\mathrm{erf}\cro{#1}} 
\def\Erfc#1{\mathrm{erfc}\cro{#1}}
\def\Erfcx#1{\mathrm{erfcx}\cro{#1}}

\def\cond{\textrm{Cond}\,}              \def\Cond#1{\cond\bigpth{#1}}

\def\sinc{{\mathrm{sinc}}\,}    \def\Sinc#1{{\mathrm{sinc}}\bigcro{#1}} \def\Sincold#1{{\mathrm{sinc}}\cro{#1}}
\def\rang{{\mathrm{rang}}\,}    \def\Rang#1{\rang\bigcro{#1}}                                   \def\Rangold#1{\rang\cro{#1}}
\def\ker{\textrm{Ker}\,}                \def\Ker#1{\ker\bigcro{#1}}                                     \def\Kerold#1{\ker\cro{#1}}
\def\img{\textrm{Im}\,}                 \def\Img#1{\img\bigcro{#1}}                                     \def\Imgold#1{\img\cro{#1}}
\def\vect{{\mathrm{Vect}}\,}    \def\Vect#1{\vect\bigcro{#1}}                                   \def\Vectold#1{\vect\cro{#1}}
\def\sgn{{\mathrm{sgn}}}                \def\Sgn#1{\sgn\bigcro{#1}}                                     \def\Sgnold#1{\sgn\cro{#1}}

\def\Pr{\mathop{\textrm{Pr}}}

\def\reel{{\mathrm{Re}}}                \def\Reel#1{{\mathrm{Re}}\cro{#1}}
\def\card{{\mathrm{Card}}}              \def\Card#1{{\mathrm{Card}}\cro{#1}}

\def\sh{{\mathrm{sh}}}                % sin hyperbolique en FR
\def\ch{{\mathrm{ch}}}                % cos hyperbolique en FR
\def\th{{\mathrm{th}}}                % th hyperbolique en FR
\def\coth{{\mathrm{coth}}}            % cot hyperbolique en FR
\def\coeffbin#1#2{\pth{\setlength{\arraycolsep}{.1em}\barr{c}#1\\#2\earr}}
\def\Div{{\mathrm{div}}}                                                                                        % divergence
\def\Rotv{\overrightarrow{\mathop{{\mathrm{rot}}}}}             % rotationnel avec fleche
\def\Gradv{\overrightarrow{\mathop{{\mathrm{grad}}}}}           % gradient avec fleche
\def\IF{\text{if\:}}             \def\SI{\text{si\:}}
\def\If{\text{If\:}}             \def\Si{\text{Si\:}}
\def\AND{\text{and\:}}           \def\ET{\text{et\:}}
\def\OR{\text{or\:}}             \def\OU{\text{ou\:}}
\def\THEN{\text{then\:}}         \def\ALORS{\text{alors\:}}
                                 \def\DOU{\text{d'o\`u\:}}
\def\WHERE{\text{where\:}}       \def\Ou{\text{o\`u\:}}
\def\WHEN{\text{when\:}}         \def\QUAND{\text{quand\:}}
\def\FOR{\text{for\:}}           \def\POUR{\text{pour\:}}
\def\FORALL{\text{for all\:}}    \def\POURTOUT{\text{pour tout\:}}
\def\ST{\text{s.t.\:}}           \def\SC{\text{s.c.\:}}
\def\SUBJTO{\text{subject to\:}} \def\SOUSC{\text{sous contraintes\:}}
\def\OTHERWISE{\text{otherwise}} \def\SINON{\text{sinon}}
\def\WITH{\text{with\:}}         \def\AVEC{\text{avec\:}}
\def\IN{\text{in\:}}             \def\DANS{\text{dans\:}}

\def\arrayp{\renewcommand{\arraystretch}{.7}\setlength{\arraycolsep}{2pt}}
\def\tabp{\renewcommand{\arraystretch}{.7}\setlength{\tabcolsep}{2pt}}

\newsavebox{\fminibox}
\newlength{\fminilength}
\newenvironment{fminipage}[1][\linewidth]
  {\setlength{\fminilength}{#1}%-2\fboxsep-2\fboxrule}%
   \begin{lrbox}{\fminibox}\begin{minipage}{\fminilength}}
  {\end{minipage}\end{lrbox}\noindent\fbox{\usebox{\fminibox}}}

\def\M{^{-1}} \def\T{^\tD} \def\+{^\dagger}
\def\I{\,|\,}           % "sachant" bien espac\'e pour les formules
\def\J{\mathop{\,;\,}}  % "point virgule" bien espac\'e pour les formules.
\def\w{,\thinspace}
\def\W{,\thickspace}
\def\ldotsv{,\,\ldots,\,}
\def\V{,}               % Virgule pour les nombres decimaux
\def\e#1{.10^{#1}}      % Notation scientifique a la francaise

\def\egdef{\stackrel{\Delta}{=}}
\def\nequiv{\not\kern-.05em\equiv}
\def\egal{\kern-.5em=\kern-.5em}        % Moins d'espace autour de "="
\def\propt{\kern-.2em\propto\kern-.2em} % Idem
\def\dans{\in\!}                        % Trop d'espace apres "appartient a"
\def\pourtt{\forall\,}                  % Pas assez d'espace
\def\wh#1{\widehat{#1}}                 % Sombrero !
\def\wt#1{\widetilde{#1}} 

\def\argmax{\mathop{\mathrm{arg\,max}}} % Mieux que \def\argmax{\arg\max}
\def\argmin{\mathop{\mathrm{arg\,min}}} % car l'indice est reparti
\def\Argmax#1#2{\displaystyle \argmax_{#1}\left\{{#2}\right\}} % Ajout \HS
\def\Argmin#1#2{\displaystyle \argmin_{#1}\left\{{#2}\right\}}  % 

\def\EL{\mathrm{EL}}
\def\Vect{\mathop{\text{Vect}}}

\def\froc#1#2{{#1/#2}}                  % Frac en toc
\def\fric#1#2{\frac1{#2}#1}
\def\fracds#1#2{\frac{\displaystyle#1}{\displaystyle#2}}
\def\diff#1#2{{\frac{d#1}{d#2}}}
\def\dd{\,d}                            % doit etre en italique en anglais
\def\derpar#1#2{{\frac{\partial #1}{\partial #2}}}
\def\derpor#1#2{{\froc{\partial #1}{\partial #2}}}
\def\parsec#1#2#3{{\frac{\partial^2 #1}{\partial #2\,\partial #3}}}
\def\parsecd#1#2{{\frac{\partial^2 #1}{{\partial #2}^2}}}
\def\porsec#1#2#3{{\froc{\partial^2 #1}{\partial #2\,\partial #3}}}
\def\porsecd#1#2{{\froc{\partial^2 #1}{{\partial #2}^2}}}
\def\pornd#1#2#3{{\froc{\partial^{#3} #1}{{\partial #2}^{#3}}}}
\def\intdouble{\int\kern-0.3em\int}
\def\inttriple{\int\kern-0.3em\int\kern-0.3em\int}
\def\prods{\mathop{\text{\footnotesize$\displaystyle\prod$}}}

\def\rond#1{\overset{\kern-0.33em~_\circ}{#1}}
\def\rondit[#1]#2{\overset{\kern#1~_\circ}{#2}}

\def\incirc#1{\pscirclebox[framesep=1pt]{\scriptsize#1}}
\def\incircp#1{\pscirclebox[framesep=1pt]{\tiny#1}}
\def\outcirc#1{\raisebox{-3pt}{\huge$\times$}\hspace*{-.45cm}\text{\incirc{#1}}}

\def\bdoc{\begin{document}}             \def\edoc{\end{document}}
\def\babs{\begin{abstract}}             \def\eabs{\end{abstract}}
\def\bcc{\begin{center}}                \def\ecc{\end{center}}
\def\ben{\begin{enumerate}}             \def\een{\end{enumerate}}
\def\bit{\begin{itemize}}               \def\eit{\end{itemize}}
\def\bpic{\begin{picture}}              \def\epic{\end{picture}}
\def\beq{\begin{equation}}              \def\eeq{\end{equation}}
\def\barr{\begin{array}}                \def\earr{\end{array}}
\def\btab{\begin{tabular}}              \def\etab{\end{tabular}}
\def\btabu{\begin{tabular}}             \def\etabu{\end{tabular}}
\def\bcc{\begin{center}}                \def\ecc{\end{center}}
\def\beqn{\begin{eqnarray}}             \def\eeqn{\end{eqnarray}}
\def\beqnx{\begin{eqnarray*}}           \def\eeqnx{\end{eqnarray*}}
\def\bfig{\begin{figure}}               \def\efig{\end{figure}}
\def\bver{\begin{verb}}						 \def\ever{\end{verb}}

\def\d#1{\,\mbox{d} #1}
\def\disp#1{\displaystyle{#1}}

\def\blue#1{{\color{blue}#1}}
\def\red#1{{\color{red}#1}}
\def\green#1{{\color{green}#1}}
\def\magenta#1{{\color{magenta}#1}}
\def\cyan#1{{\color{cyan}#1}}
\def\titre#1{\setlength{\shadowsize}{4pt}\centerline{\shadowbox{\blue{\mbox{~~\sc #1~~}}}}}

\def\iii{\int_{-\infty}^{+\infty}}
\def\izi{\int_{0}^{\infty}}
\def\izpi{\int_{0}^{\pi}}
\def\izdpi{\int_{0}^{2\pi}}
\def\intd{\int\kern-.8em\int}
\def\intt{\int\kern-.8em\int\kern-.8em\int}
\def\intg{\int\kern-1.1em\int}

\def\xh{\widehat{x}}
\def\thetah{\widehat{\theta}}
\def\betah{\widehat{\beta}}

\def\xbh{\widehat{\xb}}
\def\thetabh{\widehat{\thetab}}
\def\betabh{\widehat{\betab}}

\def\xbhk{\widehat{\xb}^{k}}
\def\thetahk{\widehat{\theta}^{k}}
\def\betahk{\widehat{\beta}^{k}}

\def\xbhkp{\widehat{\xb}^{k+1}}
\def\thetahkp{\widehat{\theta}^{k+1}}
\def\betahkp{\widehat{\beta}^{k+1}}

\def\thetabhk{\widehat{\thetab}^{k}}
\def\betabhk{\widehat{\betab}^{k}}

\def\thetabhkp{\widehat{\thetab}^{k+1}}
\def\betabhkp{\widehat{\betab}^{k+1}}

\def\thetamin{\theta_{\mbox{\tiny min}}}
\def\thetamax{\theta_{\mbox{\tiny max}}}
\def\betamin{\beta_{\mbox{\tiny min}}}
\def\betamax{\beta_{\mbox{\tiny max}}}

\def\ra{\rightarrow}
\def\la{\leftarrow}
\def\da{\downarrow}
\def\ua{\uparrow}

\def\Ra{\Rightarrow}
\def\La{\Leftarrow}
\def\Da{\Downarrow}
\def\Ua{\Uparrow}

\def\lra{\longrightarrow}
\def\lla{\longleftarrow}
\def\Lra{\Longrightarrow}
\def\Lla{\longleftarrow}

\def\lrarr{\leftrightarrow}
\def\Lrarr{\Leftrightarrow}
\def\udarr{\updownarrow}
\def\Uparr{\Updownarrow}

\def\expf#1{\exp\left[ {#1} \right]}

\def\argmax#1#2{\arg\max_{#1}\left\{ #2 \right\}}
\def\argmin#1#2{\arg\min_{#1}\left\{ #2 \right\}}

\def\Prob#1{\mbox{Pr}\left\{#1\right\}}
\def\var#1{\mbox{Var}\left\{#1\right\}}
\def\cov#1{\mbox{Cov}\left\{#1\right\}}
\def\corr#1{\mbox{Corr}\left\{#1\right\}}
\def\trace#1{\mbox{Tr}\left\{#1\right\}}
\def\rang#1{\mbox{rang}\left\{#1\right\}}
\def\det#1{\mbox{d\'et}\left\{#1\right\}}

\def\gbu{\underline{\gb}}
\def\fbu{\underline{\fb}}
\def\zbu{\underline{\zb}}
\def\epsilonbu{\underline{\epsilonb}}
\def\thetabu{\underline{\thetab}}

\def\sigmae{{\sigma_{\epsilon}}}
\def\Sigmae{{\Sigmab_{\epsilonb}}}
\def\Sigmaf{{\Sigmab_{\fb}}}
\def\Sigmag{{\Sigmab_{\gb}}}

\def\fbuh{\widehat{\fbu}}
\def\zbuh{\widehat{\fbu}}
\def\thetabuh{\widehat{\thetabu}}

\def\fbh{\widehat{\fb}}
\def\Pbh{\widehat{\Pb}}
\def\Sigmabh{\widehat{\Sigmab}}
\def\Abh{\widehat{\Ab}}
\def\Bbh{\widehat{\Bb}}
\def\thetabh{\widehat{\thetab}}
\def\Rbe{\Rb_{\epsilon}}
\def\Rbeh{\widehat{\Rbe}}

\def\Rjk{{R_j}_k}
\def\fbik{{\fb_i}_k}
\def\fbjk{{\fb_j}_k}

\def\mik{{m_i}_k}
\def\sigmaik{{\sigma_i^2}_k}
\def\Sigmaik{{\Sigmab_i}_k}
\def\alphaik{{\alpha_i}_k}
\def\betaik{{\beta_i}_k}

\def\muik{{\mu_i}_k}
\def\vik{{v_i}_k}

\def\mjk{{m_j}_k}
\def\sigmajk{{\sigma_j^2}_k}
\def\Sigmajk{{\Sigmab_j}_k}
\def\alphajk{{\alpha_j}_k}
\def\betajk{{\beta_j}_k}

\def\mujk{{\mu_j}_k}
\def\vjk{{v_j}_k}
\def\njk{{n_j}_k}

\def\alphae{\alpha^{\epsilon}}
\def\betae{\beta^{\epsilon}}
\def\alphaez{\alpha^{\epsilon}_{0}}
\def\betaez{\beta^{\epsilon}_{0}}
\def\alphaei{\alpha^{\epsilon}_{i}}
\def\betaei{\beta^{\epsilon}_{i}}
\def\alphaeiz{\alpha^{\epsilon}_{i0}}
\def\betaeiz{\beta^{\epsilon}_{i0}}

\def\mbz{{{\mb}_z}}
\def\Sigmabz{{{\Sigmab}_z}}
\def\diag#1{\mbox{diag}\left[#1\right]}

\def\sigmah{\widehat{\sigma}}
\def\fh{\widehat{f}}
\def\sigmaz{\sigma_z}

\def\blue#1{{\color{blue}#1}}
\def\red#1{{\color{red}#1}}
\def\green#1{{\color{green}#1}}

\def\sumi{\sum_{i=1}^{M} }
\def\sumj{\sum_{j=1}^{N} }
\def\lambdah{\wh{\lambda}}
\def\lambdabh{\wh{\lambdab}}

\def\det#1{\hbox{det}\left(#1\right)}
\def\defined{\,\shortstack{$\triangle$\\ =}\,}

\def\zmat{\left[\matrix{0}\right]}
\def\det#1{\hbox{det}(#1)}
\def\th{\widehat{t}}
\def\xh{\widehat{x}}
\def\lambdah{\widehat{\lambda}}
\def\muh{\widehat{\mu}}
\def\sigmah{\widehat{\sigma}}
\def\rhoh{\widehat{\rho}}
\def\betah{\widehat{\beta}}
\def\alphah{\widehat{\alpha}}
\def\tbh{\widehat{\tb}}
\def\mubh{\widehat{\mub}}
\def\sigmabh{\widehat{\sigmab}}
\def\rhobh{\widehat{\rhob}}
\def\oneb{\mbox{\bf 1}}

\def\bm#1{\mbox{\boldmath $#1$}}
\def\zerob{{\bf 0}}
\def\oneb{{\bf 1}}
\def\intg{\int\kern-1.1em\int}
\def\expf#1{\exp\left[ {#1} \right]}

\def\gbu{\underline{\gb}}
\def\fbu{\underline{\fb}}
\def\zbu{\underline{\zb}}
\def\epsilonbu{\underline{\epsilonb}}
\def\uthetab{\underline{\thetab}}

\def\sigmae{{\sigma_{\epsilon}}}
\def\Sigmae{{\Sigma_{\epsilonb}}}
\def\Sigmabe{{\Sigmab_{\epsilonb}}}
\def\Sigmaf{{\Sigmab_{\fb}}}
\def\Sigmag{{\Sigmab_{\gb}}}

\def\fbuh{\widehat{\fbu}}
\def\zbuh{\widehat{\fbu}}
\def\uthetabh{\widehat{\uthetab}}

\def\fbh{\widehat{\fb}}
\def\Sigmabh{\widehat{\Sigmab}}
\def\Abh{\widehat{\Ab}}
\def\thetabh{\widehat{\thetab}}

\def\Rjk{{R_j}_k}
\def\fbik{{\fb_i}_k}
\def\fbjk{{\fb_j}_k}

\def\mik{{m_i}_k}
\def\sigmaik{{\sigma_i^2}_k}
\def\Sigmaik{{\Sigmab_i}_k}
\def\alphaik{{\alpha_i}_k}
\def\betaik{{\beta_i}_k}

\def\muik{{\mu_i}_k}
\def\vik{{v_i}_k}

\def\mjk{{m_j}_k}
\def\sigmajk{{\sigma_j^2}_k}
\def\Sigmajk{{\Sigmab_j}_k}
\def\alphajk{{\alpha_j}_k}
\def\betajk{{\beta_j}_k}

\def\mujk{{\mu_j}_k}
\def\vjk{{v_j}_k}
\def\njk{{n_j}_k}

\def\alphae{\alpha^{\epsilon}}
\def\betae{\beta^{\epsilon}}
\def\alphaez{\alpha^{\epsilon}_{0}}
\def\betaez{\beta^{\epsilon}_{0}}
\def\alphaei{\alpha^{\epsilon}_{i}}
\def\betaei{\beta^{\epsilon}_{i}}
\def\alphaeiz{\alpha^{\epsilon}_{i0}}
\def\betaeiz{\beta^{\epsilon}_{i0}}
\def\alphaeiz{\alpha^{\epsilon}_{i0}}
\def\betaeiz{\beta^{\epsilon}_{i0}}
\def\alphaeiz{\alpha^{\epsilon}_{i0}}
\def\betaeiz{\beta^{\epsilon}_{i0}}

\def\mbz{{{\mb}_z}}
\def\Sigmabz{{{\Sigmab}_z}}

\def\vect{\mbox{vect}}
\def\lra{\longrightarrow}

\def\gir{g_i(\rb)}
\def\fjr{f_j(\rb)}
\def\zjr{z_j(\rb)}
\def\epsilonir{\epsilon_i(\rb)}
\def\gbr{\gb(\rb)}
\def\fbr{\fb(\rb)}
\def\zbr{\zb(\rb)}
\def\epsilonbr{\epsilonb(\rb)}
\def\fbhr{\fbh(\rb)}
\def\Sigmabhr{\Sigmabh(\rb)}

\def\mbzr{\mb_{z(\rb)}}
\def\Sigmabzr{\Sigmab_{z(\rb)}}
\def\Sigmagbzr{{\Sigmab_g}_{z(\rb)}}

\def\mjzjr#1{{m_{#1}}_{z_{#1}(\rb)}}
\def\sigmajzjr#1{{\sigma_{#1}}_{z_{#1}(\rb)}}

\def\Beta{B}
\def\Betab{\bm{\Beta}}
\def\Betabe{\bm{\Beta}_{\epsilonb}}

\def\fh{\widehat{f}}
\def\sigmah{\widehat{\sigma}}
\def\sigmaei{\sigma_{\epsilon_i}^2}

\def\thetah{\widehat{\theta}}
\def\sigmah{\widehat{\sigma}}

\def\Ftxm{\widetilde{F}_{X}(x_-)}
\def\Ftxp{\widetilde{F}_{X}(x_+)}

\def\NU{{\cal V}}

\def\tFx{\widetilde{F}_{X}}

\def\Fx{F_{X}(x)}
\def\fx{f_{X}(x)}

\def\Fxv{F_{X}(x)}
\def\fxv{f_{X}(x)}

\def\Fxi{{F}_{X_i}(x_i)}
\def\fxi{{f}_{X_i}(x_i)}

\def\Gnu{G_{\NU}(\nu)}
\def\gnu{g_{\NU}(\nu)}

\def\Fnu{F_{\NU}(\nu)}
\def\fnu{f_{\NU}(\nu)}

\def\Finnu#1{F^{-1}_{\NU}(#1)}

\def\Gnua#1{G_{\NU}(#1)}
\def\Gnub#1{G_{\NU}^{-1}(#1)}
\def\Gnuh{G_{\NU}^{-1}(1/2)}

\def\Ftx{\widetilde{F}_{X}(x)}
\def\ftx{\widetilde{f}_{X}(x)}

\def\Ftxi{\widetilde{F}_{X_i}(x_i)}
\def\ftxi{\widetilde{f}_{X_i}(x_i)}

\def\Ftxv{\widetilde{F}_{X}(x)}
\def\FtX#1{\widetilde{F}_{X}(#1)}
\def\ftxv{\widetilde{f}_{X}(x)}

\def\fxnu{f_{X|\NU}(x|\nu)}
\def\fxNU{f_{X|\NU}(x|\NU)}
\def\FxNU{F_{X|\NU}(x|\NU)}

\def\fxNUtheta{f_{X|\NU,\theta}(x|\NU,\theta)}
\def\FxNUtheta{F_{X|\NU,\theta}(x|\NU,\theta)}

\def\Fxnu{F_{X|\NU}(x|\nu)}
\def\Fxnui#1{F_{X|\NU}(x|\nu_#1)}

\def\FxN#1{F_{X|\NU}(#1)}
\def\FxNU{F_{X|\NU}(x|\NU)}
\def\Fxnun#1{F_{X|\NU}(x|#1)}
\def\Fxinun#1{F_{X_i|\NU}(x_i|#1)}
\def\fxnu{f_{X|\NU}(x|\nu)}
\def\fxNU{f_{X|\NU}(x|\NU)}

\def\Fxvnu{F_{X|\NU}(x|\nu)}
\def\Fxvnun#1{F_{X|\NU}(x|#1)}
\def\FxvNU{F_{X|\NU}(x|\NU)}

\def\FXvnu{F_{X|\NU}}

\def\fxvnu{f_{X|\NU}(x|\nu)}
\def\fxvNU{f_{X|\NU}(x|\NU)}

\def\FXvNU#1{{F}_{X|\NU}(#1|\NU)}
\def\FXvn#1{{F}_{X|\NU}(x|#1)}

\def\Fxtheta{F_{X|\theta}(x|\theta)}
\def\fxtheta{f_{X|\theta}(x|\theta)}

\def\Fxvtheta{F_{X|\theta}(x|\theta)}
\def\fxvtheta{f_{X|\theta}(x|\theta)}

\def\Ftxtheta{\widetilde{F}_{X|\theta}(x|\theta)}
\def\ftxtheta{\widetilde{f}_{X|\theta}(x|\theta)}

\def\Ftxvtheta{\widetilde{F}_{X|\theta}(x|\theta)}
\def\ftxvtheta{\widetilde{f}_{X|\theta}(x|\theta)}

\def\Fxnutheta{F_{X|\NU,\theta}(x|\nu,\theta)}
\def\fxnutheta{f_{X|\NU,\theta}(x|\nu,\theta)}

\def\ftheta{f_{\Theta}(\theta)}
\def\fthetaxnuz{f_{\Theta|X,\nu_0}(\theta|x,\nu_0)}
\def\fthetax{f_{\Theta|X}(\theta|x)}

\def\Fxvnutheta{F_{X|\NU,\theta}(x|\nu,\theta)}
\def\FxvNUtheta{F_{X|\NU,\theta}(x|\NU,\theta)}
\def\fxvnutheta{f_{X|\NU,\theta}(x|\nu,\theta)}
\def\Ftxvnutheta{\widetilde{F}_{X|\theta}(x|\theta)}
\def\ftxvnutheta{\widetilde{f}_{X|\theta}(x|\theta)}

\def\FxvNUm{F_{X|\NU}(x_-|\NU)}
\def\FxvNUp{F_{X|\NU}(x_+|\NU)}

\def\Fxinu{F_{X_i|\NU}(x_i|\nu)}
\def\fxinu{f_{X_i|\NU}(x_i|\nu)}

\def\Ftxi{\widetilde{F}_{X_i}(x_i)}
\def\ftxi{\widetilde{f}_{X_i}(x_i)}

\def\fxitheta{f_{X_i|\theta}(x_i|\theta)}
\def\ftxitheta{\widetilde{f}_{X_i|\theta}(x_i|\theta)}

\def\thetah{\widehat{\theta}}
\def\thetat{\widetilde{\theta}}

\def\lnuxv{l_{\nu|X}(\nu|x)}

\def\Lnuxva#1{L_{\nu|X}(#1 | x)}
\def\lnuxva#1{l_{\nu|X}(#1 | x)}

\def\Prob#1{\mbox{P}\left(#1\right)}

\def\lthetaxvnuz{l(\theta)}

\def\lthetaxvnuz{l_{\theta|x,\nu_0}(\theta | x,\nu_0)}
\def\fxnuztheta{f_{X|\nu_0,\theta}(x | \nu_0,\theta)}
\def\fxinuztheta{f_{X|\nu_0,\theta}(x_i | \nu_0,\theta)}
\def\fxvtheta{f_{X|\theta}(x | \theta)}
\def\lthetaxv{l_{\theta|X}(\theta | x)}
\def\ltthetaxv{\widetilde{l}_{\theta|X}(\theta | x)}

\def\ltheta{l(\theta )}
\def\lttheta{\widetilde{l}(\theta)}
\def\Lnu{L(\nu )}
\def\Lnui#1{L(\nu_#1)}
\def\Linu{L_i(\nu )}
\def\Linu{L_i(\nu )}
\def\LNU{L(\NU )}
\def\Lin#1{L^{-1}(#1)}
\def\L#1{L(#1)}
\def\Li#1{L_i(#1)}
\def\FNU#1{F_\NU(#1)}
\def\FinNU#1{F^{-1}_\NU(#1)}

\def\ftthetax{\tilde{f}_{\Theta|X}(\theta|x)}
\def\ER{R}

\def\Ndd#1#2#3{{\cal N}\left( #1 ;~ #2, #3 \right)}
\def\Cdd#1#2#3{{\cal C}\left( #1 ;~ #2, #3 \right)}
\def\Sdd#1#2#3#4{{\cal S}\left( #1 ;~ #2, #3, #4 \right)}
\def\Edd#1#2{{\cal E}\left( #1 ;~ #2 \right)}
\def\DEdd#1#2{{\cal DE}\left( #1 ;~ #2 \right)}
\def\Gdd#1#2#3{{\cal G}\left( #1 ;~ #2, #3 \right)}
\def\IGdd#1#2#3{{\cal IG}\left( #1 ;~ #2, #3 \right)}

\def\Xwr{X(\omega,\rb)}
\def\xwr{x(\omega,\rb)}

\def\muzw{\mu_{z}(\omega)}
\def\muzwr{\mu_{z(\rb)}(\omega)}
\def\szw{\sigma_{z}^2(\omega)}
\def\szwr{\sigma_{z(\rb)}^2(\omega)}
\def\pzw{p_{z}(\omega)}
\def\pzwr{p_{z(\rb)}(\omega)}

\def\hszw{\widehat{\sigma_{z}^2}(\omega)}
\def\hmuzw{\widehat{\mu}_{z}(\omega)}
\def\hpzw{\widehat{p}_{z}(\omega)}

\def\muzwrl{\mu_{z(\rb)}(\omega-l)}

\def\pxwrcz{p\left(\xwr ~|~ z\right)}
\def\pxwr{p\left(\xwr\right)}

\def\Xbw{\Xb(\omega)}
\def\xbw{\xb(\omega)}
\def\uXb{\underline{\Xb}}
\def\uxb{\underline{\xb}}
\def\pxbw{p(\xbw)}
\def\puxb{p(\uxb)}
\def\pzb{P(\zb)}

\def\Szw{S_z(\omega)}
\def\mubw{\mub(\omega)}

\def\Srw{S_{\rb}(\omega)}
\def\Swr{S_{\omega}(\rb)}
\def\Sbw{\Sb(\omega)}
\def\sbw{\sb(\omega)}
\def\Sbr{\Sb(\rb)}
\def\mubw{\mub(\omega)}
\def\cov#1{\mbox{cov}[#1]}

\def\pdf{\emph{pdf}}
\def\pmf{\emph{pmf}}
\def\dpdx#1#2{\frac{\partial #1}{\partial #2}}

\def\REM#1{}

\def\XXwr{X(\omega,\rb)}
\def\xxwr{x(\omega,\rb)}

\def\Xwr{X_{\omega}(\rb)}
\def\xwr{x_{\omega}(\rb)}
\def\Xrw{X_{\rb}(\omega)}
\def\xrw{x_{\rb}(\omega)}

\def\Ywr{Y_{\omega}(\rb)}
\def\ywr{y_{\omega}(\rb)}
\def\Yrw{Y_{\rb}(\omega)}
\def\yrw{y_{\rb}(\omega)}

\def\Ewr{E_{\omega}(\rb)}
\def\ewr{\epsilon_{\omega}(\rb)}
\def\Erw{E_{\rb}(\omega)}
\def\erw{\epsilon_{\rb}(\omega)}

\def\Xbr{\Xb(\rb)}
\def\xbr{\xb(\rb)}
\def\Xbw{\Xb(\omega)}
\def\xbw{\xb(\omega)}

\def\uXb{\underline{\Xb}}
\def\uxb{\underline{\xb}}
\def\uXbz{\underline{\Xb}_z}
\def\uxbz{\underline{\xb}_z}

\def\uthetab{\underline{\thetab}}
\def\uthetabz{\underline{\thetab}_z}

\def\Ywr{Y_{\omega}(\rb)}
\def\ywr{y_{\omega}(\rb)}
\def\Yrw{Y_{\rb}(\omega)}
\def\yrw{y_{\rb}(\omega)}
\def\Ybr{\Yb(\rb)}
\def\ybr{\yb(\rb)}
\def\Ybw{\Yb(\omega)}
\def\ybw{\yb(\omega)}
\def\uYb{\underline{\Yb}}
\def\uyb{\underline{\yb}}
\def\uYbz{\underline{\Yb}_z}
\def\uybz{\underline{\yb}_z}

\def\Ewr{E_{\omega}(\rb)}
\def\ewr{\epsilon_{\omega}(\rb)}
\def\Erw{E_{\rb}(\omega)}
\def\erw{\epsilon_{\rb}(\omega)}
\def\Ebr{Eb(\rb)}
\def\ebr{\epsilonb(\rb)}
\def\Ebw{Eb(\omega)}
\def\ebw{\epsilonb(\omega)}
\def\uEb{\underline{\Eb}}

\def\Erwp{E_{\rb}(\omega')}
\def\Xrwp{X_{\rb}(\omega')}

\def\xwmr{x_{\omega-1}(\rb)}
\def\muzwm{\mu_{z}(\omega-1)}

\def\rhoz{\rho_z}
\def\sez{\sigma_{\eta_z}^2}

\def\xwrp{x_{\omega}(\rb')}
\def\xrwp{x_{\rb}(\omega')}
\def\xwpr{x_{\omega'}(\rb)}
\def\xwprp{x_{\omega'}(\rb')}
\def\Xwrp{x_{\omega}(\rb')}
\def\Xwrp{X_{\omega}(\rb')}
\def\Xwpr{X_{\omega'}(\rb)}
\def\Xwprp{X_{\omega'}(\rb')}

\def\sigmaed{\sigmae^2}
\def\sigmaew{\sigmae(\omega)}
\def\sigmaedw{\sigmaed(\omega)}

\def\Sigmabe{\Sigmab_{\epsilon}}
\def\sigmabh{\widehat{\Sigmab}}
\def\xbh{\widehat{\xb}}
\def\zbh{\widehat{\zb}}
\def\qbh{\widehat{\qb}}
\def\sbh{\widehat{\sb}}

\def\argmax#1#2{\arg\max_{#1}\left\{#2\right\}}
\def\argmin#1#2{\arg\min_{#1}\left\{#2\right\}}
\def\espx#1#2{\mbox{E}_{#1}\left\{#2\right\}}
\def\esp#1{\mbox{E}\left\{#1\right\}}
\def\d#1{\mbox{~d}#1}

\def\Tr#1{\mbox{Tr}\left[#1\right]}

\def\syslins#1#2#3#4{
\bpic(150,40)
  \put(5,15){\makebox(15,0){#1}}
  \put(20,15){\vector(1,0){10}}
  \put(30,7){\framebox(30,15){#2}}
  \put(60,15){\vector(1,0){10}}
  \put(75,15){\circle{10}}
  \put(75,15){\makebox(0,0){+}}
  \put(80,15){\vector(1,0){10}}
  \put(125,15){\makebox(15,0){#4}}
  \put(75,30){\vector(0,-1){10}}
  \put(68,32){\makebox(30,0){#3}}
\epic
}
\def\syslinsa#1#2#3#4{
\bpic(90,40)
  \put(5,15){\makebox(10,0){#1}}
  \put(15,15){\vector(1,0){10}}
  \put(25,7){\framebox(20,15){#2}}
  \put(45,15){\vector(1,0){10}}
  \put(60,15){\circle{10}}
  \put(60,15){\makebox(0,0){+}}
  \put(65,15){\vector(1,0){10}}
  \put(75,15){\makebox(10,0){#4}}
  \put(60,30){\vector(0,-1){10}}
  \put(60,32){\makebox(10,0){#3}}
\epic
}
\def\inversions#1#2#3{
\begin{picture}(100,40)
  \put(0,15){\makebox(15,0){#1}}
  \put(20,15){\vector(1,0){10}}
  \put(30,7){\framebox(40,15){#2}}
  \put(70,15){\vector(1,0){10}}
  \put(85,15){\makebox(15,0){#3}}
\end{picture}
}

\def\tomoXaz{
\begin{picture}(70,68)
  \put(15,35){\makebox(20,15){$\red{f(x,y)}$}}
  \put(-1,0){\axexy}
  \put(10,10){\objet}
  \put(5,5){\vector(1,1){55}}
  \put(55,55){\makebox(5,10){$r$}}
  \put(35,28){\makebox(5,10){$\phi$}}
  \put(60,8){\makebox(10,15){$\bullet$D}}
  \put(60,0){\makebox(15,10){$\blue{g(r,\phi)}$}}
  \put(10,55){\makebox(10,10){S$\bullet$}}
  \put(62,15){\line(-1,1){45}}
  \put(40,-5){\proj}
\thinlines  \blue{\multiput(10,14)(5,0){9}{\line(0,1){40}}}
\thinlines  \blue{\multiput(8,14)(0,5){9}{\line(1,0){40}}}
\end{picture}
}

\def\tomoXay{
\begin{picture}(70,70)
  \put(15,35){\makebox(20,15){$f(x,y)$}}
  \put(-1,0){\axexy}
  \put(10,10){\objet}
  \put(5,5){\vector(1,1){55}}
  \put(55,55){\makebox(5,10){$r$}}
  \put(35,28){\makebox(5,10){$\phi$}}
  \put(60,8){\makebox(10,15){$\bullet$D}}
  \put(60,0){\makebox(15,10){$g(r,\phi)$}}
  \put(10,55){\makebox(10,10){S$\bullet$}}
  \put(62,15){\line(-1,1){45}}
  \put(40,-5){\proj}
\end{picture}
}

\def\slice{
\begin{picture}(140,70)
  \put(13,35){\makebox(20,15){\red{$f(x,y)$}}}
  \put(35,28){\makebox(5,10){$\phi$}}
  \put(80,10){\makebox(5,10){\blue{$g(r,\phi)$--FT--$G(\Omega,\phi)$}}}
  \put(0,0){\axexy}
  \put(0,0){\axers}
  \put(10,10){\objet}
  \put(35,28){\makebox(5,10){$\phi$}}
  \put(35,-5){\proj}
  \put(80,0){\axewxwy}
  \put(80,0){\axeaw}
  \put(90,30){\makebox(20,15){\red{$F(\omega_{x}, \omega_{y})$}}}
  \put(115,28){\makebox(5,10){$\phi$}}
  \put(96,15){\line(1,1){30}}
\end{picture}
}

\def\cresta{\tt    \setlength{\unitlength}{0.8pt}
\begin{picture}(120,120)
\thicklines   \multiput(0,27)(0,20){5}{\line(1,0){80}}
              \multiput(0,27)(20,0){5}{\line(0,1){80}}
              \put(61,33){\red{$f_N$}}
              \put(0,92){\red{$f_1$}}
              \put(40,72){\red{$f_j$}}
              \put(88,50){\blue{$g_{i}$}}
              \put(38,120){\blue{$H_{ij}$}}
              \put(-5,115){\line(3,-2){90}}
              \put(42,85){\blue{\line(3,-2){20}}}
\end{picture}}

\def\crestb{\tt    \setlength{\unitlength}{0.8pt}
\begin{picture}(120,120)
\thicklines   \multiput(0,27)(0,20){5}{\line(1,0){80}}
              \multiput(0,27)(20,0){5}{\line(0,1){80}}
              \put(61,33){\red{$f_N$}}
              \put(0,92){\red{$f_1$}}
              \put(40,72){\red{$f_j$}}
              \put(0,12){\blue{$g_1$}}
              \put(60,12){\blue{$g_m$}}
              \put(88,33){\blue{$g_{m+1}$}}
              \put(88,75){\blue{$g_i$}}
              \put(88,95){\blue{$g_M$}}
              \put(-5,75){\blue{\line(1,0){90}}}
\end{picture}}

\def\homedir{/home/djafari/}
\def\Picfs{\homedir Tex/Ali/Picfs/}
\def\Epsfs{\homedir Tex/Ali/Eps/}
\def\sfEps{\homedir Matlab/sf2d/Eps/}
\def\decEps{\homedir Matlab/dec1d/Eps/}
\def\imEps{\homedir Matlab/dec2d/Eps/}
\def\matlabdat{\homedir Matlab/dat/}
\def\radonEps{\homedir Matlab/Radon/Eps/}
\def\Physics{\homedir Tex/Ali/Articles/Physics/}
\def\Maxent{\homedir Tex/Ali/Congres/Maxent96/}
\def\RappEps{\homedir Tex/Ali/Rapports/Rapp95/Eps/}
\def\ProbaEps{\homedir Tex/Ali/Proba/Eps/}

\begin{document}
\pagenumbering{roman}
%\maketitle  
\begin{titlepage}
  \begin{center}
    \Huge{\textbf{USING THE NOTION OF COPULA IN TOMOGRAPHY}}
  \end{center}

  \vspace{1.5cm}

  \begin{center}
    \large {\href{http://www.aims.ac.za/~doriano}{\textbf{Doriano-Boris \textsc{POUGAZA}}}} \\
  \end{center}
  \vspace{0.3cm}

  \begin{center}
      \begin{tabular}{cc}
       \small  Master by Research - University of Cergy-Pontoise (FRANCE)
      \end{tabular}
  \end{center}
   \vspace{0.3cm}
   \begin{center}
    \begin{tabular}{c}
August 2008
    \end{tabular}
  \end{center}
  \vspace{3cm}
    %\vfill
  \begin{center}
    \begin{tabular}{c}
 \vspace{ 1 cm}
Supervisors \\

\href{http://djafari.free.fr}{A. Mohammad-Djafari and J-F. Bercher}\\
University of Paris-Sud XI (Orsay) \\
Signal and Syst\`{e}me Laboratory  \\
Inverse Problems Group  \\
UMR 8506 CNRS \\
    \end{tabular}
  \end{center}

\end{titlepage}
\chapter*{Abstract}
\addcontentsline{toc}{chapter}{Abstract}
In 1917, Johann Radon introduced the Radon transform \cite{Rad17} which is used in 1963 by Allan MacLeod Cormack for application in the context of tomographic image reconstruction. He proposed to reconstruct the spatial variation of the material density of the body from X-Ray images  (radiographies) for different directions; He implemented this method and made a test for a prototype Computarized Tomography (CT) scanner \cite{All63}. Independently, Godfrey Newbold Hounsfield derived an algorithm and built the first medical CT scanner. This was a great achievement for the twentieth century, because one can see inside an object without opening it up; Cormack and Hounsfield won the Nobel Prize of Medicine in 1979 for the development of computer assisted tomography.

Basically the idea of the X-ray CT is to get an image of the interior structure of an object by X-raying the object from many different directions. X rays go in straight lines inside the body and its energy is attenuated more and less depending on the density of the matter in his trajectory. So the simplest model relating the log ratio ($\ln I/I_{0}$) of the observed energy $I$ with emitted energy $I_{0}$ to the spatial spatial distribution $f$ of the body is a line integral. The mathematical problem is then estimating a multivariate function $f$ from its line integrals.

Four year before Cormack's idea, Abe Sklar introduced another theory in the context of Statistics called <<copula>>. He gave the theorem which now bears his name \cite{Skl59}. Succinctly stated, copulas are functions that link multivariate distributions to theirs univariate marginal functions. 

One of the problems arising in Statistics is the reconstruction of joint distribution function from its given marginal functions. It appeared that copulas captivated all dependence structure concerning the marginal functions and offer a wide range of parametric family model which could be used as a model for the joint distribution function. This problem is the same as in Tomography, because a marginal density is obtained from a line integral of its joint distribution. The analogy is then that the joint density $f$ has to be reconstructed from its marginals who are obtained by their integration over lines.

To achieve our goal based on the mathematical aspects of tomography in imaging sciences using copulas, we give some prerequisites about copulas and tomography. In the particular case of only given horizontal and vertical projections corresponding to a given two marginal functions, we link the theory of copula to tomography via the Radon transform and Sklar's theorem. 
Let us note that determining the density functions (or the object) from two projections is an ill-posed inverse problem.

Finally, to simulate an image reconstruction we use some members of copulas families (Archimedean, Elliptic) already in widespread use. We have also written a package so called <<copula-tomography>> to handle all those copulas and allow users to simulate a tomographic image reconstruction through copulas. Some preliminary results for a few number of projections are promising.

\tableofcontents
\listoffigures
\addcontentsline{toc}{chapter}{List of Figures}
%--------------------------------------------------------------------------
\newpage
\pagenumbering{arabic}
%-----------------------------------------------------------------------------
% content goes in separate files

\pagestyle{myheadings} 

\chapter{Introduction}
The word <<copula>> originates from the Latin meaning <<link, chain; union>>.
In statistical literature, according to the seminal result in the copula's theory stated by Abe Sklar \cite{Skl59} in 1959; a copula is a function that connects a multivariate distribution function to its given univariate marginal distributions. 

The main point of our report, is bridging the gap between the theory of copula and tomography. 
We consider the inverse problem which is the reconstruction in Computed Tomography (CT) of an object using the Radon transform from only two horizontal and vertical projections. This problem of reconstruction is similar, in the statistical point of view, when it is known or assumed to know the distributions of each random variables but not their joint distribution function, or their copula density.

There is an increasing interest concerning copulas, widely used in Financial Mathematics \cite{Will08}, in modelling of Environmental Data \cite{Joe94}. Recently, in Computational Biology, copulas are used for the reconstruction of accurate cellular networks \cite{Bmc08}. Copula appeared to be a new powerful tool to model the structure of dependence. Copulas are useful for constructing joint distributions, particularly with nonnormal random variables.

What are copulas and how they are related and suitable in the modelling processes of image reconstruction? 

To give more details in order to answer those questions, we organise our report as follows: in the next chapter, we recall some definitions related to multivariate distribution functions and we give the Sklar's theorem, highlighted by some methods to generate a new copula. In the chapter 3, through many illustrations, we show some parametrics family of copulas. We dedicated the chapter 4 on operation yielding to discrete copula via bistochastic matrices. 
We expose some well known statistical methods such that the Maximum Likelihood Estimation (MLE), the Inference Functions for Margins (IFM) and the Bayesian estimation, in the chapter 5. And we give also other methods from literature dealing with the good way to choose the right copulas. In the chapter 6, we start by explanation of the basic mathematical model of X-ray tomography. The top point is discussed in the chapter 7, where we link the theory of copula to tomographic image reconstruction through the Radon Transform. We focus on the case of only two given horizontal and vertical projections which correspond to a given two marginal functions. This is an ill-posed inverse problem because information about the density functions (or the object) to reconstruct is obtained from indirect and limited data. 

Andrei Nikolaevich Tikhonov (Russian mathematician, 1906-1993) said:
\begin{quote}
\texttt{For a long time mathematicians felt that ill-posed problems cannot describe real phenomena and objects. However [...] the class of ill-posed problems includes many classical mathematical problems and, most significantly, that such problems have important applications}.
\end{quote}
For our goal of image reconstruction, some Archimedean family of copula, the Elliptic class of copula in particular the Gaussian copula and mixture of Gaussian density, are used successively for simulation. We present some results we have obtained and a summary of our future work in the last chapter, also a short description concerning the <<copula-tomography>> package in appendix.
%  Introduction 
\chapter{Copula in Statistics}
\section{Multivariate distribution}
In this section, we first give a short summary of definition and properties of a multivariate distribution. We will then focus on copula. We extend the notion of increasing function of one variable to $n$ variables, so called $n$-increasing function, then we introduce the Sklar's theorem and also a method to generate a copula. We assume that our random variables are continuous if necessary.

\subsection{Joint probability density function(pdf)}
Let $X_1,\dots, X_n$ be $n$ continuous random variables \footnote{Concisely  a $n$-dimensional random variables is a function from a sample space $\mathcal{S}$ of an experiment into $\mathbb{R}^{n}.$} all defined on the same probability space. The \textbf{joint probability density function} (pdf) of $X_1,\dots, X_n$, denoted by $f_{\left(X_1,\dots, X_n\right) }(x_1,\dots,x_n)$, is the function
$f: \mathbb{R}^n \to \mathbb{R}$ such that for any domain $D\subset \mathbb{R}^n$ in the $n$-dimensional space of the values of the variables $X_1,\dots,X_n$, the probability that a realisation of the set variables falls inside the domain $D$ is

\begin{equation}
 \text{Prob}({X_1,X_2,\dots,X_n}\in D) =\displaystyle\int_{D} f(x_1,\dots,x_n) dx_1 \dots dx_n.
\end{equation}
\begin{property}
The two following properties are satisfied:
\begin{itemize}
\item $f(x_1,\dots,x_n) \geq 0$ \, , $\forall (x_1,\dots,x_n),$ 
\item $\displaystyle \int_{\mathbb{R}^n} \, f(x_1,\dots,x_n) dx_1\,\dots dx_n = 1.$
\end{itemize}
\end{property}
\subsection{Cumulative distribution function(cdf)}
The cumulative distribution function (cdf) is defined by the formula :
\[F(x_1,\dots ,x_n)=\displaystyle \int^{x_1}_{-\infty} \dots \int^{x_n}_{-\infty} \, f(y_1,\dots,y_n)\, dy_1\,\dots \, dy_n.\]
 There is a relation between the pdf and the cdf:
\begin{equation}
\label{cdfpdf}
 f(x_1,\dots ,x_n)=\dfrac{\partial^{n}F(x_1,\dots ,x_n)}{\partial x_1 \dots \partial x_n}.
\end{equation}

\subsection{Gaussian Mixture(GM) distribution}
The Gaussian mixture (GM) model is one of the most used model in statistics and modelling process (for clustering, or density estimation). It is defined via a Gaussian (or normal) distribution.

Let $\boldsymbol{X}=(X_1,\dots,X_n)^{T}$ be a $n$-dimensional random vector that is multivariate normally distributed, then the probability density function (pdf) is 
\begin{equation}
\label{jointdensitygaussian}
f(\boldsymbol{x} \mid \boldsymbol{\mu},\boldsymbol{\Sigma})
=\frac{1}{(2\pi)^{n/2}|\Sigma|^{1/2}}\exp \left(-\frac{1}{2}
 \left( \boldsymbol{x}-\boldsymbol{\mu}\right)^\top \boldsymbol{\Sigma}^{-1} \left(\boldsymbol{x} - \boldsymbol{\mu}\right) \right), 
\end{equation}
where 
\begin{itemize}
 \item $\boldsymbol{\mu}=\left[\mu_{1},\dots,\mu_{n} \right]^{T}$, is the vector of the mean values $\mu_i = \mathrm{E}(X_i)$,
\item  $\boldsymbol{\Sigma}$ is the covariance matrix, a $n\times n$ non-singular, positive definite real matrix,
with entries  $\Sigma_{ij} =\mathrm{E}\begin{bmatrix}(X_i - \mu_i)(X_j - \mu_j)\end{bmatrix}.$ 
\end{itemize}
When the distribution of $\boldsymbol{X}$ is the multivariate normal distribution, we will use the following notation:
$$\boldsymbol{X} \sim \mathcal{N}(\boldsymbol{\mu},\boldsymbol{\Sigma}).$$

One could also derive the cumulative distribution function (cdf), which is simply
\begin{equation}
\label{cumulativegaussian}
\Phi(x_{1},\dots,x_{n})
=\int^{x_{1}}_{-\infty} \dots\int^{x_{n}}_{-\infty}\frac{1}{(2\pi)^{n/2}\mid\boldsymbol{\Sigma}\mid^{1/2}}\exp [-\frac{1}{2}\left(\boldsymbol{y}-\boldsymbol{\mu}\right)^\top\boldsymbol{\Sigma}^{-1} \left(\boldsymbol{y}-\boldsymbol{\mu}\right)]\,dy_{1}\dots dy_{n}.
\end{equation}

In the 2-dimensional case, the pdf has the form:
\begin{equation}
\nonumber
  f(x_1,x_2) =
\frac{1}{2\pi\sigma_{x_1}\sigma_{x_2}\sqrt{1-\rho^2}}\exp \left(\dfrac{-1}{2 (1-\rho^2)}\left(\dfrac{(x_1-\mu_1)^2}{\sigma_{x_1}^2}+\dfrac{(x_2-\mu_2)^2}{\sigma_{x_2}^2}-\dfrac{2 \rho (x_1-\mu_1) (x_2-\mu_2)}{\sigma_{x_1} \sigma_{x_2}}
 \right)\right), 
\end{equation}
where $\rho$ is the correlation between $x_1$ and $x_2,$ and $\Sigma =
\begin{bmatrix}
\sigma_{x_1}^2 & \rho \sigma_{x_1} \sigma_{x_2}\\
\rho \sigma_{x_1} \sigma_{x_2}  & \sigma_{x_2}^2
\end{bmatrix}.$

For the GM distribution, each of the $K$ components $X_{k}$, is such that $X_{k} \sim \mathcal{N}(\boldsymbol{\mu_{k}},\boldsymbol{\Sigma_{k}})$, with the probabilistic component weights $\alpha_{k}$.
Therefore, the GM pdf has the following form in the case of discrete variables (change the sum to an integral 
for continuous variables):
\begin{equation}
 f\left(\boldsymbol{x}\right)  = \displaystyle\sum^{K}_{k =1} \alpha_{k} \mathcal{N}(\boldsymbol{\mu_{k}},\boldsymbol{\Sigma_{k}}),
\end{equation}
 where $0\leq \alpha_{k} \leq 1\, ,$ $\displaystyle\sum^{K}_{k=1} \alpha_{k}=1,$ and $\boldsymbol{x}=(x_1,\dots,x_n).$ 

\subsection{t-distribution}
The $t-$distribution (or the Student's $t$-distribution) with $\nu$ degrees of freedom has the probability density function :
\begin{equation}
 f(t)=\frac{1}{\sqrt{\nu}\operatorname{B}(\frac{\nu}{2},\frac{1}{2})}
\Big(1+\frac{t^2}{\nu}\Big)^{-\frac{\nu+1}{2}},
\end{equation}
where $\operatorname{B}(\alpha,\beta)$ is the \emph{beta function} which is defined as.
$$B(p,q) = \int_0^1 x^{p-1} (1-x)^{q-1} dx$$ for any real numbers $p,q > 0$.
We will denote $X\sim \operatorname{t}(\nu).$
We could also write the \emph{beta function} in term of the \emph{gamma function},
$$\Gamma(z) = \int_0^\infty \! e^{-t} t^{z-1} \, dt $$
for $z\in \mathcal{C}$ with $\Re(z)>0$, and by analytic continuation for the rest of the complex plane,
except for the non-positive integers, where it has simple poles.
Another equivalent definition is
$$ \Gamma(z) = \frac{e^{-\gamma z}}{z} \prod_{n=1}^\infty \left(1 + \frac{z}{n}\right)^{-1} e^{z/n}, $$
where $\gamma$ is Euler's constant. 
\begin{property}
Remind about properties of the \emph{beta} and \emph{gamma} functions:

$ B(p,q) = \dfrac{\Gamma(p) \Gamma(q)}{\Gamma(p+q)}$ for all complex numbers $p$ and $q$ for which the right-hand side is defined.

Also $B(p,q) = B(q,p)$ and $B\left( {\dfrac{1}{2},\dfrac{1}{2}}\right)  = \pi.$ \\

$\Gamma(p)=(p-1)\Gamma(p-1)$ and for any integer $n \geq 1,$ $\Gamma(n)=(n-1)!$
\end{property}
See \cite{Hav03} for more details.

Therefore, the probability density function of the Student's $t$-distribution is :
\begin{equation}
 f(x) = \frac{\Gamma(\frac{\nu+1}{2})} {\sqrt{\nu\pi}\,\Gamma(\frac{\nu}{2})} \left(1+\frac{x^2}{\nu} \right)^{-(\frac{\nu+1}{2})}.
\end{equation}
\begin{property}
We recall some useful properties of the t-distribution:
\begin{itemize}
\item $\operatorname{t}(1) = \operatorname{Cauchy}(0,1)$, the Cauchy distribution with parameters $0$ and $1$.
\item $\operatorname{t}(\nu) \sim \mathcal{N}(0,1)$ when $\nu$   $ \longrightarrow $ $\infty $.
\item If $X\sim \operatorname{t}(\nu)$, $E\left[ \mid X\mid^k\right] $ exists if and only if $k<\nu$. 
\end{itemize}
\end{property}
The last property means that, there is no mean for a $t-$distribution if $\nu = 1$. 
And for $k=1$, that is $\nu>1$, $\operatorname{E}[X] = 0$; then for the case $\nu >2$, the variance of $X$ is equal to $ \frac{\nu}{\nu-2}$. If $X_1,\dots,X_n $ are random samples from a normal distribution with mean $\mu$ and variance $\sigma^2$. If we denote $\overline{\mu}$, $\overline{\sigma}^2$ respectively the sample mean $\overline{\mu}=\frac{1}{n}\displaystyle\sum^{n}_{k =1}x_i$  and the sample variance $\overline{\sigma}^2=\frac{1}{n}\displaystyle\sum^{n}_{k =1}\left( x_i-\overline{\mu} \right)^{2} $, then
\begin{equation}
\frac{\overline{\mu}-\mu}{\overline{\sigma}} \sqrt{\nu} \quad  \sim \quad   \operatorname{t}(\nu-1).
\end{equation}
One can define the multivariate t-distribution.
Let $\boldsymbol{X}$ be a vector of $n$ variate t-distribution with $\nu$ degrees of freedom and mean vector $\boldsymbol{\mu}$, according to the previous property in the univariate case, we set the mean $\mu > 1$ and the degree of freedom $\nu >2$, therefore the covariance matrix is $\frac{\nu}{\nu-2}\boldsymbol{\Sigma}$. \\
Hence $\boldsymbol{X}$ takes the following form
$$\boldsymbol{X}=\boldsymbol{\mu}+\frac{\sqrt{\nu}}{\sqrt{S}}\boldsymbol{Z}$$ where $\boldsymbol{Z} \sim \mathcal{N}(\boldsymbol{0},\boldsymbol{\Sigma})$ and $S$ is distributed independently of $\boldsymbol{Z}$ and has a $\chi_{\nu}^{2}$ distribution, then the multivariate t-distribution is given by
\begin{equation}
 f(\boldsymbol{x}) = \frac{\Gamma\left( \frac{\nu+n}{2}\right) }{\left(\nu \pi \right)^{n/2}}\,\Gamma\left( \frac{\nu}{2}\right)  \mid \Sigma \mid^{-1/2}\left( 1+\frac{1}{\nu}\left(\boldsymbol{x}-\boldsymbol{\mu}\right)^\top \Sigma^{-1} \left(\boldsymbol{x}-\boldsymbol{\mu}\right)\right) ^{-(\frac{\nu+n}{2})}.
\end{equation}

From the graphs below with  $\nu =2$, $\nu = 20$, we could clearly see that as $\nu$ increases, asymptotically the $t-$ distribution reaches the standard normal distribution.

\begin{displaymath}
% use packages: array
\begin{array}{ll}
 \includegraphics[height=.12\textheight]{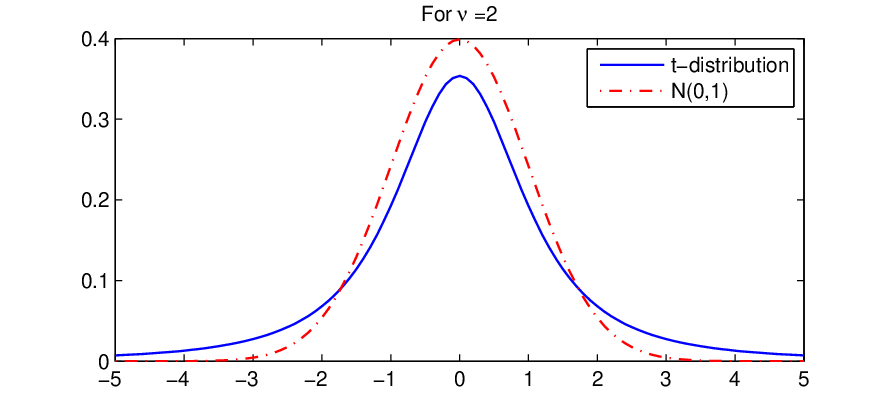}
 % AliCopulaCDF.eps: 1179667x1179664 pixel, 300dpi, 9987.85x9987.82 cm, bb=   87   262   507   578
 & 
 \includegraphics[height=.12\textheight]{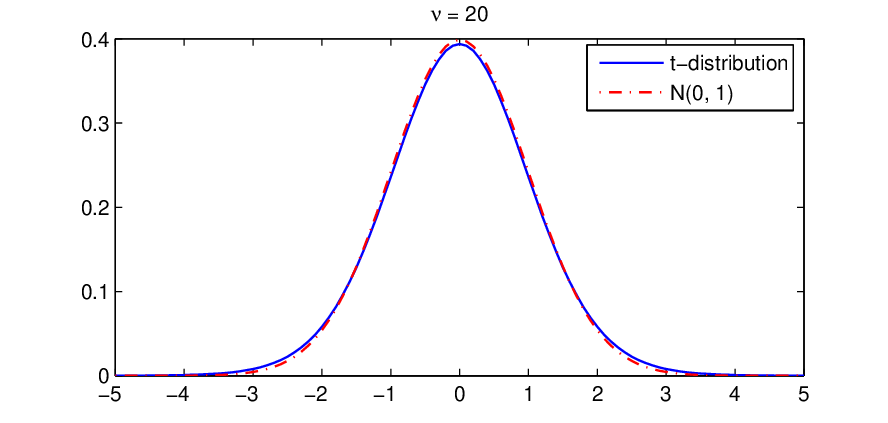}
 % AliCopulaPDF.eps: -3330x0 pixel, 300dpi, -28.19x0.00 cm, bb=   87   262   507   578
\end{array}
\end{displaymath}
\subsection{$n$-dimensional marginal distribution} 
\begin{definition}
Let $\left\lbrace X_i :i \in I\right\rbrace $  be a set of random variables such that the subset $I \subset \left\lbrace 1,2,\dots,n\right\rbrace.$ If we denote $F\left( x_1,\dots,x_n\right)$ the joint cdf defined on $\overline{\mathbb{R}}\,^{n}$. 
The marginal distribution $F_{\left\lbrace X_i : i \in I\right\rbrace}$, is obtained by summing (for discrete variables) 
\[F_{\left\lbrace X_i : i \in I\right\rbrace} \left( x_i\right) \equiv F_{i} \left( x_i\right) = \displaystyle\sum_{\left\lbrace x_i : i \notin I\right\rbrace} F_{X_1,\dots,X_n}\left( x_1,\dots,x_n\right) ,\] 
or integrating (for continuous variables) 
\[F_{\left\lbrace X_i : i \in I\right\rbrace} \left( x_i\right) \equiv F_{i} \left( x_i\right) = \displaystyle\int_{\left\lbrace x_i : i \notin I\right\rbrace }F_{X_1,\dots,X_n}(u_1,\dots,u_n)\prod_{ \left\lbrace u_i : i \notin I\right\rbrace}du_i \] 
over all values of the other variables.
\end{definition}
\begin{remark}
 $F_{1}(x_1)=F(x_1,\infty)$ and $F_{2}(x_2)=F(\infty,x_2)$, for continuous bivariate distribution.
\end{remark}

\bcc
\btabu{cc}
\includegraphics[height=0.18\textheight]{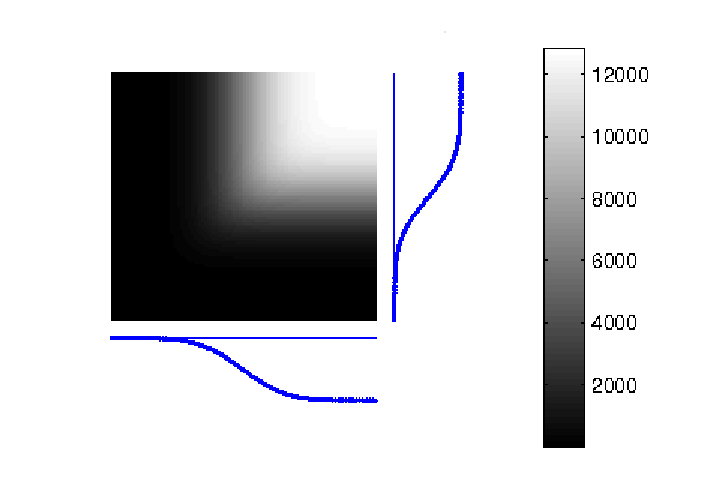}&
\includegraphics[height=0.18\textheight]{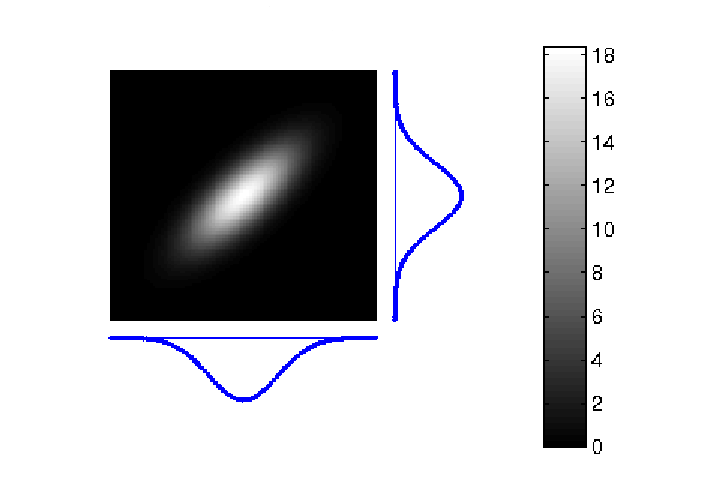}\\
$\Phi(x_1,x_2)$ with marginal cdf's & 
Gaussian pdf with marginal pdf's \\ 
\includegraphics[height=0.18\textheight]{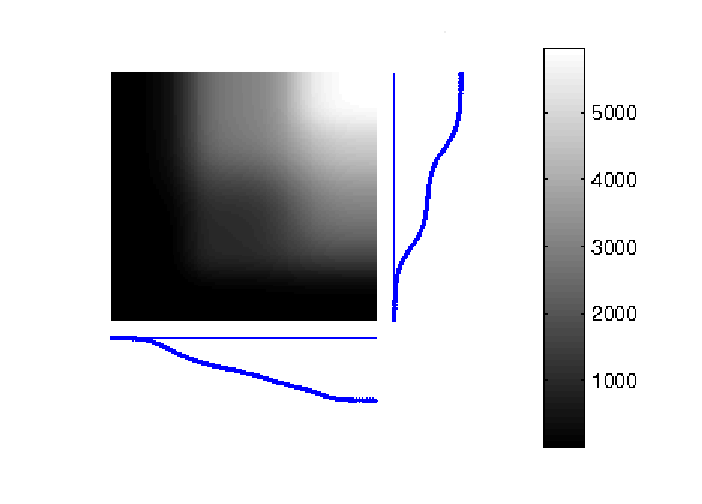}&
\includegraphics[height=0.18\textheight]{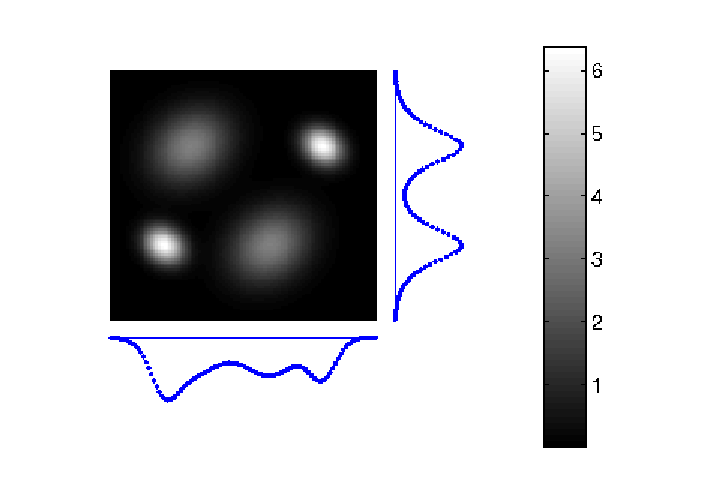}\\
GM cdf with marginal cdf's  & GM pdf with marginal pdf's \\ 
\includegraphics[height=0.18\textheight]{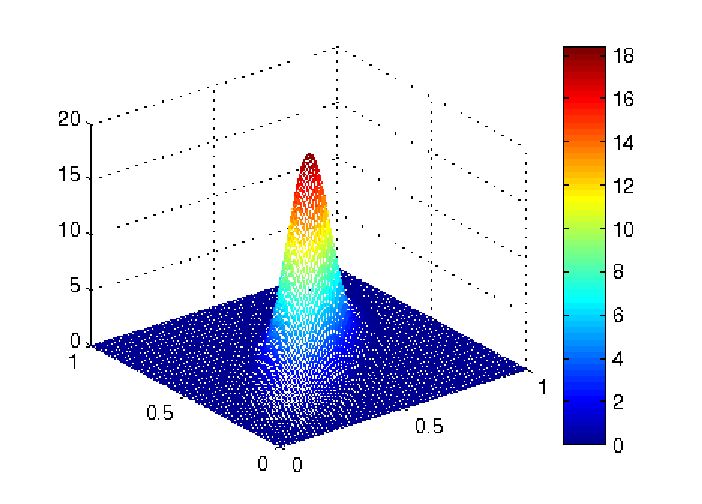}&
\includegraphics[height=0.18\textheight]{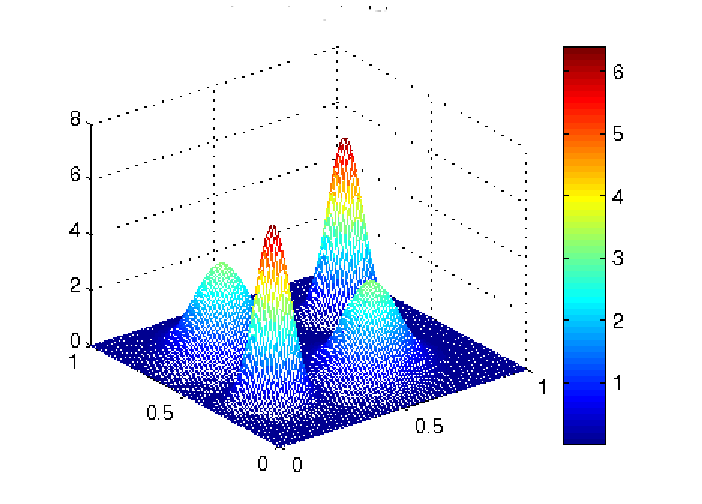}\\
Gaussian pdf mesh plot & GM pdf mesh plot \\
\etabu
\ecc

\section{Sklar's Theorem} 
\subsection{Multivariate Copula}
A $n-$copula (or copula) is a multivariate joint distribution defined on the $n$-dimensional unit cube $ [0,1]^n $ such that every marginal distribution is uniform on the interval  $[0,1]$ .
\begin{definition}
A multivariate copula, or a $n-$copula denoted by $C$, is a function from $[0,1]^n $ to
 $[0,1] $ with the following properties : 
\begin{enumerate}
 \item $ \forall \textbf{u}=(u_1,\dots,u_n) \in \left[ 0,1 \right]^{n}$,
\begin{itemize}
 \item $ C(u_1,\dots,u_n)=0 $, if at least one component of $\textbf{u}$ is equal to zero ,
\end{itemize}
 \item $ \forall \textbf{u}=(u_1,\dots,u_n) \in \left[ 0,1 \right]^{n}$,
\begin{itemize}
 \item $C(1,\dots,1,u_i,1,\dots,1)=u_i $, if all components of $\textbf{u}$ are equal to $1$ except $u_i$ ;
\end{itemize}
\item $ \forall (u_{1}^{(1)},\dots,u_{n}^{(1)}) \in \left[ 0,1 \right]^{n}$, and $\forall (u_{1}^{(2)},\dots,u_{n}^{(2)}) \in \left[ 0,1 \right]^{n} $  such that $u_{i}^{(1)} \leq u_{i}^{(2)}$ \, $ \forall j $
\begin{itemize}
\item $  \displaystyle\sum^{2}_{i_{1}=1} \dots \sum^{2}_{i_{n}=1} \left( -1\right)^{\displaystyle\sum^{n}_{j=1}i_{j}} C(u_{1}^{(i_{1})},\dots,u_{n}^{(i_{n})}) \geq 0.$
\end{itemize}
\end{enumerate}
The last property below is a $n$ dimensional analogue version of an univariate increasing function. Therefore any copula $C(u_1,\dots,u_n)$ is $n-$ increasing function. We can also express this idea in the following way. 
\subsection{C-Volume}
Let $ \textbf{a}=\left(a_1, a_2, \dots, a_n \right)$ and $ \textbf{b}=\left(b_1, b_2, \dots, b_n \right)$.
An $n$-Box, denoted by $B$ is the Cartesian product $\left[a_1, b_1\right] \times \left[a_2, b_2\right] \times \dots \left[a_n, b_n\right] $. Let also denote by $\textbf{c}=\left(c_1, c_2, \dots, c_n \right)$ the vertices of $B$, i.e. $c_k$ is equal to $a_k$ or $b_k$, and $\textbf{a} \leq \textbf{b}$ be equivalent to $a_{k} \leq b_{k}$ \, $\forall$ $k$.
\begin{definition}
Let $S_1, S_2, \dots, S_n$ be nonempty subsets of $\overline{\mathbb{R}}$, and $C$ an $n-$place real \\
function \footnote{function whose domain, $Dom\,C$, is a subset of $\overline{\mathbb{R}}\,^{n}$ and whose range, $Ran\,C$, is a subset of $\mathbb{R}$.} such that $Dom\,C= S_1 \times S_2,\dots , S_n$. Let $B=\left[\textbf{a} ,\textbf{b} \right] $ be an $n-$box  such that all of its vertices are in $Dom\,C.$ Then the $C-$volume  of $B$ is given by 
\begin {equation}
 V_{C}(B)= \sum_{\textbf{c}\in B} sgn(\textbf{c})C(\textbf{c}),
\end {equation}
where  
\begin{equation}
\nonumber
sgn(\textbf{c})=  
\begin{cases}
1,  & \mbox{if  }c_{k}= a_{k} \mbox{  for an even number of } k'\mbox{s}\\
-1, & \mbox{if  }c_{k}= a_{k} \mbox{  for an odd number of  } k'\mbox{s}. 
\end{cases}
\end{equation}
\end{definition}
\end{definition}

We can also express, the $C-$volume of an $n-$box $B$ in the term of the $n$th order difference of $C$ on $B$
\[ V_{C}(B)=\Delta_{\textbf{a}}^{\textbf{b}}C(\textbf{t})=\Delta_{a_{n}}^{b_{n}}\Delta_{a_{n-1}}^{b_{n-1}}\dots \Delta_{a_{2}}^{b_{2}}\Delta_{a_{1}}^{b_{1}}C(\textbf{t}),\]
where the $n$ first order difference of $C$ is 
\[\Delta_{\textbf{a}_{k}}^{\textbf{b}_{k}}C(\textbf{t})=C\left(t_1,\dots,t_{k-1}, b_k,t_{k+1},\dots,t_{n} \right)-C\left(t_1,\dots,t_{k-1}, a_k,t_{k+1},\dots,t_{n} \right).\]

\begin{definition}
\label{Lipschiz}
$C$ is an $n$-increasing function if the $C-$volume of $B$, $V_{C}(B) \geq 0.$ 
\end{definition}
A copula $C$ induces a probability measure on $[0,1]^n $ via $V_{C}([0,1]\times[0,1] \dots [0,1])=C(u_1,\dots,u_n)$ (Carath\'{e}odory's theorem (measure theory)) see \cite{Sch83}.
As consequence of definition \ref{Lipschiz}, any copula $C$ satisfies the following inequality (see \cite{Nel99}).

\fbox{%
\begin{minipage}{\textwidth}
\begin{theorem}
Let $C$ be an $n$-copula. 
Then for every $\textbf{u}=(u_1,\dots,u_n)$ and $\textbf{v}=(v_1,\dots,v_n)$ in $\left[ 0,1 \right]^{n}, $ 
\[\vert C(\textbf{u})-C(\textbf{v}) \vert \leq  \sum^{n} _{k=0}\vert u_{k}- v_{k} \vert .\]
Therefore $C$ is uniformly continuous on $\left[ 0,1 \right]^{n}.$
\end{theorem}
\end{minipage}
}

\subsection{Bivariate Copula}
A bivariate copula, or shortly a copula is a function from  $\left[ 0,1 \right]^{2}$ to $\left[ 0,1 \right]$
\begin{property}
with the following properties from the previous with $n=2.$
\begin{enumerate}
\item $\forall u,v \in \left[0,1\right], \quad C(u,0) = 0 = C(0,v)$; 
\item $\forall u,v \in \left[0,1\right], \quad C(u,1) = u \quad \mbox{and} \quad C(1,v) = v $;
\item $ \forall u_{1},u_{2},v_{1},v_{2} \in \left[ 0,1 \right] \quad \mbox{such that} \quad u_{1} \leq u_{2} \quad \mbox{and} \quad v_{1} \leq v_{2},$ \\
 $V_{C}\left( \left[u_1,u_2 \right]\times\left[v_1,v_2 \right]\right) = C(u_2,v_2)-C(u_2,v_1)-C(u_1,v_2)+C(u_1,v_1) \geq 0.$
\end{enumerate}
\end{property}
To show the last property in the bivariate case, one have to set $u_{1}^{(i)}= u_{i} \quad \mbox{and} \quad u_{2}^{(i)}=v_{i},\, i \in \left\lbrace 1,2 \right\rbrace $ in the last property for the above multivariate case.
And to make sure that this relation is the generalisation of the definition of an increasing univariate function, one can set for example $v_{1}=0$ and $v_{2}=1.$ 
\subsection{Sklar's Theorem}
The Sklar's theorem watertight the theory of copula. The proof of the 2-dimensional case can be found in \cite{Skl59}.
We refer to \cite{Skl96}, for the proof of the $n-$ dimensional version of this theorem. It is the most important result concerning copulas and widely used in all applications.

\fbox{%
\begin{minipage}{\textwidth}
\begin{theorem}
\emph{(Sklar's Theorem)}

Let $F$ be a two-dimensional cumulative distribution function with marginal distributions functions $F_1$ and $F_2.$
Then there \textbf{exists} a copula $C$ such that: 
\begin{equation}
\label{sklar}
 F(x_1,x_2)=C(F_1(x_1),F_2(x_2)).
\end{equation}
Furthermore, if the marginal functions are continuous, then the copula $C$ is \textbf{unique}, and is given by 
\begin{equation}
\label{useful}
 C(u_1,u_2)=F(F_1^{-1} (u_1),F_2^{-1} (u_2)).
\end{equation}
Otherwise, if $F_{1}$ or $F_{2}$ is discontinuous, then $C$ is uniquely determined on $Ran\,F_1 \times Ran\,F_2.$

\textbf{Conversely}, for any univariate distribution functions $F_1$ and $F_2$ and any copula $C$, the function $F$ is a two-dimensional distribution function with marginals $F_1$ and $F_2,$ given by (\ref{sklar}).

\end{theorem}
\end{minipage}
}

Along this report, relation (\ref{useful}) appeared to be very useful in practice.

\fbox{%
\begin{minipage}{\textwidth}
\begin{theorem}
\emph{($n$-dimensional Sklar's Theorem)}

Let $F$ be a joint distribution function with marginals cumulative distribution functions $F_1,\dots,F_n $. Then there
\textbf{exists} a copula $C$ such that  for all $x_1, \dots, x_n$ 
\begin{equation}
\label{sklarn}
 F(x_1,\dots,x_n)=C(F_1(x_1),\dots,F_d(x_n)).
\end{equation}
Furthermore, if $F_1,\dots,F_n $ are continuous functions, then the copula $C$ is \textbf{unique} and for all $\left( u_1, \dots, u_n\right) \in \left[ 0,1\right]^{n},$ 
\begin{equation}
\label{derivationfordensity}
C(u_1,\dots,u_n)=F(F_1^{-1}(u_1),\dots,F_n^{-1} (u_n)).
\end{equation}
Otherwise, if there is at least one marginal discontinuous, then $C$ is uniquely determined on $Ran\,F_1 \times \dots \times Ran\,F_n.$
\textbf{Conversely}, suppose $C$ is a copula and that $F_1(u_1),\dots,F_n(u_n)$ are the univariate cumulative distribution functions. The function $F$ defined as follows:
\begin{equation}
F(x_1,\dots,x_n)=C(F_1(x_1),\dots,F_d(x_n))
\end{equation} 
is a joint distribution function with marginals cumulative distribution functions $F_1(u_1)$,$\dots$,$F_n(u_n).$
\end{theorem}
\end{minipage}
}

From the Sklar's Theorem, one can understand already why copula can be seen as a powerfull tools in modelling dependences of several random variables.
\subsection{Copula Density}
From (\ref{cdfpdf}) and differentiating (\ref{derivationfordensity}) gives the density of a copula
\begin{equation}
\label{density}
c(u_1,\dots,u_n)=\dfrac{\partial^{n}C}{\partial u_1 \dots \partial u_n  }=\dfrac{f\left[F_1^{-1} (u_1),\dots,F_n^{-1} (u_n) \right]}{\displaystyle \prod^{n}_{i=1}f_{i}\left[ F_i^{-1}(u_{i})\right] },
\end{equation}
where $f$ is the joint probability density function of the cdf $F$ and each $f_{i}$ is the marginal density functions
of the marginal cdf  $F_i$. \\
The pdf could be expressed as :
\begin{equation}
\nonumber
f\left[F_1^{-1} (u_1),\dots,F_n^{-1} (u_n) \right]=c(u_1,\dots,u_n)\displaystyle \prod^{n}_{i=1}f_{i}\left[ F_i^{-1}(u_{i})\right],
\end{equation}
Equivalently, since $x_i=F_i^{-1}(u_{i})\longrightarrow u_i=F_i(x_{i})$, it follows that 
\begin{equation}
\label{tomographyequation}
\boxed{f\left(x_1,\dots,x_n\right)=c\left( F_{1}(x_{1}),\dots,F_{n}(x_{n})\right) \displaystyle \prod^{n}_{i=1}f_{i}\left(x_{i}\right).}
\end{equation}
Equation (\ref{tomographyequation}) is central to attend the goal of this report.

\section{Generating Copulas by the Inversion Method}
Given $X_1$ and $X_2$ two random variables. $F(x_1,x_2)$ a joint distribution function and its margins $F_{1}(x_1)$ and $F_{2}(x_2)$, all assumed to be continuous. The corresponding copula can be constructed using the unique inverse  transformations (Quantile transform) $X_1= F^{-1}_{1}(U_1)$, $X_2= F^{-1}_{2}(U_2)$,
where $U_1$, and $U_2$ are uniformly distributed on $\left[0,1 \right]$.\\ 
This method based directly on Sklar's theorem, generates a copula from the following equation \\
\begin{equation} 
\label{inversionmethod} 
 C(u_1,u_2)=F(F^{-1}_{1}(u_1),F^{-1}_{2}(u_2))= F(x_1,x_2),\, \mbox{where} \, u_1, u_2 \, \mbox{are uniform on} \, \left[0,1 \right]. 
\end{equation}

Now let us give an example to illustrate, how to construct a parametric family of copula using inversion method.

Starting with the joint distribution function:
$$F_{\alpha}(x_1,x_2)=\left[(1+e^{-x_1}+ e^{-x_2}+(1-\alpha) e^{-x_1-x_2}\right]^{-1} \quad \mbox{for} \quad x_1,x_2 \in \mathbb{\overline{R}}.$$
By taking limit when $x_2$ and $x_1$ goes to infinity successively in the above expression leads to the marginal distribution functions, which are \\
$$F_{1}(x_1)=F_{\alpha}(x_1,\infty)=(1+e^{-x_1})^{-1}$$ and $$F_{2}(x_2)= F_{\alpha}(\infty,x_2)=(1+e^{-x_2})^{-1}.$$ \\

Simple algebra gives us the inverse transformations  
$$x_1= F^{-1}_{1}(u_1)=\ln\left( \frac{u_1}{1-u_1}\right), $$ and
$$x_2= F^{-1}_{2}(u_2)=\ln\left( \frac{u_2}{1-u_2}\right) .$$\\

Finally after substituting in (\ref{inversionmethod}) yields to:
\begin{align}
\nonumber
  C(u_1,u_2) &=\left[ 1+\frac{1-u_1}{u_1} +\frac{1-u_2}{u_2}+(1-\alpha) \frac{(1-u_1)(1-u_2)}{u_1u_2}\right]^{-1}, \\
\nonumber
  C(u_1,u_2) &= \left[ \frac{1-\alpha(1-u_1)(1-u_2)}{u_1u_2}\right]^{-1}, \\
\label{Ali}
  C(u_1,u_2) &= \frac{u_1u_2}{1-\alpha(1-u_1)(1-u_2)}\quad \mbox{where} \quad \alpha \in \left[-1,1 \right) .
\end{align}

The copula (\ref{Ali}) we have constructed is known as the \textit{Ali-Mikhail-Haq} copula. It was introduced in 1978, see Ali \textit{et al}.\cite{Ali78}.

\begin{remark} 
Construction of new families of copulas by different other existing methods (algebraic or geometric) is an important domain of research. For example as a tool of constructing a new copula, the geometric method takes account only of the definition of copula, without using any given distribution function. In order that the third property (see Definition \ref{Lipschiz}) about copula holds, prior knowledge about the geometric representation is used. Those informations could be the support of the function, the graphical shape of its different sections (horizontal, vertical or diagonal).
\end{remark}

%  Copula in Statistics 
\chapter{Some Families of Copulas}
There are many copulas and each of them has a specific property. In this section, we list few of the widely
used families. We will end by definition of discrete copula associated to a bistochastic matrix, and some statistical methods followed by discussion on the way to choose the right copula. Some graphical representations of copulas are listed in Appendix \ref{app2}. 
\section{Usual Copulas}
The \textbf{product copula} (or independent copula) is the simplest copula, has the form \\
$$\Pi(u_1,\dots,u_n) =u_1\dots u_n \quad \mbox{for all} \quad u_i \in \left[0,1\right],$$ corresponds to independence, therefore it is important as a benchmark. \\
The \textbf{Fr\'echet-Hoeffding upper bound copula} (or comonotonicity copula) is : \\
$$ M(u_1,\dots,u_n)=\min \left\lbrace u_1,\dots, u_n\right\rbrace \quad \mbox{for all} \quad u_i \in \left[0,1\right].$$
And the \textbf{Fr\'echet-Hoeffding lower bound} (or countermonotonicity copula) :
$$ W(u_1,\ldots,u_n) = \max\left\{1-n+\sum\limits_{i=1}^n {u_i}, 0 \right\} \quad \mbox{for all} \quad u_i \in \left[0,1\right].$$
One can easily check that all properties about copula are satisfied by $W$, $\Pi$ and $M.$
\begin{property}
Any copula $C(u_1,\dots,u_n)$, satisfied the inequality called the Fr\'echet-Hoeffding bound inequality
$$W(u_1,\ldots,u_n) \le C(u_1,\ldots,u_n) \le M(u_1,\ldots,u_n).$$
\end{property}
\begin{remark}
The upper bound is a copula for any value of $n$ and the lower bound is a copula for $n=2$. 
For $n > 2$ the lower bound may be a copula under some condition given in $Theorem$ $3.6$ from the book \cite{Joe97}.
\end{remark}

\bcc
\btabu{ccc}
\includegraphics[height=0.15\textheight]{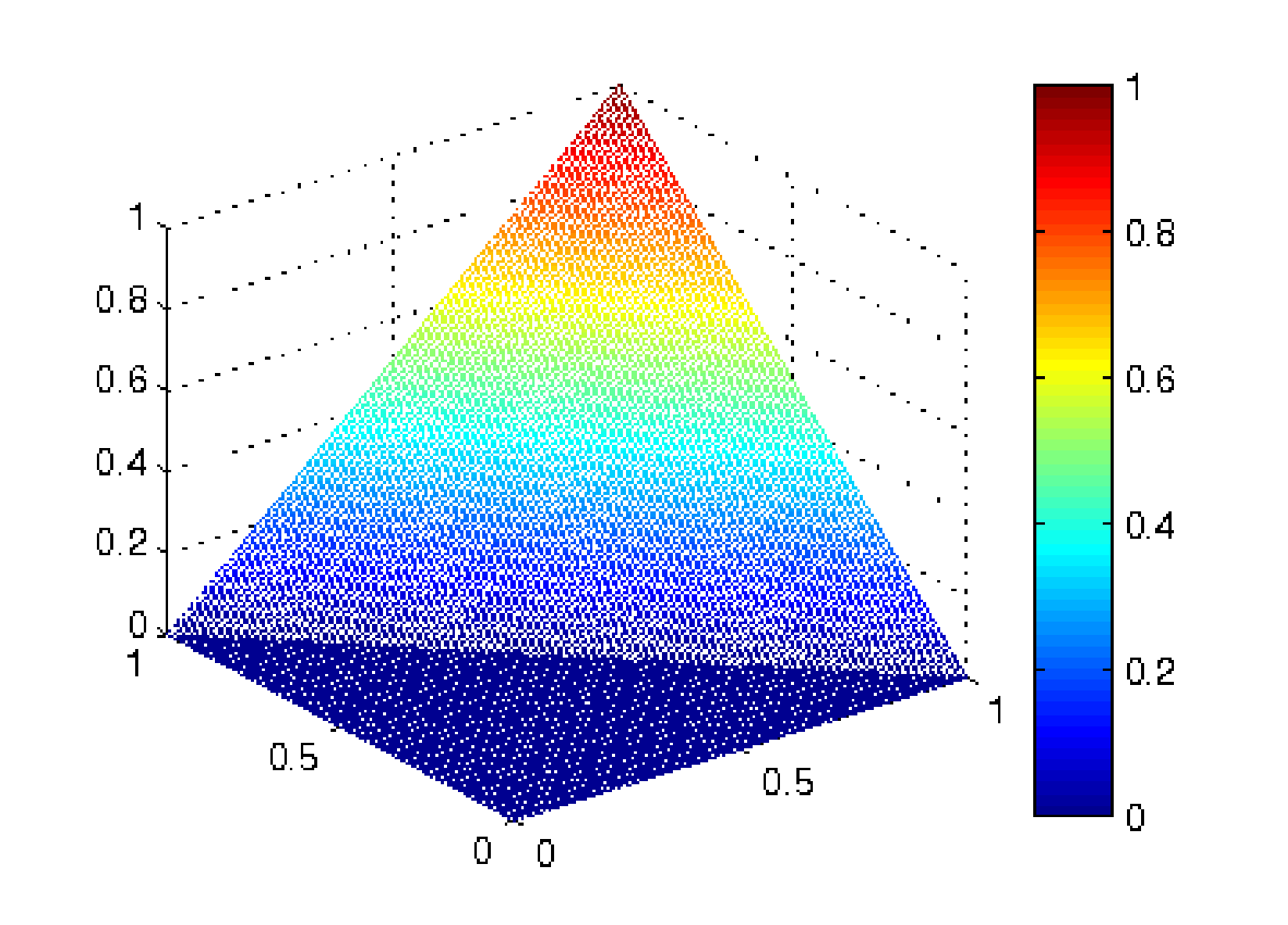}&
\includegraphics[height=0.15\textheight]{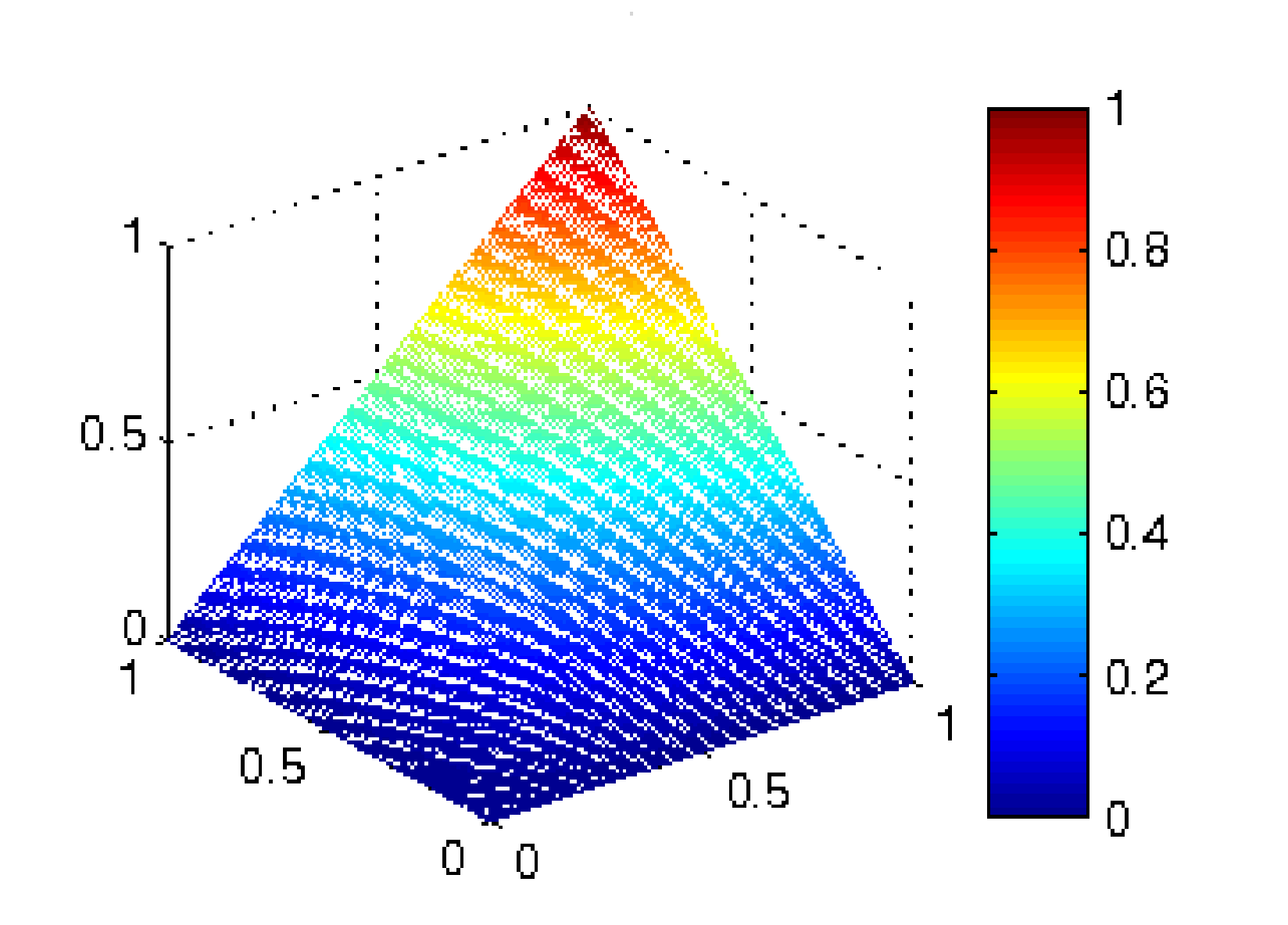}&
\includegraphics[height=0.15\textheight]{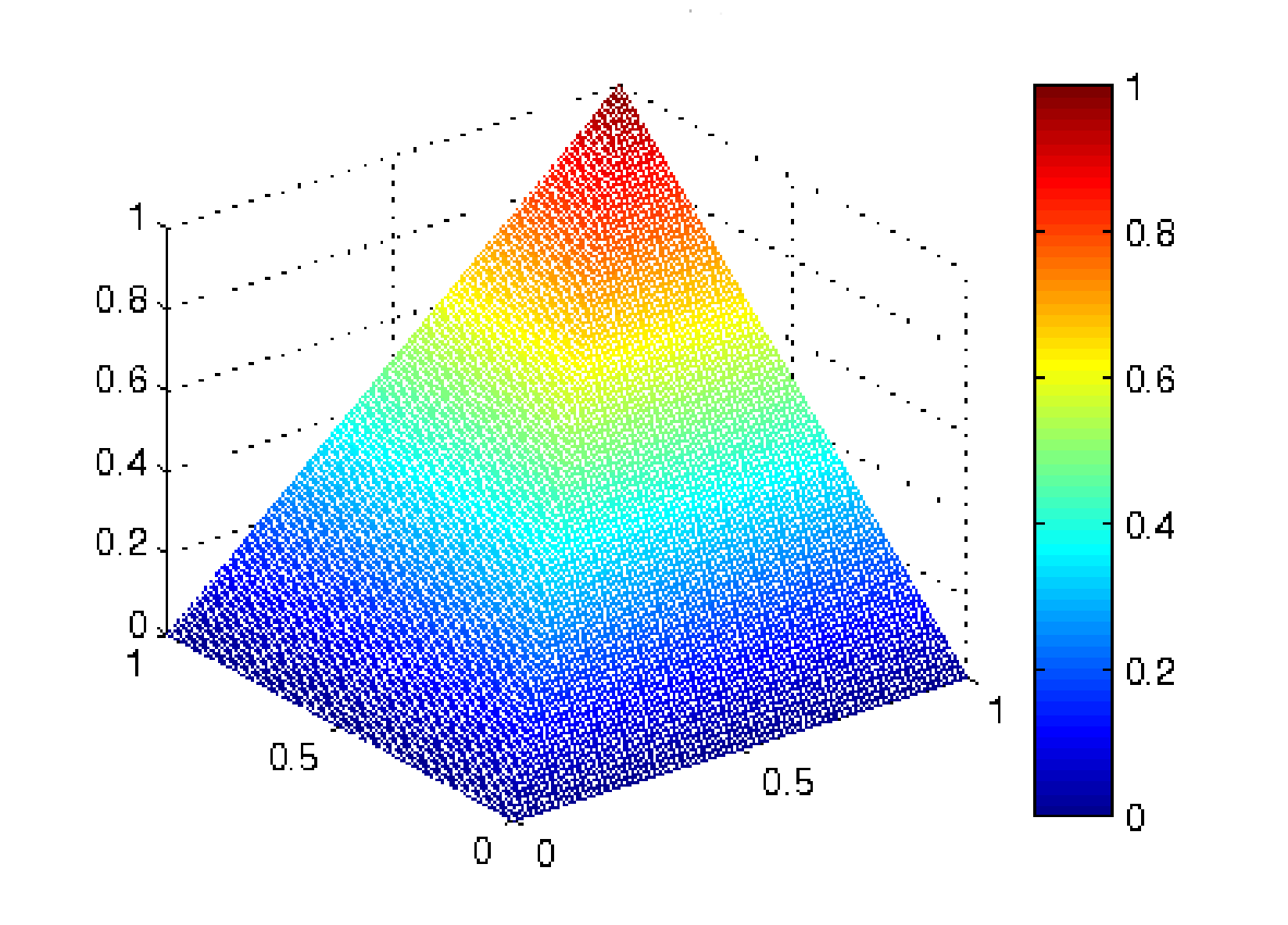}\\
 Lower bound copula $W$ & Independent copula $\Pi$ & Upper bound copula $M$\\ 
\includegraphics[height=0.15\textheight]{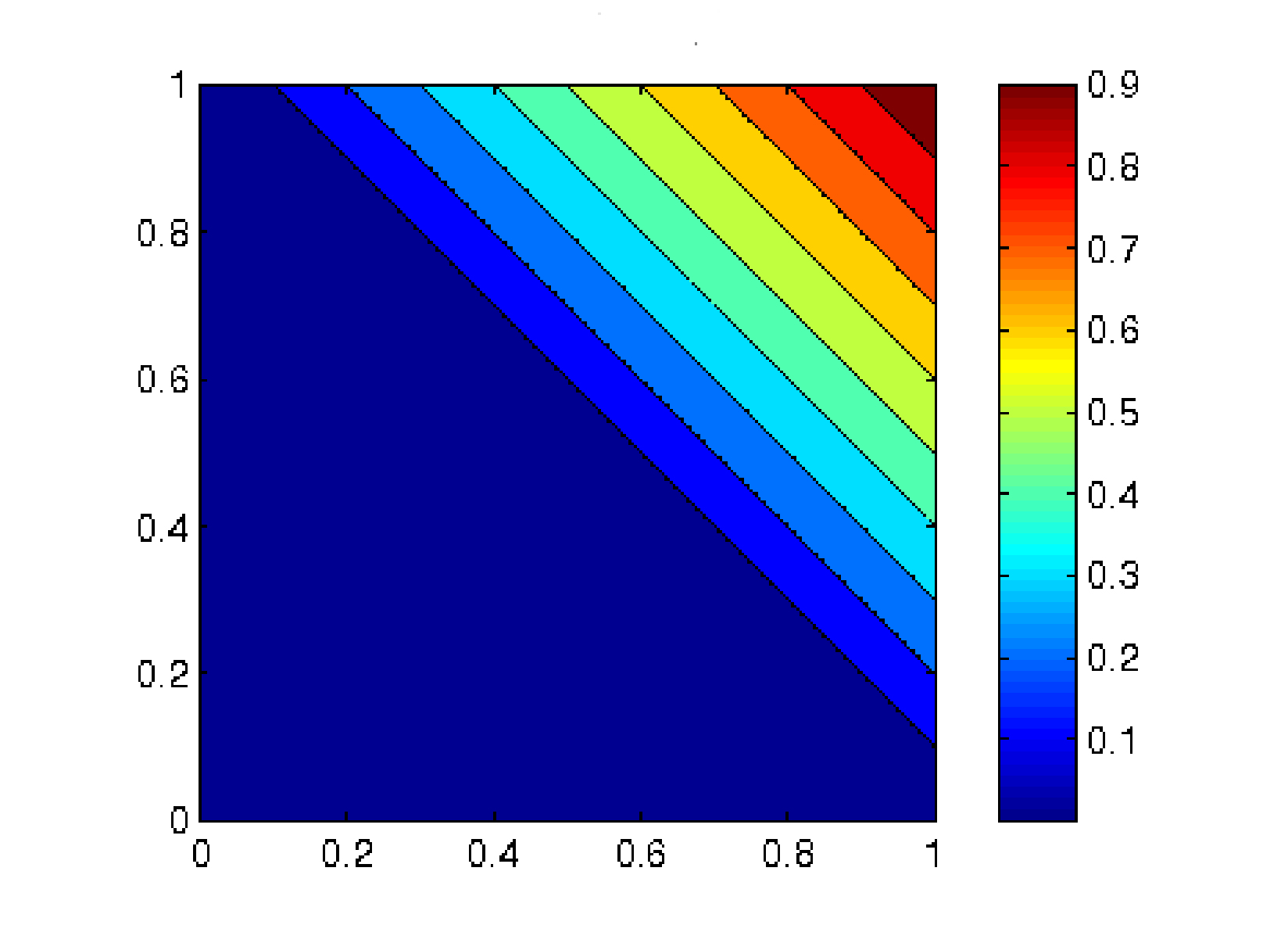}&
\includegraphics[height=0.15\textheight]{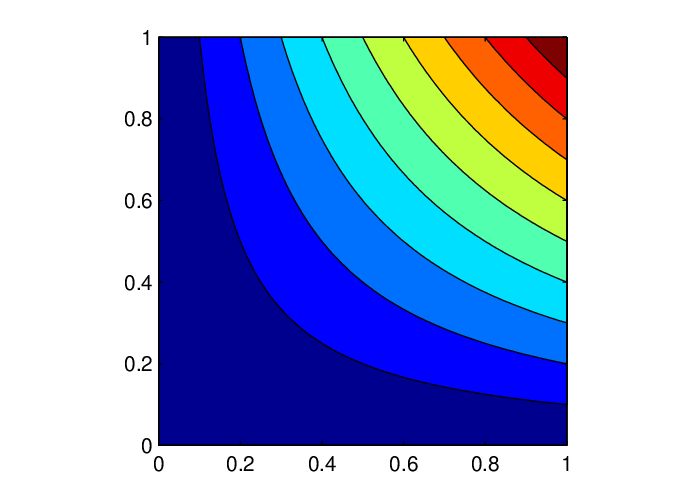}&
\includegraphics[height=0.15\textheight]{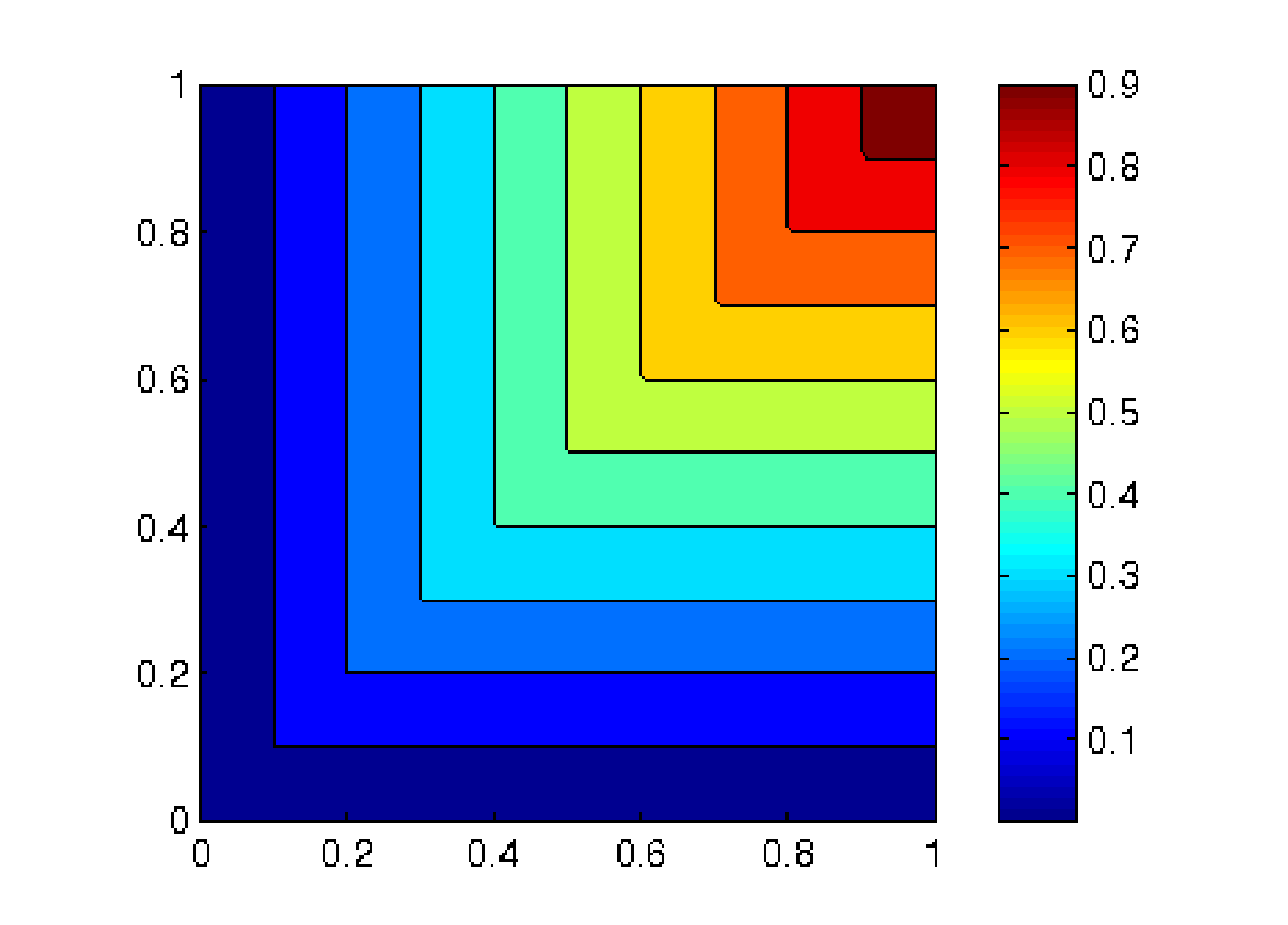}\\
$W$ contour plot & $\Pi$ contour plot & $M$ contour plot \\
\etabu
\ecc

\section{Archimedean Copulas}
The Archimedean copulas is one of the important classes of copulas which has many applications. More details about this section can be found in \cite{Nel99} page $89$.
\begin{definition}
Let $\varphi $ be a continuous, strictly decreasing function from $\left[0,1 \right]$ to $\left[0,\infty \right]$ such that 
$ \varphi(1)=0 $. The \textit{pseudo-inverse} of $ \varphi $ is the function $ \varphi^{\left[-1 \right] } $ with 
$ Dom \, \varphi^{\left[-1 \right] }= \left[0,\infty \right]$ and $ Ran \, \varphi^{\left[-1 \right] }= \left[0,1 \right]$ given by
\begin{eqnarray}
\nonumber
\varphi^{\left[-1 \right] }=\left\lbrace \begin{array}{c}
                   \varphi^{-1}(t)\, \hspace{2cm} 0\leq t \leq \varphi(0) ,\\
\nonumber
 0  \hspace{2cm}  \varphi(0)\leq t \leq \infty . \end{array} \right. 
\end{eqnarray}
Note that $\varphi^{\left[-1 \right] }$ is continuous and decreasing on $\left[0,\infty \right]$, and strictly decreasing on 
$\left[0, \varphi(0)\right].$ Furthermore, $\varphi^{\left[-1 \right] }\left( \varphi(u_1) \right) = u_1 $ on $\left[0,1 \right]$ , and 
\begin{eqnarray}
\nonumber
\varphi \left( \varphi^{\left[-1 \right] }\right) & = &\left\lbrace \begin{array}{c}
                   t\,  \hspace{2cm}   0\leq t \leq \varphi(0) ,\\
\nonumber
 \varphi(0)  \hspace{2cm}  \varphi(0)\leq t \leq \infty . \end{array} \right. \\
& = & \min\left( t, \varphi(0)\right). 
\end{eqnarray}
Finally, if $\varphi(0)= \infty $, then $\varphi^{\left[-1 \right] }=\varphi^{-1 }.$
\end{definition}
\fbox{%
\begin{minipage}{\textwidth}
\begin{theorem}
Let $\varphi $ be a continuous, strictly decreasing function from $\left[0,1 \right]$ to $\left[0,\infty \right]$ such that 
$ \varphi(1)=0 $, and let $ \varphi^{\left[-1 \right] } $ be the \textit{pseudo-inverse} of $ \varphi $. Let $C$ be the function from $\left[0,1 \right]^{2}$ to $\left[0,1 \right]$ given by 
\begin{equation}
\label{archimedean}
 C(u_1,u_2)= \varphi^{\left[-1 \right] }\left( \varphi(u_1)+ \varphi(u_2)\right).
\end{equation}
Then $C$ is a copula if and only if $\varphi$ is convex.
\end{theorem}
\end{minipage}
}

Archimedean copulas are in the form (\ref{archimedean}) and the function $\varphi$ is called the generator of the copula. $\varphi$ is a strict generator if $\varphi(0)=\infty$, then $\varphi^{\left[-1 \right] }=\varphi^{-1}$ and the relation \[C(u_1,u_2)= \varphi^{-1 }\left( \varphi(u_1)+ \varphi(u_2)\right)\] gives a strict Archimedean copula.

\begin{property}
The following algebraic properties are satisfied by any Archimedean copula $C$, those properties distinguish this class
of copula from all other copula.
\begin{enumerate}
 \item $C(u_1,u_2)=C(u_2,u_1)$ meaning that $C$ is symmetric $\forall \, u_1,u_2 \in \left[0,1 \right];$
 \item $C$ is associative $\forall \,u_1,u_2,u_3 \in \left[0,1 \right]$ i.e. $C(C(u_1,u_2),u_3)=C(u_1,C(u_2,u_3));$
 \item If $a > 0$ is any constant then $a\varphi$ is again a generator of $C$.
\end{enumerate}
\end{property}
From the relation (\ref{archimedean}), we clearly see that all information about 
the dependence structure of the multivariate copula is reduced only to the study of the univariate generator $\varphi$.
This characteristic of the Archimedean family made them more attractive. \\

The following theorem gives a technique to find a generator of an Archimedean copula. \\
\fbox{%
\begin{minipage}{\textwidth}
\begin{theorem}
Let $C$ be an Archimedean copula with generator $\varphi$ in $\Omega$.
Then for almost all $u_1$ and $u_2$ in $\left[ 0,1\right]$,
\begin{equation}
 \varphi^{'}(u_1)\dfrac{\partial C(u_1,u_2)}{\partial u_2}=\varphi^{'}(u_2)\dfrac{\partial C(u_1,u_2)}{\partial u_1}.
\end{equation}
\end{theorem}
\end{minipage}
}

\begin{definition}
If $F(x_1,x_2,\cdots,x_n)$, and $F_i(x_i)$ denoted respectively the multivariate distribution and its marginal functions, one particularly simple form of a $n-$dimensional Archimedean is
$$ F(x_1,x_2,\cdots,x_n) = \varphi^{-1}\left( \sum_{i=1}^n\varphi(F_i(x_i)) \right) \, ,$$ 
where $\varphi $ is the generator function such that $\varphi (1) = 0, \, \varphi(0) = \infty$;  and satisfies the convexity properties $ \varphi^{'}(x) < 0, \, \varphi^{''} (x) > 0. $ 
\end{definition}
One easy way to compute the bivariate copula density function $c(u_1,u_2)$ of the copula $C(u_1,u_2)$, using the generator function $\varphi$ under some conditions is given by:
\begin{equation}
 c(u_1,u_2)=-\dfrac{\varphi^{''} (C(u_1,u_2))\varphi^{'} (u_1)\varphi^{'} (u_2)}{\left[\varphi^{'} (C(u_1,u_2)) \right]^{3}}.
\end{equation}

\begin{remark}
Other rigorous mathematics way to define the Archimedean copula is related to the Laplace transform (for details and beauty of this method, we refer to \cite{Mar88}). \\
Let $\Lambda$ be a distribution function with support $\mathbb{R}_{+}$ and $\varphi$ its Laplace transform,
$$\varphi(t)= \int_{0}^{\infty} \exp(-tx)\,\Lambda(dx),$$ 
$\varphi$ is strictly nondecreasing function, $ \varphi(0)=1, \, \varphi(+\infty)= 0,$ then the following
relation define a copula
$$C(u_1,\dots, u_n) = \varphi\left( \sum_{i=1}^n\varphi_{i}^{-1}(u_i)) \right).$$ 
\end{remark}
\subsection{Independent copula}  
For all $u_1, u_2 \in \left[ 0,1\right],$ $\quad$  $\Pi(u_1,u_2)=u_1u_2,$  with the generator $\varphi(x) = -\ln(x).$
And its copula density is identically equal to one.

\subsection{Gumbel copula}
This copula was originally studied by Gumbel in 1960 ( see \cite{Gum60}) \\
$$ C_{\alpha}(u_1,u_2)= \exp\left(-\left[ \left(-\ln u_1 \right)^{\alpha} + \left(-\ln u_2 \right)^{\alpha}\right]^{ \frac{1}{\alpha}} \right),$$ with the generator $\varphi(x)= (-\ln(x))^\alpha \,$ and $\alpha \in \left[1,\infty \right).$  \\
And the density has the following form:
\begin{equation}
\nonumber
\begin{split}
c_{\alpha}(u_1,u_2)& = \\ 
 &(-\ln u_1)^{-1+\alpha} \left[-1+\alpha+( -\ln u_1)^{\alpha}+  (-\ln u_2 )^{\alpha}\right] ^{\frac{1}{\alpha}}
\left[ (-\ln u_1 )^{\alpha} + (-\ln u_2)^{\alpha} \right] ^{-2+\frac{1}{\alpha}}\left(-\ln u_2 \right)^{-1+\alpha} \\ 
&\times \frac{1}{\exp \left\lbrace (-\ln u_1 )^{\alpha} + (-\ln u_2 )^{\alpha} \right\rbrace^{1/\alpha}u_1u_2 }.
\end{split}
\end{equation}

\subsection{Clayton copula}
The following copula family was discussed in 1978 by Clayton \cite{Cla78} : \\ 
\[ C_{\alpha}(u_1,u_2)=\max\left(\left[u^{-\alpha}_{1}+u^{-\alpha}_{2}-1\right]^\frac{-1}{\alpha},0\right),\alpha \in \left[-1,\infty\right)\diagdown\left\lbrace  0\right \rbrace. \] 
The generator is $\varphi(t)=\frac{1}{\alpha}\left(t^{-\alpha}-1\right)$. \\
We compute its density which is given by: 
\[ c_{\alpha}(u_1,u_2)=\left(1+ \alpha \right) u^{-1-\alpha}_{1}u^{-1-\alpha}_{2}\left(-1+u^{-\alpha}_{1}+u^{-\alpha}_{2}\right)^{-2-\frac{1}{\alpha}}.\]

For $\alpha = 0$ the random variables are statistical independent, since $C_{0}(u_1,u_2)= \Pi.$ \\
But for large value of $\alpha$, we have $C_{\infty}(u_1,u_2)= M$ and the limiting case $\alpha = -1$, yields to $C(u_1,u_2)= W.$

\subsection{Ali-Mikhail-Haq copula}
Ali-Mikhail-Haq family:
\[C_{\alpha}(u_1,u_2)= \dfrac{u_1u_2}{1 -\alpha(1-u_1)(1-u_2) },\quad \alpha \in \left[ -1,1 \right),\]
and the generator function is $\varphi(x)= \ln(\frac{1-\alpha(1-x)}{x}).$
Its density is given by :
\[c_{\alpha}(u_1,u_2)= \dfrac{-1+\alpha^{2} \left(-1+u_1+u_2-u_1u_2\right)-\alpha\left(-2+u_1+u_2+u_1u_2\right)}{\left[ -1+\alpha\left(-1+u_1\right)\left(-1+u_2\right)\right] ^{3}}.\]

\subsection{Frank copula}
Frank copula has the following form (see \cite{Fra79}):
\[C_{\alpha}(u_1,u_2)=-\frac{1}{\alpha} \ln \left( \dfrac{1}{1-e^{-\alpha}}\left[ (1-e^{-\alpha})-(1-e^{-\alpha u_1})(1-e^{-\alpha u_2}) \right] \right),\] where $ \alpha \neq 0 $ and the generator is 
$\varphi(x)= \ln\left( \frac{e^{\alpha x} -1}{e^{\alpha} -1}\right).$ \\
Its density is 
\[c_{\alpha}(u_1,u_2)= \dfrac{\alpha \exp \left[\alpha \left(1+u_1+u_2 \right)\right]\left[-1+ \exp(\alpha) \right] }{    \left\lbrace \exp(\alpha) -\exp\left[ \alpha (1+u_1)\right] - \exp\left[ \alpha (1+u_2)\right]+ \exp \left[ \alpha (u_1+u_2)\right] \right\rbrace^{2} } .\]

\subsection{Joe Copula}
Joe copula \cite{Joe93} has the form of 
\[C_{\alpha}(u_1,u_2)= 1-\left[\left(1-u_1 \right) ^{\alpha}+\left(1-u_2 \right) ^{\alpha}-\left(1-u_1 \right) ^{\alpha} \left(1-u_2 \right) ^{\alpha}   \right] ^{\frac{1}{\alpha}}, \, \mbox{where} \, \alpha \in \left[ 1,\infty \right) ,\]
and the generator function is $\varphi(x)= -\ln(1-(1-x)^\alpha).$

Its corresponding density is :
\begin{equation}
\nonumber
\begin{split}
c_{\alpha}(u_1,u_2)&= \left(1-u_1 \right) ^{-1+\alpha}\left\lbrace \alpha- \left[-1+\left(1-u_1\right)^{\alpha} \right] \left[-1+\left(1-u_2\right)^{\alpha} \right] \right\rbrace \\
&\times \left[\left(1-u_1 \right) ^{\alpha}+ \left(1-u_2 \right) ^{\alpha}- \left(1-u_1 \right) ^{\alpha}\left(1-u_2 \right) ^{\alpha}\right]^{-2+\frac{1}{\alpha}} \left(1-u_2 \right) ^{-1+\alpha}.
\end{split}
\end{equation}

\section{The Farlie-Gumbel-Morgenstern family}
Farlie-Gumbel-Morgenstern (shortly ``FGM'') copula was introduced in the basic functional following form by Eyraud \cite{Eyr38} in 1938 and it was also discussed by Morgenstern  \cite{Mor56} in 1956 :
\[C_{\alpha}(u_1,u_2)=u_1u_2 \left(1 +\alpha(1-u_1)(1-u_2) \right),\quad \mbox{where} \quad -1 \leq \alpha \leq 1. \] 
For $\alpha \neq 0 $, the FGM copula felt to be associative, then this family is not an Archimedean copula.
One can check, for example 
$$ C_{\alpha}\left(\frac{1}{4},C_{\alpha}(\frac{1}{2},\frac{1}{3})\right) \neq C_{\alpha}\left( C_{\alpha}(\frac{1}{4},\frac{1}{2}),\frac{1}{3}\right).$$
FGM copula could be seen as a perturbation of the product copula, because the special case $\alpha=0$, leads to  $\Pi$ the only Archimedean member.

Its density is given by  \[c_{\alpha}(u_1,u_2)= 1+ \alpha (1-2u_1)(1-2 u_2) .\]

\section{Copula with cubic section}
Copula with cubic section \footnote{meaning that for given $u_1$ the function $C(u_1,u_2)$ is cubic in $u_1$ and vice versa.} is defined by Nelsen \cite{Nel99} and has the form:
\[C_{\alpha,\beta}(u_1,u_2)=u_1u_2 + u_1u_2(1-u_1)(1-u_2)\left[\beta+\alpha(1-2u_1)(1-2u_2) \right], \] 
where $\alpha,\beta$ are real constants such that $\alpha \in \left[-1,2 \right]$ and satisfy the conditions 

$\mid \beta \mid \leqslant \alpha +1$ for $\alpha \in \left[-1,1/2 \right],$ and 
$\mid \beta \mid \leqslant (6\alpha -3\alpha^{2})^{\frac{1}{2}}$ for $\alpha \in \left[1/2,2 \right].$

In order to have a nice representation, we consider the simple case where $\beta =0,$
\[C_{\alpha}(u_1,u_2)=u_1u_2 + \alpha u_1u_2(1-u_1)(1-u_2)(1-2u_1)(1-2u_2).\] 
Its density is given by
\[c_{\alpha}(u_1,u_2)= 1+ \alpha (1-6u_1+6u^{2}_{1})(1-6u_2+6u^{2}_{2}) .\]

\section{Elliptical copulas}
Elliptical copulas are the copula of elliptical distributions. We consider the examples of Gaussian distribution and t-distribution.
\subsection{Gaussian Copula}
The  \textbf{multivariate Gaussian copula } is the copula of $n$-dimensional random vector that is multivariate normally distributed. This copula was proposed in 1983 by Lee \cite{Lee83}
\begin{equation}
\label{gaussianCopula}
 C_{\rho}(u_1,\cdots, u_n)  = \Phi_{\Sigma} \left(\Phi^{-1}(u_1),\cdots, \Phi^{-1}(u_n) \right), \\ 
\end{equation}
where $\Phi_{ \Sigma}$ is the multivariate cdf (\ref{cumulativegaussian}) and 
\begin{equation}
\Phi(x)
=\displaystyle\int^{x}_{-\infty} \frac
 {1}
 {(2\pi)^{1/2}}
\exp
\left(
 -\frac{1}{2}
  t^{2}\right) \, dt
\end{equation} 
The bivariate Gaussian copula with the covariance matrix.
$$\Sigma=
\begin{pmatrix}
  1 & \rho \\
  \rho & 1 
\end{pmatrix} $$ with Pearson's product-moment correlation coefficient $\rho$, therefore we have explicitly
\begin{equation}
\label{gaussianCopulaexample} C_{\rho}(u_1,u_2)=\displaystyle\int^{\Phi^{-1}(u_1)}_{-\infty}\displaystyle\int^{\Phi^{-1}(u_2)}_{-\infty}\frac{1}{2\pi\sqrt{1-\rho^{2}}} \exp \left\lbrace \dfrac{-(x^{2}-2\rho x y+ y^{2})}{2(1-\rho^{2})}\right\rbrace dx\,dy.
\end{equation}
The limits cases $\rho =-1, 0, 1 $ correspond respectively to the $W(u_1,u_2)$ , $\Pi(u_1,u_2)$ and $M(u_1,u_2)$ copulas.
Let us show the easy case $\rho=0$.
\begin{eqnarray}
\nonumber C_{\rho}(u_1,u_2)&=&\displaystyle\int^{\Phi^{-1}(u_1)}_{-\infty}\displaystyle\int^{\Phi^{-1}(u_2)}_{-\infty}\frac{1}{2\pi} \exp \left\lbrace \dfrac{-x^{2}-y^{2}}{2}\right\rbrace dx\,dy \\
\nonumber
&=&\frac{1}{\sqrt{2\pi}}\displaystyle\int^{\Phi^{-1}(u_1)}_{-\infty}\exp \left\lbrace \dfrac{-x^{2}}{2}\right\rbrace dx \, \frac{1}{\sqrt{2\pi}}\displaystyle\int^{\Phi^{-1}(u_2)}_{-\infty} \exp \left\lbrace \dfrac{-y^{2}}{2}\right\rbrace dy \\
\nonumber
&=& \Phi\left( \Phi^{-1}(u_1)\right) \, \Phi\left( \Phi^{-1}(u_1)\right) =u_1u_2=\Pi(u_1,u_2).
\end{eqnarray}

In the multivariate case, from (\ref{density}) the multivariate \textbf{Gaussian copula density} is
\begin{equation}
\nonumber
c_{\rho}(u_1,\dots,u_n)=|\Sigma|^{-\frac{1}{2}}\exp\left\lbrace \frac{1}{2}\boldsymbol{\Psi}^\top\left(\boldsymbol{I}_{n}- \boldsymbol{\Sigma}^{-1}\right)\boldsymbol{\Psi}\right\rbrace, 
\end{equation}
where $\boldsymbol{\Psi}=\left( \Phi^{-1}(u_1),\dots,\Phi^{-1}(u_n) \right)^\top$ and $ \boldsymbol{I}_{n}$ is an $ n\times n$ identity matrix.\\
The density in the especial case of a bivariate normal copula has the form
\begin{equation}
\nonumber
c_{\rho}(u_1,u_2)=\frac{1}{\sqrt{1-\rho^{2}}}\exp(\frac{[\Phi^{-1}(u_1)]^2+[ \Phi^{-1}(u_2)]^2}{2})\exp\left( \frac{2\rho\Phi^{-1}(u_1)\Phi^{-1}(u_2)-[ \Phi^{-1}(u_1)]^2-[\Phi^{-1}(u_2)]^2}{2(1-\rho^{2})}\right)  
\end{equation}

\subsection{t-copula}
The multivariate \textbf{Student copula or t-copula} with two parameters  ${\rho}$ and ${\nu}$ is also inducted in the elliptical family copula, it is defined through the multivariate t-distribution as 
\begin{equation}
  C_{\rho,\nu}(u_1,\dots, u_n)  = t_{\Sigma,\nu}\left(t_{\nu}^{-1}(u_1), \dots, t_{\nu}^{-1}(u_n) \right) 
\end{equation}
where
\begin{equation}
\nonumber
t_{\nu}(x)
=\displaystyle\int^{x}_{-\infty} \frac{\Gamma(\frac{\nu+1}{2})} {\sqrt{\nu\pi}\,\Gamma(\frac{\nu}{2})} \left(1+\frac{t^2}{\nu} \right)^{-(\frac{\nu+1}{2})} \, dt
\end{equation}
\begin{equation}
\label{cummulativestudent}
t_{\Sigma,\nu}(x_{1},\dots,x_{n})
=\displaystyle\int^{x_{1}}_{-\infty} \dots \displaystyle\int^{x_{n}}_{-\infty} \frac{\Gamma(\frac{\nu+n}{2})}{(\nu \pi)^{n/2}\,\Gamma(\frac{\nu}{2})}\left(1+\frac{\boldsymbol{y}^\top \boldsymbol{\Sigma}\,\boldsymbol{y} }{\nu} \right)^{-(\frac{\nu+n}{2})}
\, dy_{1}\,\dots\,dy_{n} .
\end{equation}
In the bivariate case,
\begin{equation}
\nonumber
 C_{\rho}(u_1,u_2)=\displaystyle\int^{t_{\nu}^{-1}(u_1)}_{-\infty}\displaystyle\int^{t_{\nu}^{-1}(u_2)}_{-\infty}\frac{1}{2\pi\sqrt{1-\rho^{2}}} \exp \left\lbrace 1+ \dfrac{x_{1}^{2}-2\rho x_{1} x_{2}+ x_{2}^{2}}{\nu (1-\rho^{2})}\right\rbrace^{-(\frac{\nu+2}{2})} dx_{1}\,dx_{2}.
\end{equation}
The corresponding multivariate \textbf{t-copula density} is given by
\begin{equation}
\nonumber
  c_{\rho,\nu}(u_1,\dots, u_2)  = \frac{\Gamma(\frac{\nu+n}{2})}{(\nu \pi)^{n/2}\,\Gamma(\frac{\nu}{2})}\left(1+\frac{\boldsymbol{Y}^\top \Sigma \boldsymbol{Y} }{\nu} \right)^{-(\frac{\nu+n}{2})} \,\mbox{where}\,\boldsymbol{Y}=\left(t_{\nu}^{-1}(u_1),\dots,t_{\nu}^{-1}(u_n) \right)^\top.
\end{equation}
For $\nu <3 $, the variance does not exist and for $\nu<5$, the fourth moment does not exist.\\
As $\nu \rightarrow \infty $, the t-copula $C_{\rho,\nu}(u_1,\dots, u_2)$ is asymptotically equal to the Gaussian copula $C_{\rho}(u_1,\cdots, u_n)$.

\subsection{Advantages of copula in Modelling}
Copulas provide greater flexibility in that they allow us a much wider range of possible dependence structures. 
Imagine we have a set of marginals of a given type (e.g., normal). The classical representation only allows us one possible type of dependence structure, a multivariate version of the corresponding univariate distribution (e.g., a multivariate normal, if our marginals are normal). However, copulas still allow us the same dependence structure if we wish to apply it (i.e., through a Gaussian copula), but also allow us a great range of additional dependence structures (e.g., through Archimedean copulas). 
These advantages imply that copulas provide a superior approach to the modelling of multivariate statistical problems.
%  Some Families of Copulas 
\chapter{Copula And Algebra}
Here we present a relation between copula and matrices operations. For more discussion and rigorous proof about this section, we refer to \cite{Dar92}.
\section{The *-product Of Copulas}
The notion of $*$ product of copula was introduced by Darsow \textit{et  al.} in the context of Markov processes.

Let $C=C(u,v)$ be a copula and $D_{1}C=\dfrac{\partial C}{\partial u},\,$ $D_{2}C=\dfrac{\partial C}{\partial v}$ are two-first order partial derivatives.
\begin{definition}
Let $C_{1}$ and $C_{2}$ be two copulas and define the $*-$product of $C_{1}$ and $C_{2}$ as: 
\begin{eqnarray} 
\nonumber
 C_{1}*C_{2} : \left[0,1 \right]^{2} &\longrightarrow &\left[0,1 \right] \\
\nonumber
            (u,v)   & \longmapsto & \left(  C_{1}*C_{2}\right)(u,v) = \int_{0}^{1}D_{2}C_{1}(u,t).D_{1}C_{2}(t,v)dt
\end{eqnarray}
\end{definition}
\fbox{%
\begin{minipage}{\textwidth}
\begin{theorem}
 Given $C_{1}$, $C_{2}$ two copulas, the product $C_{1}*C_{2}$ is a copula.
\end{theorem}
\end{minipage}
}
\begin{proof}
The two first properties about copulas are obvious since, it suffices to apply the definition of the product of copula. To prove the third property, we compute :
\begin{align}
\nonumber
V_{(C_{1}*C_{2})}\left(\left[u_{1},u_{2}\right]\times\left[v_{1},v_{2} \right] \right)&= \left\lbrace (C_{1}*C_{2})\left(u_{2},v_{2} \right)-(C_{1}*C_{2})\left(u_{2},v_{1}\right)\right\rbrace \\
\nonumber
&- \left\lbrace  (C_{1}*C_{2})\left(u_{1},v_{2} \right)-(C_{1}*C_{2})\left(u_{1},v_{1}\right)\right\rbrace  \\
\nonumber
&=\int_{0}^{1}\left( D_{2}C_{1}(u_2,t).D_{1}C_{2}(t,v_2)-D_{2}C_{1}(u_2,t).D_{1}C_{2}(t,v_1)\right) dt \\ 
\nonumber
& -\int_{0}^{1}\left( D_{2}C_{1}(u_1,t).D_{1}C_{2}(t,v_2)-D_{2}C_{1}(u_1,t).D_{1}C_{2}(t,v_1)\right) dt \\
\nonumber
&=\int_{0}^{1} D_{2}C_{1}(u_2,t)\left(D_{1}C_{2}(t,v_2)-D_{1}C_{2}(t,v_1)\right) dt \\
\nonumber
&-\int_{0}^{1}D_{2}C_{1}(u_1,t)\left(D_{1}C_{2}(t,v_2)-D_{1}C_{2}(t,v_1)\right) dt \\
\nonumber
&= \int_{0}^{1} \left[ \left(D_{1}C_{2}(t,v_2)-D_{1}C_{2}(t,v_1)\right)\right] \left[ D_{2}C_{1}(u_2,t)-D_{2}C_{1}(u_1,t)\right] dt \\
\nonumber
&= \int_{0}^{1} \left[ D_{1}\left(C_{2}(t,v_2)-C_{2}(t,v_1)\right)\right] \left[D_{2}\left( C_{1}(u_2,t)-C_{1}(u_1,t)\right) \right] dt,
\end{align}
since $\left(C_{2}(t,v_2)-C_{2}(t,v_1)\right)$ and $\left( C_{1}(u_2,t)-C_{1}(u_1,t)\right)$ are both positive,
we deduce that $$V_{C_{1}*C_{2}}\left(\left[u_{1},u_{2} \right]\times\left[v_{1},v_{2} \right] \right)\geq 0,$$ meaning that the $ (C_{1}*C_{2})-$volume of the rectangle $\left[u_{1},u_{2} \right]\times\left[v_{1},v_{2} \right] $ is positive.
\end{proof}
\begin{remark}
For any copula $C$,
\begin{itemize}
\item  $\Pi*C = C*\Pi=\Pi,$ which means that the copula $\Pi$ is the null element
\item  $M*C =C*M=C,$  meaning that copula $M$ is the identity element
\item  the $*$ product is an associative operation,
\item  the $*$ product of copula is not commutative (for example $W*C\neq C*W$).
\end{itemize}
\end{remark}
This binary operation could be seen as a continuous analog of matrix multiplication. 
We have also $W*W = M.$ 

\section{Discrete Copulas and bistochastic matrices}
In this part we give only basic definition, property and proposition of discrete copula. A more extensive study
can be found in \cite{Kol06}, where Koles\'{a}rov\'{a} \textit{et al.}  have introduced the product of discrete copulas and their links to bistochastic matrices.

Let us denote the grid of the unit square by $\mathcal{I}_{n}:=\left\lbrace 0,\frac{1}{n},\frac{2}{n},\dots,\frac{n-1}{n},1 \right\rbrace.$
\begin{definition}
 A function $C_{n,m}:\mathcal{I}_{n} \times \mathcal{I}_{m} \longrightarrow [0,1]$ is called a discrete copula on  $\mathcal{I}_{n} \times \mathcal{I}_{m}$ if it satisfies the following conditions: \\
For all $i \in \left\lbrace 0,\dots,n\right\rbrace $ and $j \in \left\lbrace 0,\dots,m\right\rbrace$ 
\begin{enumerate}
 \item $ C_{n,m}\left( \frac{i}{n},0\right) =0=C_{n,m}\left(0,\frac{j}{m}\right),$
 \item $ C_{n,m}\left( \frac{i}{n},1\right) =\frac{i}{n}$ and $C_{n,m}\left(1,\frac{j}{m}\right)= \frac{j}{m},$ \\ 

while for all $i \in \left\lbrace 1,\dots,n\right\rbrace $ and $j \in \left\lbrace 1,\dots,m\right\rbrace$
 \item  $C_{n,m}\left( \frac{i}{n},\frac{j}{m}\right)+C_{n,m}\left( \frac{i-1}{n},\frac{j-1}{m}\right)\geq C_{n,m}\left( \frac{i-1}{n},\frac{j}{m}\right)+C_{n,m}\left( \frac{i}{n},\frac{j-1}{m}\right).$
\end{enumerate}
\end{definition}
Now focusing on the particular case, where $n=m$ and writing $C_{n,n}$ as $C_{n}.$ 
\begin{property}
A bistochastic matrix is a matrix $ A=\left(a_{ij}\right)_{1\leq i\leq n; 1\leq j\leq n} $ such that \\
for all $k \in \left\lbrace1,\dots,n \right\rbrace,$ $a_{ij} \geq 0 $ and 
$\displaystyle\sum^{n}_{k =1}a_{ik} = \displaystyle\sum^{n}_{k =1}a_{kj}= 1.$
\end{property}
The relation between a bistochastic matrix and the representation of a discrete copula is stated by the proposition.
\begin{proposition}
 For a function $ C_{n}:\mathcal{I}_{n} \times \mathcal{I}_{n} \longrightarrow [0,1]$ the following statements are equivalent:
\begin{enumerate}
 \item $C_{n}$ is a discrete copula;
 \item there is a bistochastic matrix  $A=\left(a_{ij}\right)_{1\leq i\leq n; 1\leq j\leq n}$ such that for
$i,j \in \left\lbrace 0,1,2 \dots,n \right\rbrace,$ 
\begin{equation}
\label{entries}
 c^{(n)}_{i,j}:=C_{n}\left(\dfrac{i}{n},\dfrac{j}{n} \right)=\dfrac{1}{n} \displaystyle\sum^{i}_{k =1}\displaystyle\sum^{j}_{m =1}a_{km}
\end{equation}
\end{enumerate}
\end{proposition}
The discrete copula $C_{n}$ could then be rewritten in the term of $n \times n$ matrix, with entries given by (\ref{entries}):
 $$C_{n}=\left( c^{(n)}_{i,j}\right).$$
\begin{definition}
 Let $C^{1}_{n}$ and $C^{2}_{n}$ be discrete copulas both defined on $\mathcal{I}_{n} \times \mathcal{I}_{n}$
and let $A_{1}$ and $A_{2}$ be the $n \times n$ bistochastic matrices corresponding respectively to them.
Then the discrete copula $C_{n}$ associated to the bistochastic matrix $A=A_{1}.A_{2}$ is the product of 
$C^{1}_{n}$ and $C^{2}_{n}$. This product is denoted by 
$$C_{n}=C^{1}_{n}\star C^{2}_{n}.$$
\end{definition}
For some special class of copulas there is a relationship between the $\star$ product of discrete copulas
and the $*$product of copulas in the sense of Darsow.%  Copula And Algebra
\chapter{Copula for Modeling and Parameters Estimation}
Let us consider, the following statistics problem. Suppose that we have for example the following set samples $\left\lbrace x^{i}_{1}, x^{i}_{2}\right\rbrace$ for $ i = {1, \dots, T}.$ And we want to choose the copula which fits them. We expose here some of the statistical methods used in that case. First we consider the case where we have selected a joint probability density function depending on parameters $\theta$ and we want to estimate these parameters. We expose these methods : MLE, IFM and Bayesian approach. Then we consider the problem of model selection
where we want to select between a few copulas the one which fits the best data in the sense to be specified and revise different criteria which have been proposed for this. 
 
\section{Maximum Likelihood Estimation(MLE)}
The Maximum likelihood estimation (MLE) is a statistical method used for fitting a mathematical model to some data. 
Modeling real world data by estimating maximum likelihood offers a way of tuning the free parameters $\theta$ of the model to provide a good fit. 
Commonly, one assumes that the data drawn from a particular distribution are independent, identically distributed (iid) with unknown parameters. This considerably simplifies the problem because the likelihood can then be written as a product of $n$ univariate probability densities. From (\ref{tomographyequation}), if we suppose that the copula, the marginal functions and the densities functions dependent on the parameter $\theta$ ( which could be a vector of parameters). 
\begin{itemize}
\item The likelihood $\mathrm{L}(\theta)$ is written as :
 \[\mathrm{L}(\theta) = \displaystyle\prod^{T}_{i=1}c\left(F_{1}(x^{i}_{1};\theta),F_{2}(x^{i}_{2};\theta);\theta\right)\displaystyle \prod_{j}^{2}f_{j}(x^{i}_{j};\theta), \]
\item and computing the log-likelihood $\mathcal{L}(\theta)$ yields to: 
\begin{equation}
\label{loglikelihood}
 \mathcal{L}(\theta)= \displaystyle\sum^{T}_{i=1} \log c\left(F_{1}(x^{i}_{1};\theta),F_{2}(x^{i}_{2};\theta);\theta\right) +\displaystyle\sum^{T}_{i=1}\displaystyle\sum^{2}_{j=1}\log f_{j}(x^{i}_{j};\theta).
\end{equation}

\end{itemize}
This method estimates  $\theta$ by finding the value of $\theta$ that maximises $\mathcal{L}(\theta).$ The value of  $\theta$ is then performed through the following MLE estimator: 
\begin{equation}
 \boxed{\hat{\theta} = \arg \max_{\theta} \mathcal{L}(\theta).}
\end{equation}

\section{Inference Function for Margin (IFM)}
For purpose of algorithm implementation when the number of parameters is large, the IFM is mostly applied.
Instead of using MLE to estimate in one step the parameter $\theta$, one can use the Inference Function for Margin (IFM) method; for more discussion about this method see \cite{Joe97}.

Basically it is a two steps method: 

\textbf{First step}: One estimates parameters $\theta_{j}$ (often a vector of parameters) for each margin function. 

The likelihood, the log-likelihood and the estimator of $\theta_{j}$ are successively 
\[\mathrm{L}_{j}(\theta_{j}) = \displaystyle\prod^{T}_{i=1}f_{j}(x^{i}_{j};\theta_{j})\,, \quad j=1,2 \]
\[\mathcal{L}_{j}(\theta_{j})= \log \mathrm{L}_{j}(\theta_{j})= \displaystyle \sum^{T}_{i=1} \log f_{j}(x^{i}_{j};\theta_{j}).\]

 \begin{equation}
 \boxed{\hat{\theta}_{j} = \arg\max_{\theta_{j}} \mathcal{L}_{j}(\theta_{j}) \quad  \mbox{for}  \quad j=1,2 .}
 \end{equation}

 \textbf{Second step}: Estimate the parameter $\theta_c$ of the joint density function  

\[\mathcal{L}(\theta_c,\hat{\theta}_{1},\hat{\theta}_{2})= \displaystyle\sum^{T}_{i=1} \log f(x^{i}_{1},x^{i}_{2}; \hat{\theta}_{1},\hat{\theta}_{2}, \theta_c) \] 

Now the IFM estimator of $\theta_c$ is 
\begin{equation}
\boxed{ \hat{\theta}_{c} = \arg \max_{\theta_{c}} \mathcal{L}(\theta_{c},\hat{\theta}_{1},\hat{\theta}_{2}).}
\end{equation}

To look how those methods work; one could assume that the two marginal functions are normals with known parameters,
and that the bivariate distribution is also Gaussian with another known parameter. From those true parameters, one could then generate a Gaussian copula data. From both MLE and IFM, compute the estimate parameters and compare to the original true parameters. This could be done using a comparison table and also through several scenarios with any other bivariate distribution functions. See \cite{Yan07} for a full treatment of these topics, and easy implementation for MLE and IFM methods, or to set up a table comparison between MLE and IFM for different copulas families.

\section{Bayesian Approach}
As in the MLE  approach, if we have a set of samples $\left\lbrace x^{i}_{1}, x^{i}_{2}\right\rbrace$ for $i = {1, \dots, T}$ for which we have chosen a parametric copula family $c(x\mid \theta)$ and a likelihood function  \[f(\textbf{x}\mid \theta) = \displaystyle \prod^{T}_{i=1}c\left(F_{1}(x^{i}_{1} \mid \theta),F_{2}(x^{i}_{2} \mid \theta)\right)\displaystyle \prod_{j=1}^{2}f_{j}(x^{i}_{j}\mid \theta),\] and if we also have some prior knowledge on the unknown parameter $\theta$ in the form of a prior probability $\pi(\theta)$, then the Bayesian approach consists in computing the posterior probability
\[
 f(\theta \mid \textbf{x}) = \dfrac{\pi(\theta)f(\textbf{x}\mid \theta)}{f(\textbf{x})} \\
      =\dfrac{\pi(\theta)f(\textbf{x}\mid \theta)}{\displaystyle \int \pi(\theta)f(\textbf{x}\mid \theta)\,d\theta}
\]
and then choosing an estimate for $\theta$ from this posterior. The general approach is to choose an utility function
$u(\theta,\tilde{\theta})$, compute its expected value \[\bar{u}(\tilde{\theta})=\displaystyle \int u(\theta,\tilde{\theta})f(\theta \mid \textbf{x} ) \, d\theta\] and choose as a point estimator 
\begin{equation}
\boxed{\hat{\theta} = \arg \min_{\tilde{\theta}}\left\lbrace \bar{u}(\tilde{\theta})\right\rbrace.}
\end{equation}
Interestingly for different choices of $u$ we find different classical estimators for $\theta.$

The Expected A Posteriori (EAP) estimation:
\begin{equation}
\nonumber
u(\theta,\tilde{\theta})=\parallel \theta-\tilde{\theta} \parallel^{2} \longrightarrow \hat{\theta}_{EAP} = \displaystyle \int \theta f(\theta \mid \textbf{x} ) \, d\theta.
\end{equation}
The Maximum A Posteriori(MAP) estimation:
\begin{equation}
\nonumber
u(\theta,\tilde{\theta})= \delta(\theta-\tilde{\theta} )\longrightarrow \hat{\theta}_{MAP}
 = \arg \max_{\theta}f(\theta \mid \textbf{x})
\end{equation}
In this second case, we can see easily the link with MLE, because 
\begin{equation}
 \boxed{\hat{\theta} = \arg \max_{\theta}f(\theta \mid \textbf{x})=\arg \max_{\theta} \left\lbrace \mathcal{L}(\theta)+\log \pi(\theta)\right\rbrace.}
\end{equation}
where $\mathcal{L}(\theta)$ is the likelihood given in (\ref{loglikelihood}).
 
\section{Choosing The Right Copula}
We have seen that copulas provide greater flexibility in that they allow us to fit any marginals we like to different random variables, and these distributions might differ from one variable to another. We might fit a normal distribution to one variable and another distribution to the second, and then fit any copula we like across the marginals. 
But one of the great difficulty is to find the right copula since there is a huge number of different copulas proposed. 
The range of parameters for each given copula family is different, then it is difficult to make a comparison between them. 

The most used procedure of copula model selection is the one that has larger likelihood. There is a comparison between different families of copulas (see \cite{Ari08}), made by computing the \textit{Kullback-Leibler distance} between copulas with densities $c_1$, $c_2$ and $E_{c_1}$ the expectation value of 
$c_1$:
\begin{equation}
K(c_1,c_2)= E_{c_1}\left[\log\left(\dfrac{c_1(u,v)}{c_2(u,v)} \right)\right]= \displaystyle \int^{1}_{0}\displaystyle \int^{1}_{0} \log \left(\dfrac{c_1(u,v)}{c_2(u,v)} \right)c_1(u,v)\,du\,dv . 
\end{equation}
More precisely for symmetry reasons, the \textit{Jeffreys' divergence measure} defined by
\begin{equation}
J( c_1,c_2)= \displaystyle \int^{1}_{0}\displaystyle \int^{1}_{0} (c_1(u,v)-c_2(u,v)) \log \left(\dfrac{c_1(u,v)}{c_2(u,v)} \right)\,du\,dv , 
\end{equation}
was used, since $$J( c_1,c_2)=K( c_1,c_2)+K( c_2,c_1).$$

Referring to \cite{Dav06} where the Bayesian method to select the most probable copula family among a given set is discussed. The prior information of choice is based on Kendall's tau ($\tau$) measure of association, defined for continuous variables $X$ and $Y$ as
\begin{equation}
 \tau (X,Y)= 4 \displaystyle \int^{1}_{0}\displaystyle \int^{1}_{0} C(u,v)dC(u,v)-1.
\end{equation}
As shown by Genest and Mackay \cite{Jock86}, for Archimedean copulas with generator $\varphi$
\begin{equation}
 \tau (X,Y)= 4 \displaystyle \int^{1}_{0} \dfrac{\varphi(t)}{\varphi^{'}(t)}dt +1.
\end{equation}
Spearman Rho ($\rho$) is another measure which uniquely dependent on the structure of $C$, but not on the behaviour of the margins function:
\begin{equation}
 \rho (X,Y)= 12 \displaystyle \int^{1}_{0}\displaystyle \int^{1}_{0} uv\,dC(u,v)-3= 12 \displaystyle \int^{1}_{0}\displaystyle \int^{1}_{0}C(u,v)\,dv\,du-3.
\end{equation}
Kendall's Tau measure and Spearman Rho take advantage of the classical Pearson's Rho (or Pearson product-moment correlation) which
reflects the degree of \textit{linear relationship} between two random variables $X$,$Y$ with expected values $E(X)=\mu_{X}$,$E(Y)=\mu_{Y}$ and finite nonzero standard deviations $\sigma_{X}$,$\sigma_{Y}$ and defined as :
 $$r(X,Y)={\mathrm{cov}(X,Y) \over \sigma_X \sigma_Y} ={E((X-\mu_X)(Y-\mu_Y)) \over \sigma_X\sigma_Y},$$
and we may also write
$$r(X,Y)=\frac{E(XY)-E(X)E(Y)}{\sqrt{E(X^2)-E^2(X)}~\sqrt{E(Y^2)-E^2(Y)}}.$$

For any increasing functions $f$ and $g$, we have :
\begin{eqnarray}
 \rho (X,Y)=\rho (f(X),g(Y)), \\
 \tau (X,Y)= \tau (f(X),g(Y)) \\
  r(X,Y)\neq r(f(X),g(Y)) 
\end{eqnarray}
For example, the bivariate Gaussian copula with correlation $\theta$ has $\tau=\frac{2}{\pi} \arcsin \theta $.
There is also another interesting measure in the theory of extreme value copula \cite{Joe97}, helping to compare different copula families, so called \textit{tail dependence measure} as we define below:
\begin{definition}
 Let $\overline{C}(u,v)=1-u-v+C(u,v)$ be the joint survival function for two uniform (0,1) random variables whose joint distribution function is the copula $C$. If $C$ is such that 
$\displaystyle \lim_{u\rightarrow 1}\dfrac{\overline{C}(u,u)}{1-u}=\lambda _{U} $ exists, then $C$ has an upper tail dependence if $\lambda _{U} \in \left(0,1 \right]$  and no upper tail dependence if $\lambda _{U}=0$. Similarly, if 
$\displaystyle \lim_{u\rightarrow 0}\dfrac{C(u,u)}{u}=\lambda _{L} $ exists, then $C$ has an lower tail dependence if $\lambda _{L} \in \left(0,1 \right]$  and no lower tail dependence if $\lambda _{L}=0.$
\end{definition}
Discussion and technical way to make a good choice of copula could be found also in \cite{Dur00}. 
Other approach of choosing a suitable copula to model dependence based on exponential family is proposed recently \cite{Will08}.

Definitely << How does one choose a copula ? >>, this is one question among many other, asked by Dr. Mikosch in his paper  \cite{Question}, when he is raising doubt on the fundamental basis of Copula's Theory in comparison with the Theory of Stochastic Processes. 

Prof. Genest answered \cite{Response}:
\begin{quotation}
Model selection is a broad question for which a completely satisfying answer does not yet exist, even in the univariate case.
The same strategies can be used here as in many other modeling exercise, i.e choices can be guided by model properties and characterisations diagnostics tools, cross-validation, predictive, accuracy , etc. Given that copula modeling is still in a relatively early stage of development, we concur that much remains to be done in this regard. For a state-of-art illustration
of methodology currently available see Genest and Favre \cite{Fav07}.
\end{quotation}%  Copula for Modeling and Parameters Estimation 
\chapter{Tomographic Image Reconstruction}
Is it possible to see the interior structure of an object without cutting it open ? The answer is yes if we can expose this object to a ray (for example X rays) and to measure its interaction with it (for example the X rays radiographies).
In this section, we give an introduction to this important subject of imaging sciences which helps to solve the previous problem and much more arising from different area of science. There is a long list of research domain where tomography technique is applied, for example in archaeology, biology, geophysics, oceanography, materials science, medical imaging, astrophysics. 

\section{Tomography technique}
More modern variations of tomography involve gathering projection data from multiple directions and feeding the data into a tomographic reconstruction software algorithm processed by a computer. Different types of signal acquisition can be used in similar calculation algorithms in order to create a tomographic image. There are  several types of tomography technique associated to a specific physical phenomenon. The following table give a list of some methods mostly used:
\begin{center}
% use packages: array
\begin{tabular}{| c | c |}
\hline
\textbf{Physical phenomenon} & \textbf{Type of tomography} \\ 
\hline
X-rays & X-rays Computed tomography (CT) \\ 
\hline
gamma rays & Single Photon Emission Computed Tomography(SPECT) \\ 
\hline
electron-positron annihilation & Positron Emission Tomography (PET) \\
\hline 
nuclear magnetic resonance & Magnetic Resonance Imaging( MRI) \\ 
\hline
ultrasound & Medical sonography (ultrasonography) \\ 
\hline
electrons & 3D Transmission Electron Microscopy (TEM)\\
\hline
\end{tabular}
\end{center}
In our case, we focus on X-rays Computerized Tomography (CT) which used x-ray, through the Radon transform.
In Medicine for example, Computed tomography (CT) is a diagnostic procedure that uses special x-ray equipment to obtain cross-sectional images of the body. The CT computer displays these pictures as detailed image of organs, bones, and other tissues. This procedure is also called CT scanning, computerized tomography, or computerized axial tomography (CAT).
One vital application is the diagnostic of cancer; CT is used in several ways:
\begin{itemize}
\item To detect or confirm the presence of a tumor;
\item To provide information about the size and location of the tumor and whether it has spread;
\item To guide a biopsy (the removal of cells or tissues for examination under a microscope);
\item To help plan radiation therapy or surgery; and
\item To determine whether the cancer is responding to treatment.
\end{itemize}

There are also different geometrical view about the object. The representation can be in $2D$, $3D$ (static tomography) or $4D$ when adding time parameter, leading to a dynamic tomography ($4D$ model representation is used also as a heart model in Radiology).

\section{Computerized Tomography and Radon Transform}
The simplest and easiest way to visualise this method is the classical system of parallel projection, where the data to be collected as considered to be a series of parallel rays, at position $r$, across a projection at angle $\theta$. 
This is repeated for various angles.

\begin{center}
\begin{tabular}{|l|}
\hline
\psfrag{A}[c]{$s$} % usage: \psfrag{text}[posn][psposn][scale][rotate]{formula} 
\psfrag{B}[l,t]{$y$} % replace 'C' by $x_2$ ([c] means 'center')
\psfrag{C}[c]{$dl$} % replace 'B' by the formula $x_2$ rotated 0 deg
\psfrag{D}[c]{$x$} 
\psfrag{E}{$\bf \red{p_{\theta}(r)}$} 
\psfrag{F}[c]{$r$} 
\psfrag{G}[c]{$r$}
\psfrag{H}[c]{$\theta$}
\psfrag{I}[c][5]{$\bf \red{f(x,y)}$}
\psfrag{j}[r]{Detector}
\psfrag{k}[l]{Source}
\includegraphics[width=0.5\textwidth]{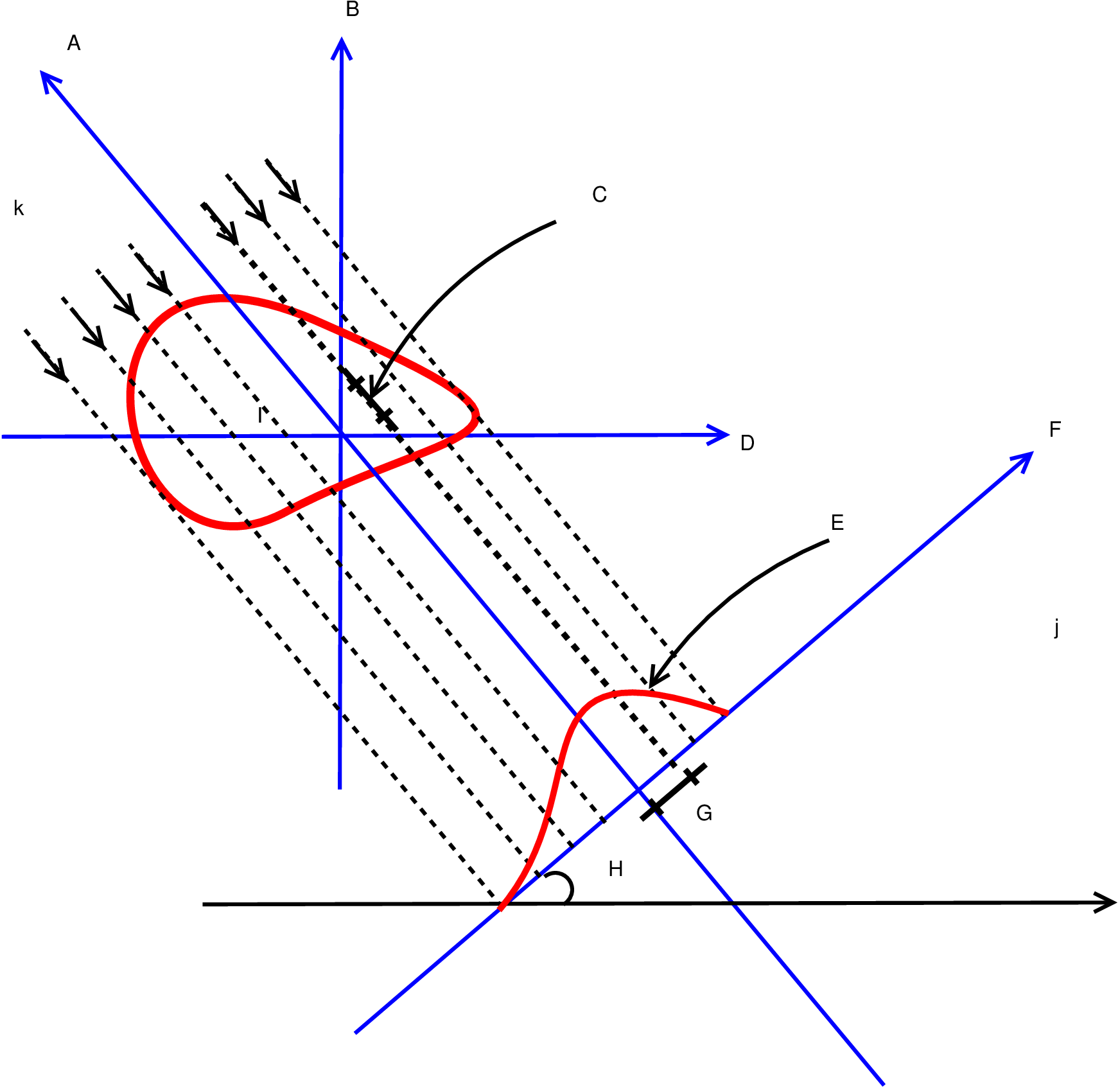} \\ % the picture to be processed 
{\bf Figure~1~: Parallel beam geometry } \\
\hline
\end{tabular}
\end{center}
Basically the idea of the X-ray CT is to get images of the interior structure of an object by X-raying the object from many different directions. X rays go in straight lines inside the body and its energy is attenuated more and less depending on the density of the matter in his trajectory. Attenuation occurs exponentially in tissue :
\begin{equation}
 I(r) = I_0\exp\left({-\int_{\mathrm{L}_{r,\theta}} f(x,y)\,dl}\right),
\end{equation}
where $f$ is the attenuation coefficient at position $(x,y)$ along the ray path and 
\[\mathrm{L}_{r,\theta}=\left\lbrace x,y: r= x \cos \theta +y \sin \theta  \right\rbrace.\]

So the simplest model relating 
the log ratio $\ln (I/I_{0})$ of the observed energy $I$ with emitted energy $I_{0}$ to the spatial spatial distribution $f$ of the body is a line integral:
\begin{equation}
\label{line}
 p_{\theta}\left(r\right) = -\ln (I/I_0) =\int_{\mathrm{L}_{r,\theta}} f(x,y)\,dl.
\end{equation}
Therefore generally the total attenuation\footnote{for continuous value of $\theta$ in $0 \leq \theta \leq \pi $ and varying $r$ within $0 \leq r < \infty $, we will denote $p_{\theta}(r)$ by $p(r,\theta)$} $p_{\theta}(r)$ of a ray at position $r$, on the projection at angle $\theta$, is given by the line integral (\ref{line}).

\subsection{Forward problem}
The forward problem is the one dealing with the mathematical expression of the projections. 
\begin{equation}
\nonumber
\boxed{\mbox{Given}\quad f(x,y) \quad  \mbox{find the projections} \quad p_{\theta}(r) \quad \mbox{where} \quad \theta \in \left[0, \pi \right] .}
\end{equation}
This is a well-posed problem. According to what we mentioned earlier, in the situation when the densities function $f$ is known for all positions $(x,y)$, the projections are given by :
\begin{equation}
\label{radon}
\boxed{
\mbox{Radon $\Rc$ ~:~~~~~~} p_{\theta}(r)=\displaystyle \iint_{\mathbb{R}^{2}} f(x,y) \, \delta(r-x\cos\theta-y\sin\theta) \d{x} \d{y}.}
\end{equation}

The relation (\ref{radon}) is known as the Radon Transform (or sinogram) of the $2D$ object $f(x,y)$, consists of X-ray projections along all possible lines in the plane. Each line has a specific direction and each direction is uniquely identified by the angle $\theta.$

$\delta$ is the Dirac's delta function defined as :
\begin{equation}
\nonumber
 \left\{
  \begin{array}{rcr}
    \delta(t)  =  0, & ~ \mbox{for} ~ t\neq 0, ~ \mbox{and} \\
\displaystyle \int_{-\infty}^{\infty} \delta(t)  =  1. & \\
  \end{array}
\right.
\end{equation}
The Dirac's delta function has the fundamental nice property that:
\[\displaystyle \int f(r) \, \delta(r-r_{0})\,dr=f(r_{0}).\]

\fbox{Geometric transformation} \\
If we denote with $(x,y)$ for Cartesian coordinates. And we define $(r,s)_{\theta}$ as the Cartesian coordinates through rotating $(x,y)$ by an angle $\theta$ along the counterclockwise direction (see {Figure~1~}). The transformation between those two coordinates can be described by the following relations:
\[
\dbinom{r}{s}= 
\begin{pmatrix}
  \cos \theta &\sin \theta \\
 -\sin \theta & \cos \theta
\end{pmatrix}\dbinom{x}{y} 
. \]
For any function $\zeta(x,y)$, the transformation \[\zeta_{\theta}(r,s)=\zeta(r \cos\theta-s \sin \theta,r \sin \theta + s \cos \theta)\]  denotes the same function in the coordinates $(r,s)_{\theta}$.

\subsection{Reconstruction problem}
Using the result for the relation (\ref{radon}) we give the typical problem of image reconstruction. The problem is to estimate a multivariate function $f(x,y)$ from its line integrals $p_{\theta}(r)$, that is called an inverse problem and can be formulated in the following ways:
\begin{equation}
\nonumber
\boxed{\mbox{Given}\quad  p_{\theta}(r) \quad \mbox{for different} \quad \theta \in \left[0, \pi \right] \quad \mbox{find} \quad f(x,y).}
\end{equation}
It is the same problem, when given the Radon transform (or projections) of a unknown object, and trying to find the object.

The \textit{projection-slice theorem} tells us that if we had an infinite number of one-dimensional projections of an object taken at an infinite number of angles, we could perfectly reconstruct the original object.

One obvious solution is to find the analytic expression of the inverse of the Radon transform.
\begin{equation}
\label{inverse}
\boxed{\mbox{Inverse Radon  $\Rc^{-1}$ ~:~~~} f(x,y)= \frac{1}{2\pi^2} \displaystyle \int_{0}^{\pi} \displaystyle \int_{0}^{\infty}\dfrac{\frac{\partial }{\partial r} p(r,\theta)}{\left( r- x \cos \theta -y \sin \theta \right) }dr \,d\theta .}
\end{equation}
In polar coordinates 
\begin{equation}
\nonumber
\boxed{\mbox{Inverse Radon  $\Rc^{-1}$ ~:~~~} f(x,y)= f(\xi\cos\phi,\xi\sin\phi)=\frac{1}{2\pi^2} \displaystyle \int_{0}^{\pi} \displaystyle \int_{0}^{\infty}\dfrac{\frac{\partial }{\partial r} p(r,\theta)}{\left(\xi \cos (\theta -\phi)- r \right) }dr \,d\theta .}
\end{equation}

The book \cite{Mar06} page 108, present the proof given by Radon himself from \cite{Rad17} about the inversion of the Radon Transform in $\mathbb{R}^{2}$ and further results.

However, the inverse of Radon transform proves to be extremely unstable with respect to noisy data. In practice, a stabilised and discretized version of the inverse Radon transform is used, known as the \textit{filtered backprojection} algorithm which we described below.
\section{Analytical Methods}
\subsection{Backprojection and Filtered Backprojection}
Let us remind briefly some operators used to express the formula of the \textit{backprojection} and the \textit{filtered backprojection}. For more detail about this approach, see the book <<Principles of Computerized Tomographic Imaging >> \cite{Kak88}.
\begin{definition}
The Fourier transform (FT) of a function  $ f(x)\in{\bf \mathbb{C}}$, $ x \in(-\infty,\infty)$ is defined as
\begin{equation}
\nonumber
\mbox{ $\Fc$~:~~~~~~}  F(\omega) = \displaystyle \int_{-\infty}^{\infty} f(x)e^{- j \omega x}\,dx,
\end{equation}
and the inverse Fourier transform (IFT) is given by 
\begin{equation}
\nonumber
 \mbox{ $\Fc^{-1}$~:~~~~~~}  f(x)= \dfrac{1}{2 \pi}\displaystyle \int_{-\infty}^{\infty} F(\omega)e^{ j \omega x}\,d\omega.
\end{equation}
\end{definition}
Conditions for the existence of the Fourier transform are complicated to state in general\cite{Cha87}, but it is sufficient for $f(x)$ to be absolutely integrable, i.e., $\displaystyle \left\Vert\,f\,\right\Vert_1 \triangleq  \displaystyle \int_{-\infty}^\infty \left\vert f(x)\right\vert dx < \infty.$ This requirement can be stated as $f \in L^1(\mathbb{R})$ meaning that $f$ belongs to the set of all functions having a finite $L^{1}\mbox{-Norm}$.\\
It is similarly sufficient for $ f(x)$ to be square integrable, $\displaystyle \left\Vert\,f\,\right\Vert _2^2 \triangleq  \displaystyle \int_{-\infty}^\infty \left\vert f(x)\right\vert^2 dx < \infty.$\\
More generally, it suffices to show $f \in L^p$ for $ 1\leq p\leq 2$.\\

\begin{definition}
The Hilbert transform is a linear operator which takes a function, $s(t)$, to another function, $\mathcal{H}(s)(t)$, with the same domain. In signal processing, this operator is used to derive the analytic representation of a signal $s(t)$.
The exact definition of the Hilbert Transform using the Cauchy principal value (denoted here by $p.v.$) is
\begin{equation}
  \mbox{ $\Hc$~:~~~~~~} \widetilde{s}(t)= p.v.\, \frac{1}{\pi}\int_{-\infty}^{\infty}\frac{s(\tau)}{t-\tau}\, d\tau\,.
\end{equation}
\end{definition}
Computationally one can write the Hilbert transform as the convolution:
\begin{equation}
  \mbox{ $\Hc$~:~~~~~~}  \widetilde{s}(t)= \frac{1}{\pi t }*s(t).
\end{equation}
which by the convolution theorem of Fourier transforms\footnote{The Fourier transform of a convolution is the product of the Fourier transforms}, may be evaluated as the product of the transform of $s(\tau)$ with $-i \, \sgn(\tau )$, where:
\[ 
\sgn(\tau) = 
\begin{cases} 
  +1,  & \mbox{if } \tau > 0 \\
   0,  & \mbox{if } \tau = 0\\
  -1,  & \mbox{if } \tau < 0. 
\end{cases}
\]

Therefore to get $f(x,y)$ back, from (\ref{radon}) means finding the inverse Radon transform using the backprojection (BP) or the filtered backprojection methods, the above operators are applied.\\

The backprojection operator is defined as 
\[
\mbox{Backprojection $\Bc$:~~~~~~}
\disp{b(x,y)= \frac{1}{2\pi} \izpi p(x\cos\theta+y\sin\theta,\theta) \d{\theta}}.
\]

Instead of dealing directly with the expression (\ref{inverse}), one computes the inverse of the Radon transform sequentially using the following steps: 

\[
\mbox{Derivative $\Dc$:~~~~~~} 
\disp{\overline{p}(r,\theta)=\dpdx{p(r,\theta)}{r}} 
\]
\[
\mbox{Hilbert Transform $\Hc$:~~~~} 
\disp{\widetilde{\overline{p}}(r',\theta)
=\frac{1}{\pi} \izi \frac{\overline{p}(r,\theta)}{(r-r')} \d{r}}
\]
\[
\mbox{Backprojection $\Bc$:~~~~~~}
\disp{f(x,y)= \frac{1}{2\pi} \izpi \widetilde{\overline{p}}(x\cos\theta+y\sin\theta,\theta) \d{\theta}}.
\]
The model of the backprojection is then defined as: 
\begin{equation}
\boxed{\disp{f(x,y) = \Bc \; \Hc \, \Dc \, p(r,\theta)}.} 
\end{equation}

\fbox{Filtered Backprojection (FBP)} 

\begin{center}
\begin{tabular}{|l|}
\hline
\psfrag{A}[c]{$F(\omega)$} % usage: \psfrag{text}[posn][psposn][scale][rotate]{formula} 
\psfrag{A3}[c]{$\sgn(\omega)\omega F(\omega)=|\omega|F(\omega)$} % replace 'C' by $x_2$ ([c] means 'center')
\psfrag{A4}[c]{$\omega F(\omega)$} % replace 'C' by $x_2$ ([c] means 'center')
\psfrag{B}[c]{$FT$} % replace 'B' by the formula $x_2$ rotated 0 deg
\psfrag{D}[c]{$\mathcal{D}$} 
\psfrag{H}{$\mathcal{H}$} 
\psfrag{f}[c]{$f(x)$} 
\psfrag{f1}[c]{$\widetilde{f}(x)$}
\psfrag{f2}[c]{$\frac{\partial f}{\partial x}$}
\includegraphics[height=0.32\textheight]{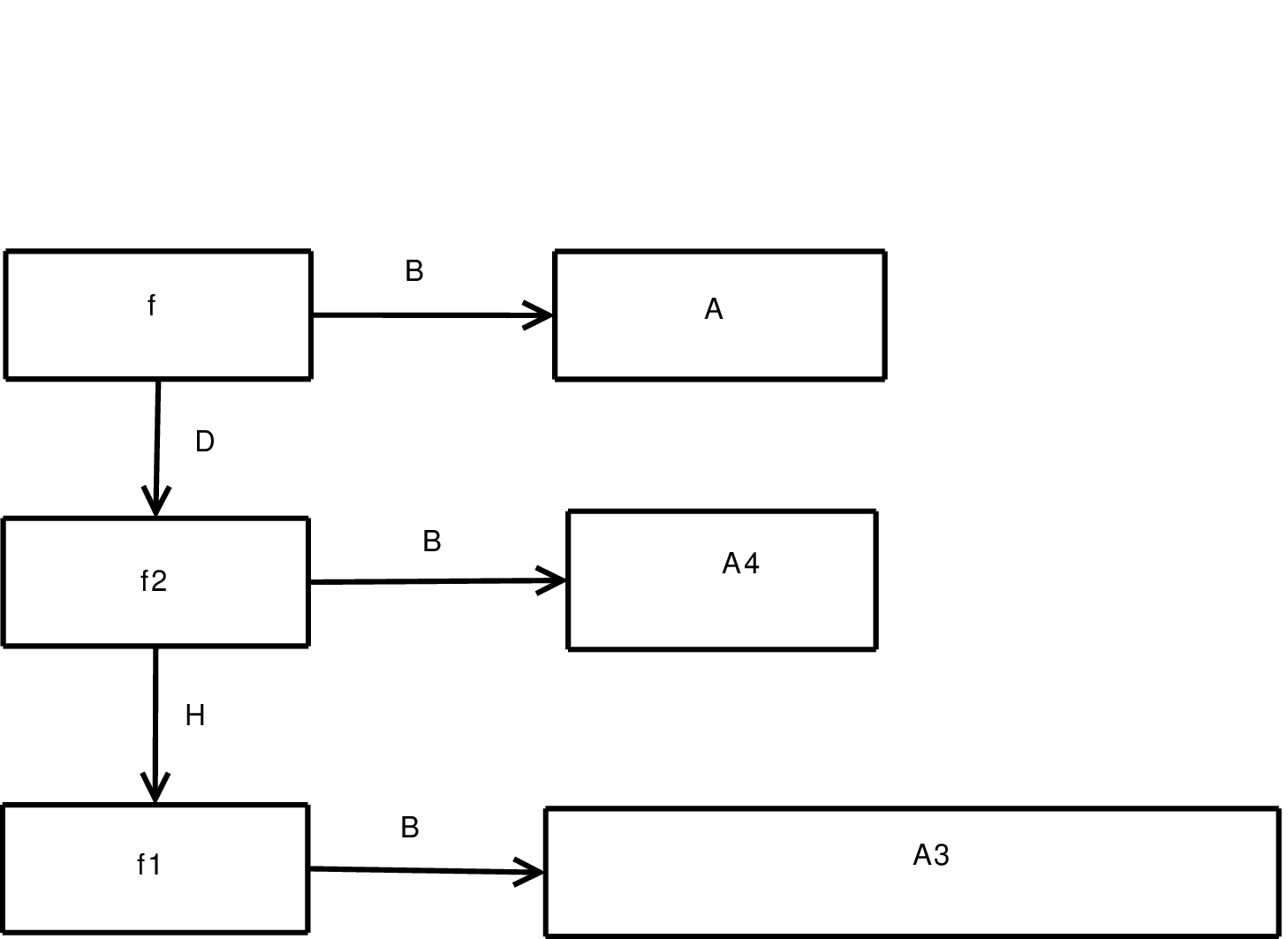} \\ % the picture to be processed 
\hline
\end{tabular}
\end{center}
From the above table where we describe the relationship between $\Dc$,\,FT and $\Hc$, defining: 
\begin{equation}
 \nonumber
 \mbox{$\Fc$~:~~~~~~}  P(\omega,\theta) =\displaystyle \int_{-\infty}^{\infty} p(r,\theta)e^{- j \omega r}\,dr. 
\end{equation}
Therefore the properties between $\Dc$, FT, $\Hc$ yields to 
\[\overline{P}(\omega,\theta)=\omega p(\omega,\theta),\]
and
\[\widetilde{\overline{P}}(\omega,\theta)=\sgn(\omega)\,\omega \,\overline{P}(\omega,\theta)=|\omega|P(\omega,\theta).\]
 
We have demonstrated the following model of the filtered backprojection (FBP)commonly used in X-ray CT,
\[
\stackrel{p(r,\theta)}{\lra} 
\fbox{\btabu{@{}c@{}} {\small \textsc{FT}}\\${\cal F}$\etabu}\lra
\fbox{\btabu{@{}c@{}} {\small Filter}\\$|\omega|$\etabu}\lra
\fbox{\btabu{@{}c@{}} {\small \textsc{IFT}}\\${\cal F}^{-1}$\etabu}
\stackrel{\widetilde{\overline{p}}(r,\theta)}
{\lra} 
\fbox{\small\btabu{@{}c@{}} {\small Backprojection}\\ $\cal B$\etabu}
\stackrel{f(x,y)}{\lra}
\]
which can be rewritten as:
\begin{equation}
 \boxed{\disp{f(x,y)= \Bc \; \Fc^{-1} \, |\omega| \, \Fc \, p(r,\theta)}.}
\end{equation}
Notice that * denotes a 2D convolution, it is also shown that: 
\[
\mbox{Backprojection $\Bc$:~~~~~~}
\disp{b(x,y)= \dfrac{1}{\sqrt{x^{2}+y^{2}}}*f(x,y)}.
\]
In polar coordinates
\[
\mbox{Backprojection $\Bc$:~~~~~~}
\disp{b(\xi,\phi)= \dfrac{1}{|\xi|}*f(\xi,\phi)}.
\]
%  Tomographic Image Reconstruction
\chapter{Bridging the gap between Copula and Tomography}
After having reviewed copulas and tomography outlined in the previous chapter, we are now able to describe our mathematical approach to the problem of tomographic image reconstruction in more detail. We consider only the case of two projections: horizontal $\theta =0$ and vertical $\theta =\pi/ 2$.
\section{Horizontal and vertical projections}
\begin{center}
\begin{tabular}{|l|}
\hline
\psfrag{A}[c]{$\bf \red{f(x,y)}$} % usage: \psfrag{text}[posn][psposn][scale][rotate]{formula} 
\psfrag{B}[c]{$\bf \red{f_{1}(x)}$} % replace 'C' by $x_2$ ([c] means 'center')
\psfrag{C}[c]{$\theta =0$} % replace 'B' by the formula $x_2$ rotated 0 deg
\psfrag{D}[l]{$\theta = \frac{\pi}{2}$} 
\psfrag{E}{$\bf \red{f_{2}(y)}$} 
\includegraphics[width=0.5\textwidth]{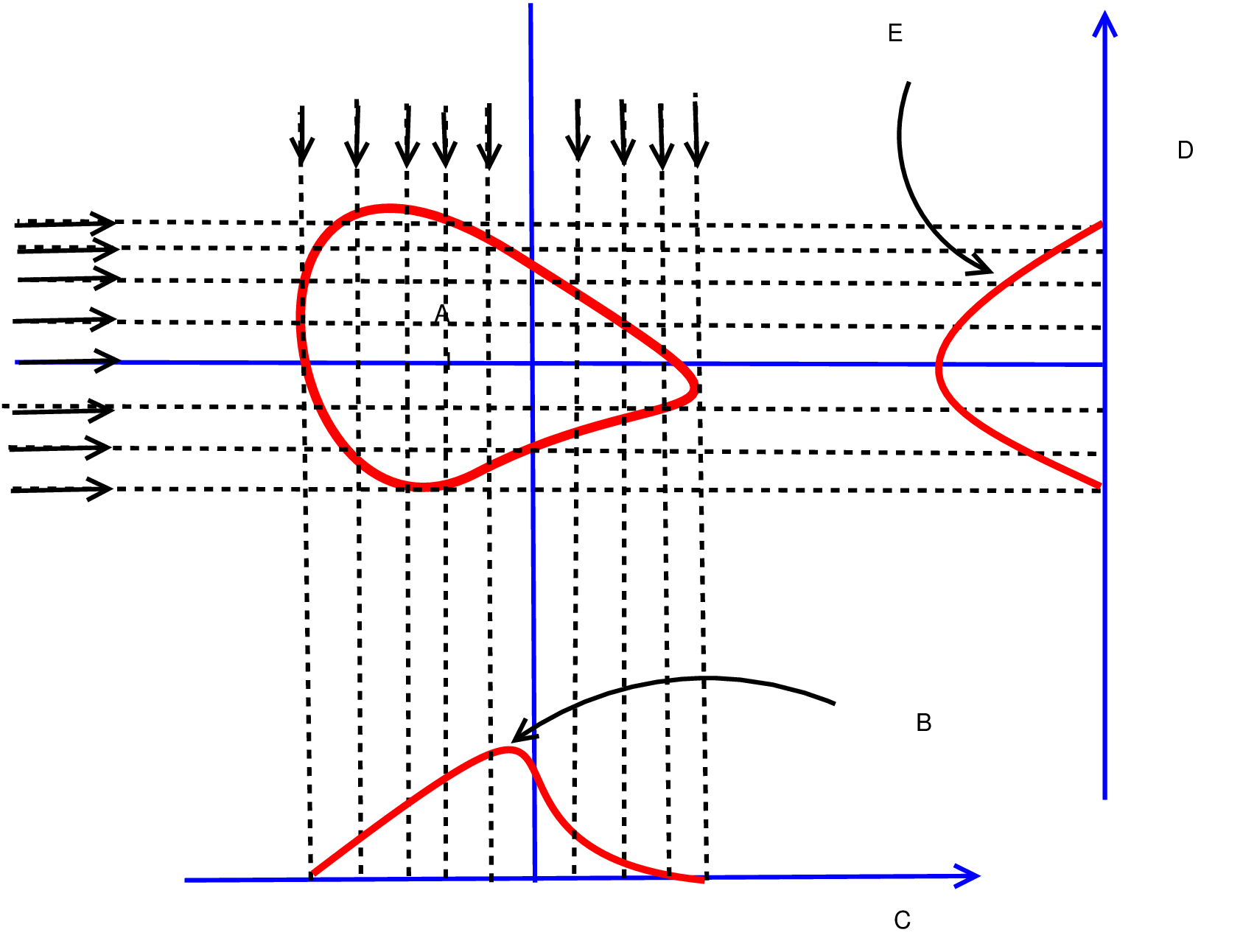} \\% the picture to be processed
{\bf Figure ~2~: Horizontal and vertical projections } \\
\hline
\end{tabular}
\end{center}
The particular case where we have only two projections $\theta=0$ (horizontal) and $\theta=\pi/2$ (vertical).
We substitute in (\ref{radon}) to obtain:
\begin{equation}
\label{projectionradon}
  \addtolength{\fboxsep}{5pt}
   \boxed{
   \begin{gathered}
\theta=0, ~~~~ p_0(r) = \displaystyle \iint f(x,y) \delta(r-x) \d{x}\d{y} \\
\theta=\frac{\pi}{2},~~~~ p_{\pi/2}(r) = \displaystyle \iint f(x,y) \delta(r-y) \d{x}\d{y}.
  \end{gathered}
   }
\end{equation}
If now, we denote $p_0=f_1$ and $p_{\pi/2}=f_2$ the following relation also holds:
\begin{equation}
\label{marginalradon}
  \addtolength{\fboxsep}{5pt}
   \boxed{
   \begin{gathered}
      \theta=0, ~~~~ f_{1}\left(x\right)= \displaystyle\int_{\mathbb{R}}f(x,y)\,dy \\
      \theta=\frac{\pi}{2},~~~~ f_{2}\left(y\right)=\displaystyle \int_{\mathbb{R}}f(x,y)\,dx.
   \end{gathered}
   }
\end{equation}
It is important to point out that projections(\ref{projectionradon}) are theoretically the same as marginal distributions (\ref{marginalradon}) if we assume the positivity and normalisation. This result in the general case was shown by Cram\'{e}r and Wold in 1936, when they inverted the Radon Transform in the context of mathematical statistics(see \cite{Cra36}).

\bfig
\bcc
\btabu{cc}
\includegraphics[width=5cm]{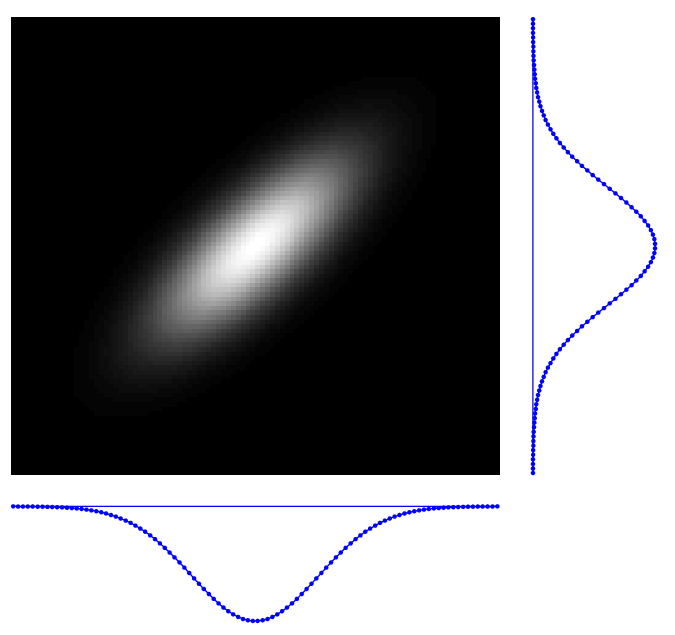}&
\includegraphics[width=5cm]{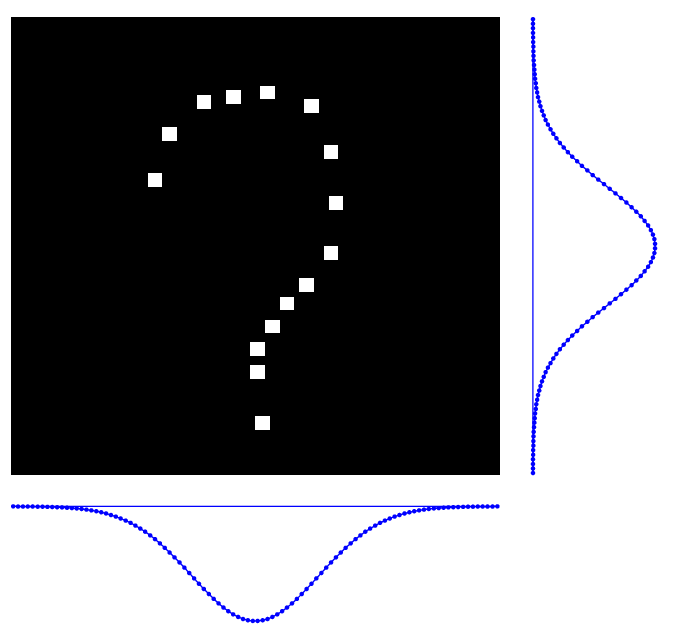}\\ 
Forward problem: & Inverse problem: \\ 
Given $f(x,y)$ find $f_1(x)$ and $f_2(y)$ &
Given $f_1(x)$ and $f_2(y)$  find $f(x,y)$ 
\etabu
\ecc
\efig

Now we clearly identity the following situation as an \textbf{inverse problem} typically \textbf{ill-posed}:

\begin{equation}
\nonumber
 \boxed{\mbox{Given}\quad f_{1}(x)\quad \mbox{and} \quad f_{2}(y) \quad \mbox{find} \quad f(x,y).}
\end{equation}

In fact one of the three conditions for a well-posed problem suggested by the French mathematician Jacques Hadamard (existence, uniqueness, stability of the solution) is not satisfied \cite{Had02}.

It is clear that given $f_{1}(x)$ and $f_{2}(y)$, there are \textbf{infinitely many solutions} for $f(x,y)$. 

If we look at the \textbf{Backprojection}(BP) solution and the \textbf{Filtered Backprojection}(FBP) solution, for $\theta=0$ and $\theta=\pi/2$, we have 
\begin{eqnarray}
\nonumber
 f(x,y)& = &\frac{1}{2 \pi} \displaystyle \int_{0}^{\pi}f_1(x\cos\theta +y \sin\theta, \theta) d\theta \\
\nonumber
& \thickapprox& \frac{1}{2} \left( f_{1}(x) + f_{2}(y)\right).
\end{eqnarray}
This implies that the BP solution, resulting from the trapezoidal rule is:
\begin{equation}
 \boxed{\widehat{f}(x,y) =\frac{1}{2} \left( f_{1}(x) + f_{2}(y)\right).}
\label{BP}
\end{equation}
And the FBP solution is:
\begin{equation}
\nonumber
 \widehat{f}(x,y) =\frac{1}{2}\izi \frac{\frac{\partial f_{1}}{\partial x}(x')}{x'-x} \d{x'} + \frac{1}{2} \izi \frac{\frac{\partial f_{2}}{\partial y}(y')}{y'-y} \d{y'},
\end{equation} 
which can also be implemented in the Fourier domain as 
\begin{equation}
 \boxed{ \widehat{f}(x,y) =\frac{1}{2}\displaystyle \int_{\mathbb{R}} \mid \omega_1 \mid f_{1}(\omega_1)e^{j\, \omega_1 \,x}\,d\omega_1 +\frac{1}{2}\displaystyle \int_{\mathbb{R}} \mid \omega_2 \mid f_{2}(\omega_2)e^{j\, \omega_2 \, y}\,d\omega_2 .}
\end{equation} 
\section{Looking at Inverse problem in a different way}

Let us consider the following problem in a different way. Between all the solution which satisfied the constraints 
\[
\mbox{$\Cc1$:~~~~~~} 
\disp{ \displaystyle \int f(x,y)\d{y}=f_{1}(x)}
\]
\[
\mbox{$\Cc2$:~~~~~~} 
\disp{ \displaystyle \int f(x,y) \d{x}=f_{2}(y)}
\]
\[
\mbox{$\Cc3$:~~~~~~}
\disp{\displaystyle \int \int f(x,y) \d{x}\d{y}=1.}
\]

Choose the one which minimise a criterion $\Omega(f)$ such as:

\[
\Omega_1(f)=\disp{\displaystyle \iint \mid f(x,y)\mid^{2} \d{x}\d{y}}
\]
\[
\Omega_2(f)=\disp{\displaystyle \int \int -f(x,y) \ln ( f(x,y)) \d{x}\d{y}.}
\]
Interestingly, if we choose $\Omega_1(f)$ we obtain the Backprojection solution
\beq
\label{omega1}
\hat{f}(x,y)=f_1(x) + f_2(y) 
\eeq
and if we choose $\Omega_2(f)$ we obtain
\beq
\hat{f}(x,y)=\dfrac{1}{Z} \exp(-f_1(x))\,\exp(-f_2(y)), 
\eeq
where $Z$ is a constant such that $\Cc 3$ is satisfied.

If we choose to minimise 
\[
\Omega_3(f)=\disp{\displaystyle \int \int \left( f(x,y) \ln \left( \dfrac{f(x,y)}{f_{1}(x)f_{2}(y)}\right)- f(x,y)\right)  \d{x}\d{y}}
\]
we obtain 
\beq
\label{omega3}
\hat{f}(x,y)=f_1(x) f_2(y). 
\eeq

The above ways to look at the same tomographic image problem, we have discussed is based on information entropy. We have used the method of Lagrange multipliers\footnote{\scriptsize{$\mathcal{L}_{g}(f,\lambda_i)=\lambda_1\Omega_j(f)+ \lambda_2\left(f_{1}(x)- \displaystyle\int f(x,y)\d{y}\right)+\lambda_3\left(f_{2}(y)-\displaystyle\int f(x,y)\d{y}\right)+\lambda_4\left(1-\displaystyle\iint f(x,y)\d{x}\d{y}\right)$}\\ where $i=1,2,3,4$ and $j=1,2,3$ then solve $
\disp{\dpdx{\mathcal{L}_{g}(f,\lambda_i)}{f}=0\,\mbox{and}\,\dpdx{\mathcal{L}_{g}(f,\lambda_i)}{\lambda_i}=0 \,\, \mbox{to find}\,\, \lambda_i}. 
$} (denoted $\mathcal{L}_{g}$) to optimise the solution under criteria $\Omega_1$ (denoted the $L_{2}-Norm$),$\,$ $\Omega_2$ (denoted the Shannon Entropy). 

We have also sat the criterion $\Omega_3$ by modifying the KullBack-Leibler distance in order to obtain (\ref{omega3}).

From those preliminary result, we conclude also that the notion of entropy gives some solution we have already found and we will find in the next section about copula\footnote{compare eq.(\ref{omega3}) to eq.(\ref{MBP}); and also eq.(\ref{omega1}) to eq.(\ref{BP})}.

\section{Simulation Results Using Copula}
Those results we propose in this section are a starting point for further research in this area.

\begin{figure}
\begin{center}
\begin{tabular}{|c|c|c|}
\hline
\includegraphics[height=0.2\textheight]{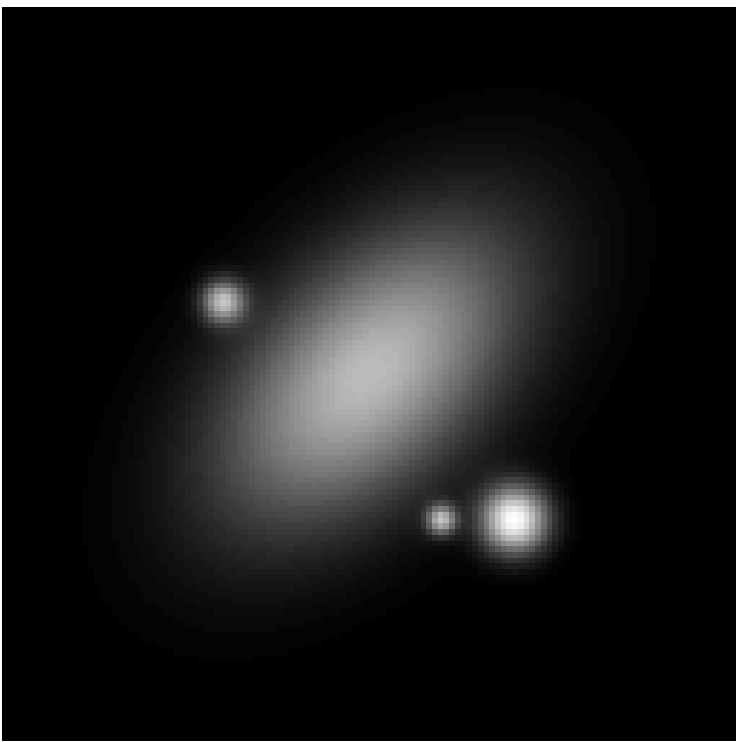}&
\includegraphics[height=0.2\textheight]{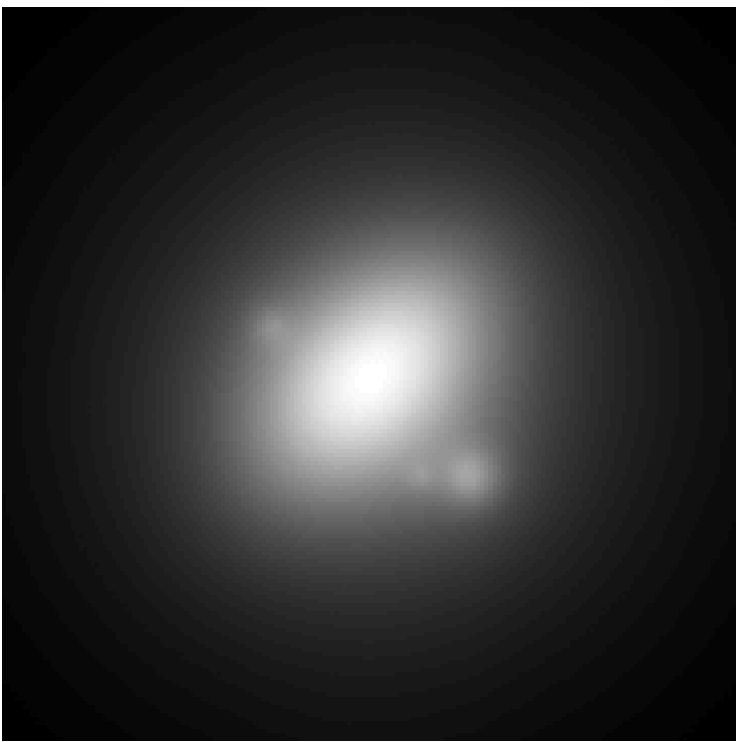}&
\includegraphics[height=0.2\textheight]{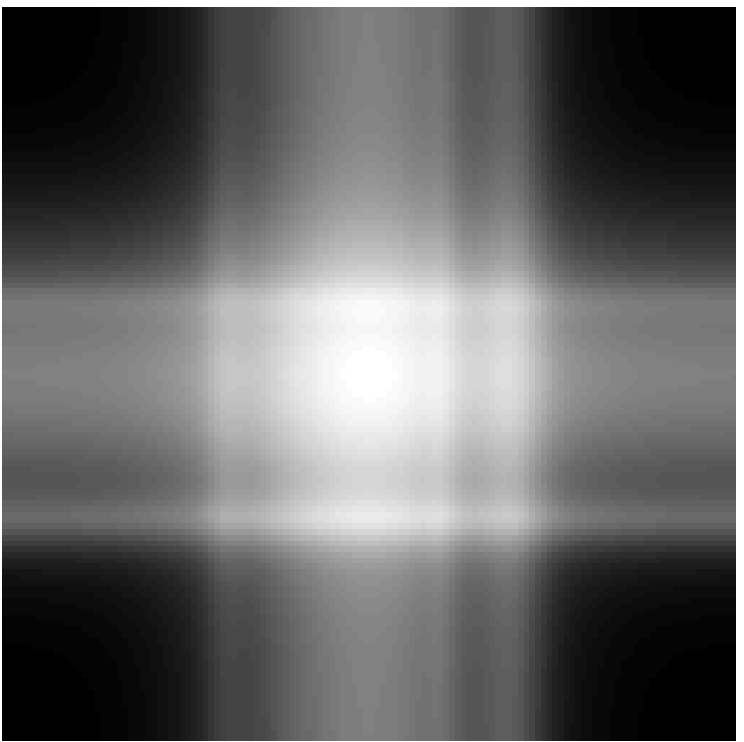}\\
Original $f(x,y)$ & BP: 128 projections & BP: 2 projections\\ 
\includegraphics[height=0.2\textheight]{imagethesis/Tomo/f004}&
\includegraphics[height=0.2\textheight]{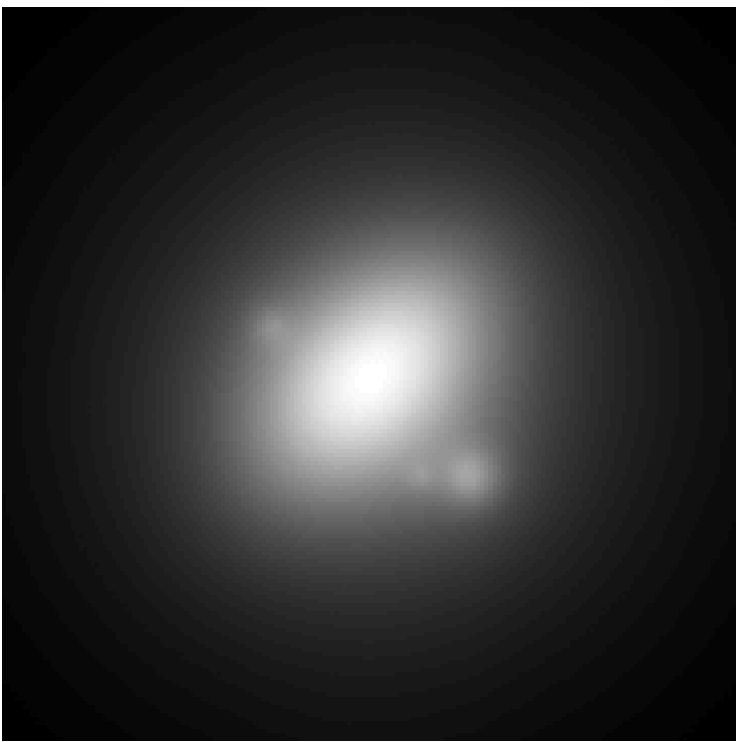}&
\includegraphics[height=0.2\textheight]{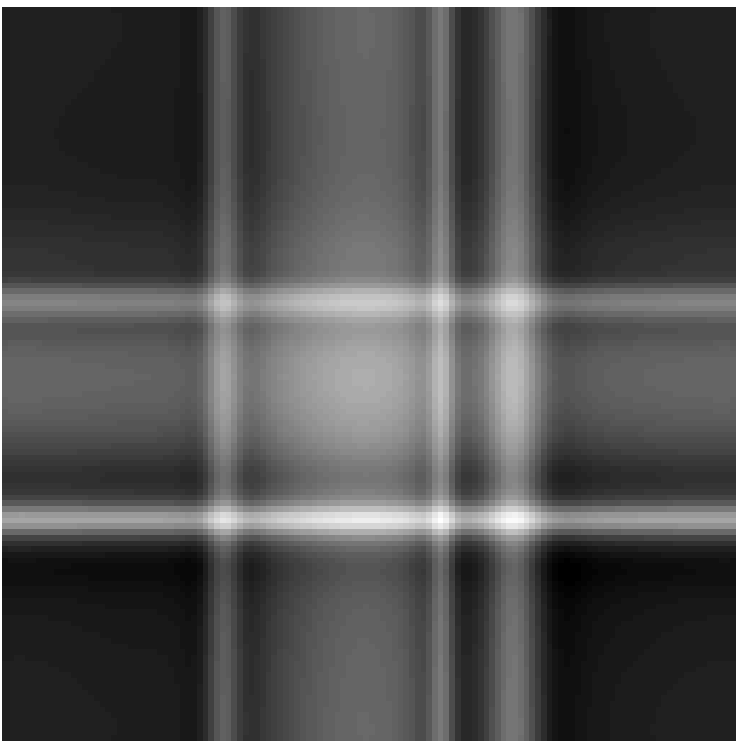}\\
 Original $f(x,y)$ & FBP: 128 projections & FBP: 2 projections\\ 
\hline
\end{tabular}
\end{center}
\caption{BP and FBP: using 128 and 2 projections.}
\label{comparative} 
\end{figure}

In X-ray CT, if we have a large number of projections uniformly distributed without noise in $\left[0,\pi \right]$ angles, the BP and the FBP are good solutions to the inverse problem. But when we have a few number of projections, BP and FBP images felt to be sufficient solutions Figure \ref{comparative}.

The definition and the notion of copula give us the possibility to propose another way to look at X-ray CT method. 

Let first consider the case of two projections. In this case, immediately, we can propose a first use which correspond to the copula $\Pi$. We call this method \emph{Multiplicative Backprojection (MBP)}. This name comes naturally if we compare the two equations (\ref{BP}) and (\ref{MBP}). 
In practice however, we have to normalise each projection in such a way that they can be assimilated to a pdf. 

{\bf MBP:} 
\beq
\boxed{f(x,y)=f_1(x) \, f_2(y)}
\label{MBP}
\eeq
Figure\ref{comparative1} shows two examples of comparisons between BP and MBP on a few simulated case. 
As we can see with only two projections, there is not any hope to reconstruct a complex shape object. 
We need more projections. 

Now let us have a closer look at expression (\ref{tomographyequation}), in an attempt to find a simple expression of $f(x,y)$ and to extend (\ref{MBP}). We need one more important property of copula.

\fbox{Invariance property}

The dependence captured by a copula is invariant with respect to increasing and continuous transformations $\Lambda$ of the marginal distributions \cite{Sch83}. This means that the same copula may be used for, the joint distribution of $(X_1,X_2)$ as $\left(\Lambda(X_1), \Lambda(X_2)\right) $ or $\left( \ln X_1,\ln X_2 \right) $, and thus whether the marginals are expressed in terms of natural units or logarithmic values does not affect the copula.

The general case and extension of (\ref{MBP}) using the invariance property can be formally written as:

{\bf General MBP:} 
\beq
\boxed{f(x,y)=f_1(x) \, f_2(y) \,c(x,y).}
\label{GMBP}
\eeq

\bit
\item <<Choose>> any copula density $c(x,y)$ (one of the list we have given) 
\item Normalise each projection in such a way to satisfy 
$p_{\theta}(r)\ge 0$ and $\displaystyle \int p_{\theta}(r)\d{r} = 1$.
\item For each projection, compute a backprojected image, and just multiply them pointwise, in place of adding them up.
\eit

Some few results from the copula-tomography package (see Appendix \ref{app1}) to simulate different methods of tomographic images reconstruction (BP, FBP, MBP and much more in future) are shown in the next section.

\bfig[h]
\label{comparative1}
\bcc
\btabu{ccc}
\includegraphics[height=0.2\textheight]{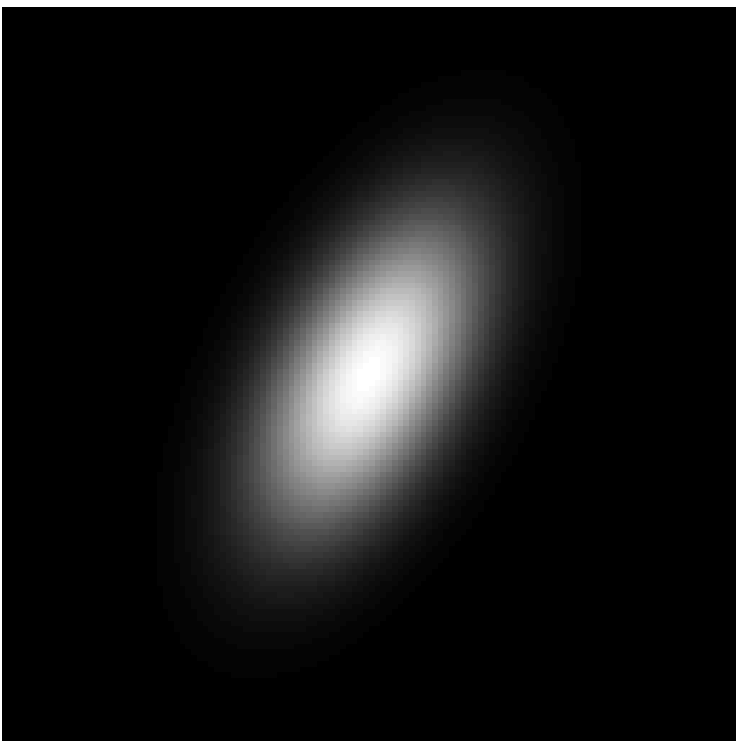}&
\includegraphics[height=0.2\textheight]{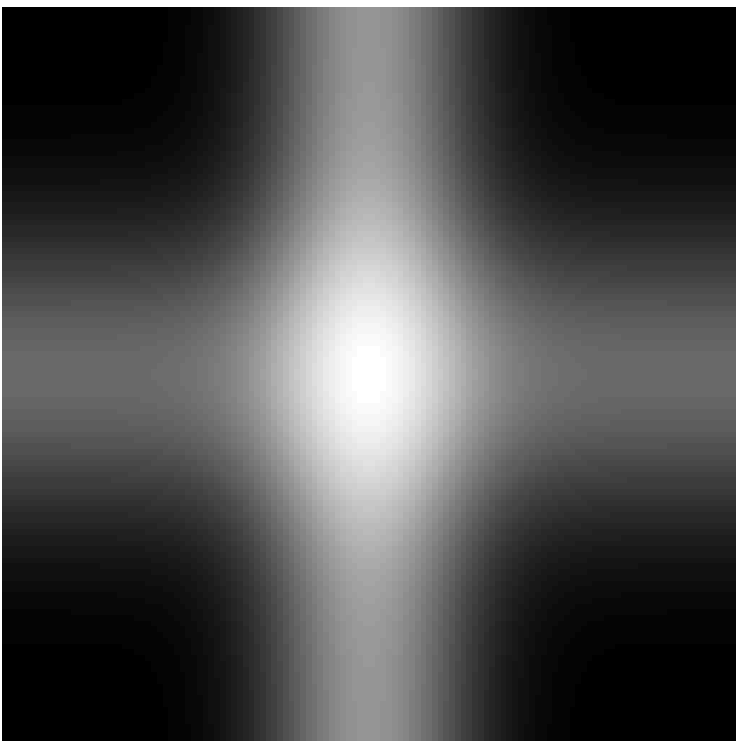}&
\includegraphics[height=0.2\textheight]{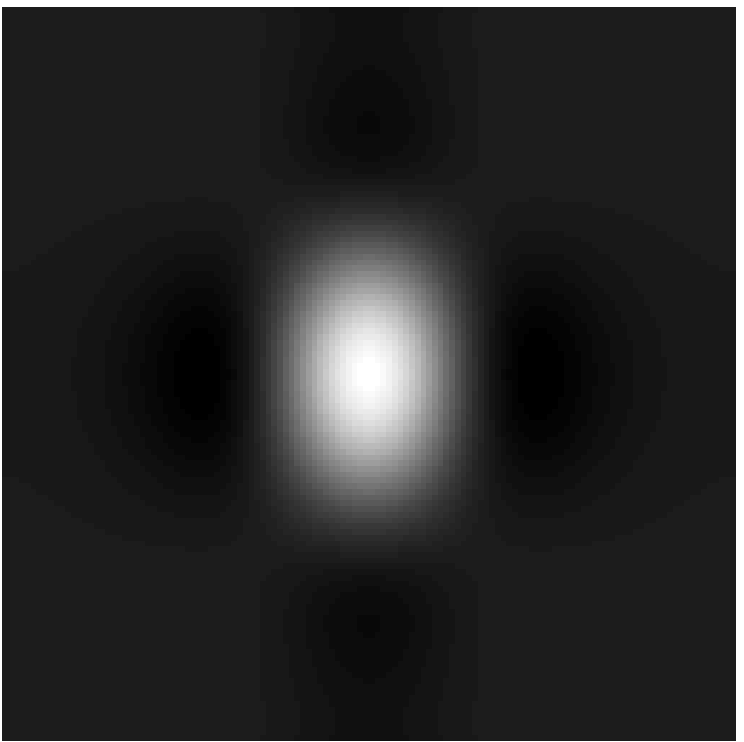}\\
Original $f(x,y)$ & BP $\widehat{f}(x,y)$ & MBP $\widehat{f}(x,y)$\\ 
\includegraphics[height=0.2\textheight]{imagethesis/Tomo/f004}&
\includegraphics[height=0.2\textheight]{imagethesis/Tomo/fh004_bp}&
\includegraphics[height=0.2\textheight]{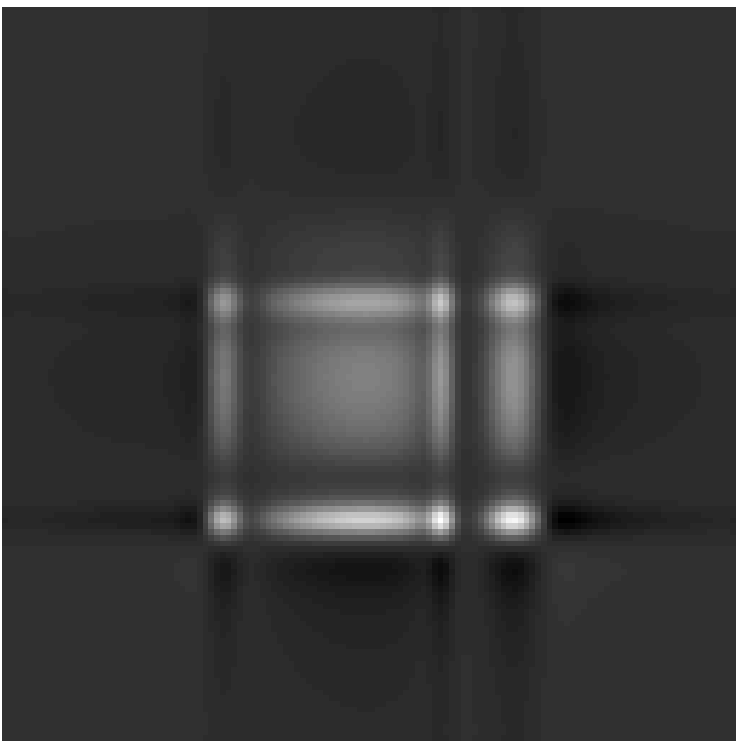}\\
Original $f(x,y)$ & BP $\widehat{f}(x,y)$ & MBP $\widehat{f}(x,y)$\\ 
\includegraphics[height=0.2\textheight]{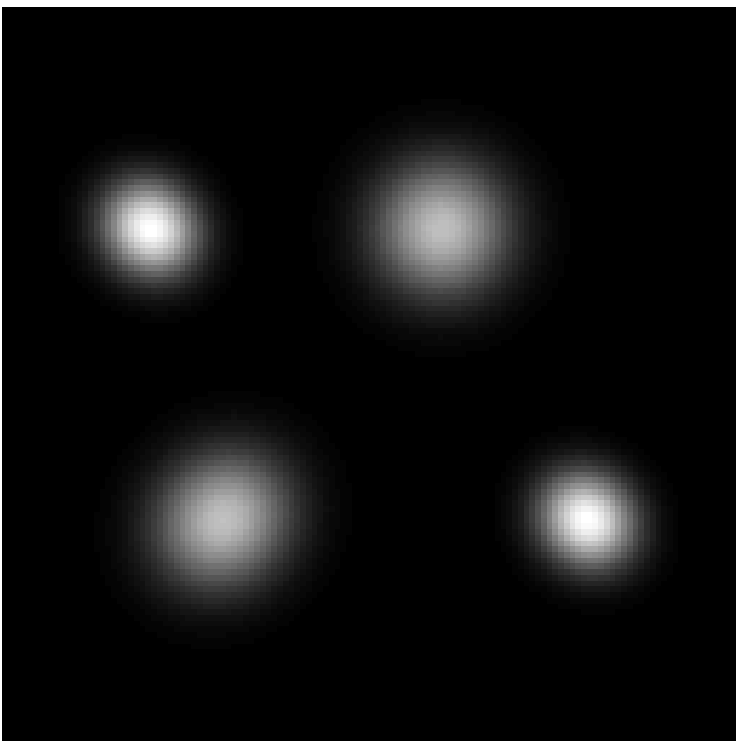}&
\includegraphics[height=0.2\textheight]{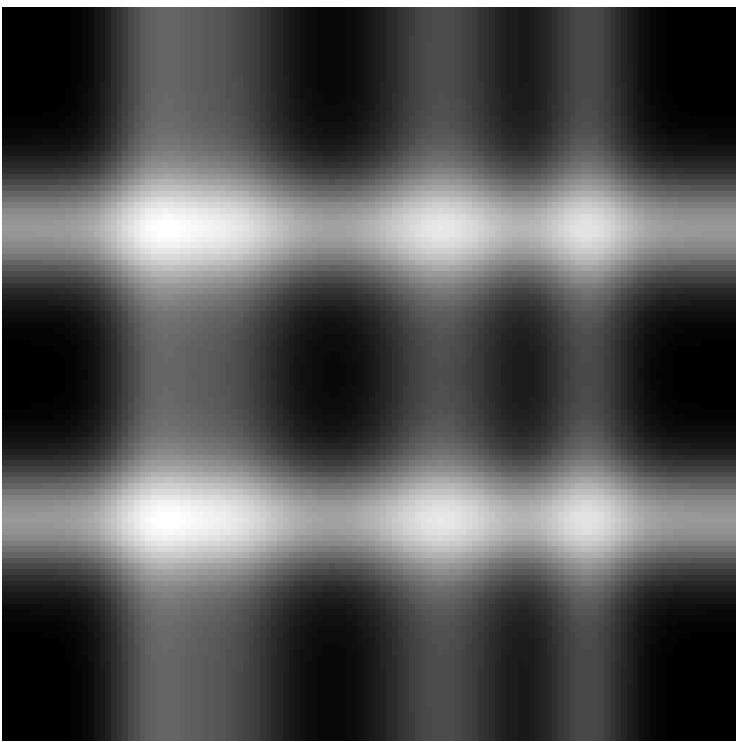}&
\includegraphics[height=0.2\textheight]{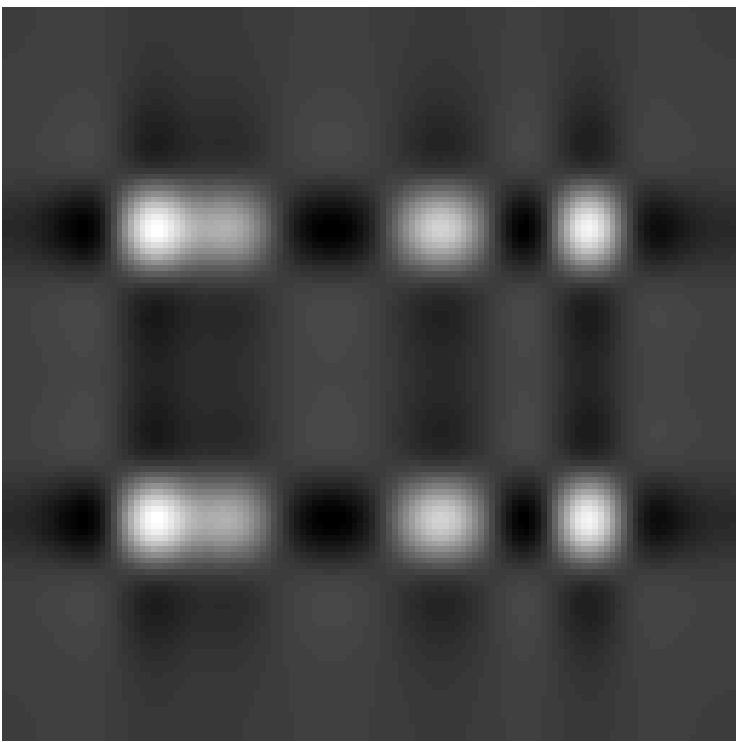}\\
Original $f(x,y)$ & BP $\widehat{f}(x,y)$ & MBP $\widehat{f}(x,y)$\\ 
\etabu
\ecc
\caption{BP and MBP on three synthetic examples: using only 2 projections.}
\efig

\bfig
\label{comparative2}
\bcc
\btabu{ccc}
\includegraphics[height=0.2\textheight]{imagethesis/Tomo/f001}&
\includegraphics[height=0.2\textheight]{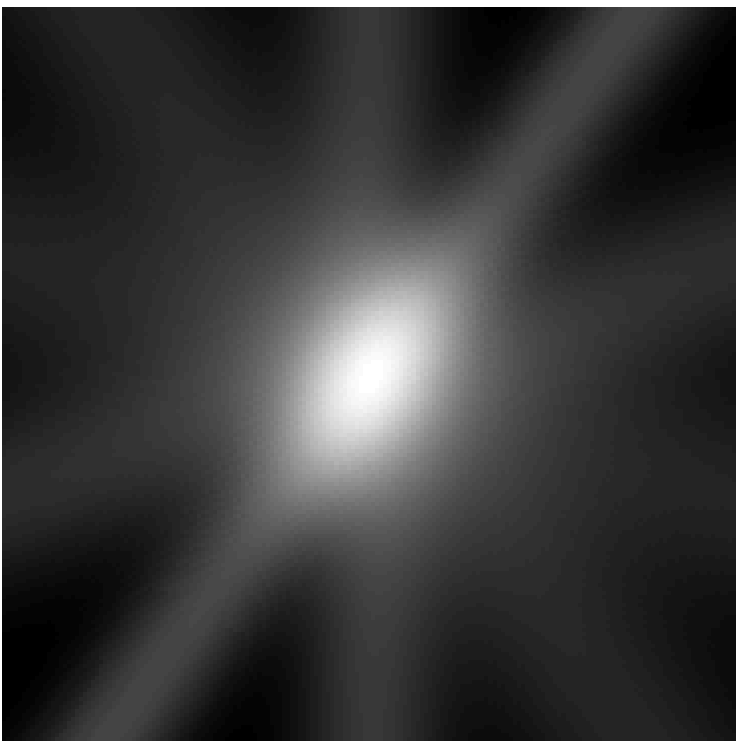}&
\includegraphics[height=0.2\textheight]{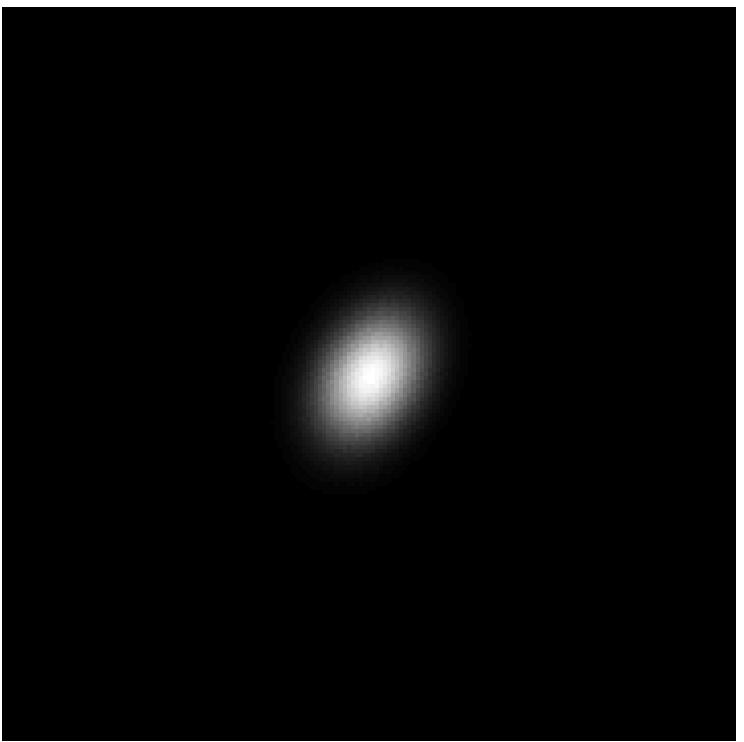}\\
a)Original $f(x,y)$ & b) BP $\widehat{f}(x,y)$ & c) MBP $\widehat{f}(x,y)$\\ 
\includegraphics[height=0.2\textheight]{imagethesis/Tomo/f004}&
\includegraphics[height=0.2\textheight]{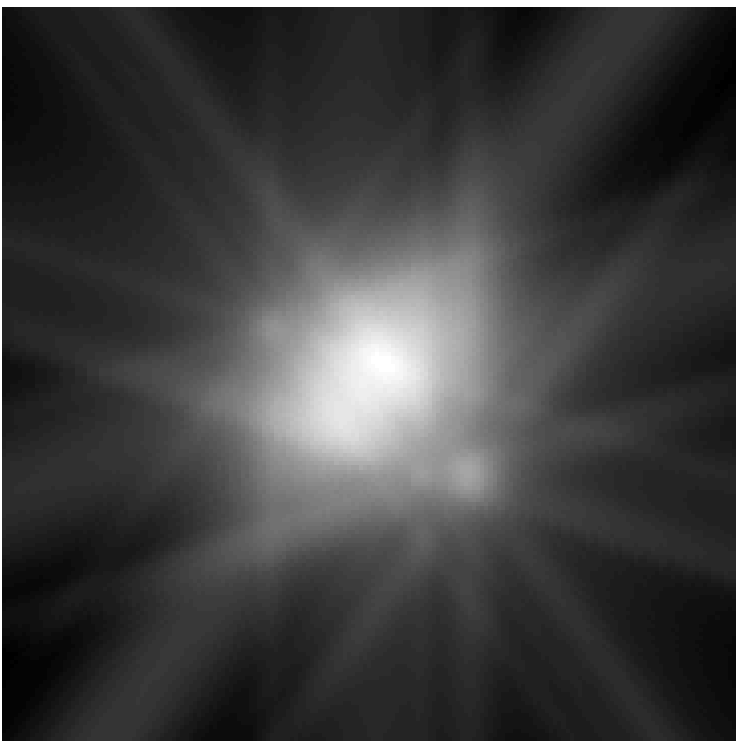}&
\includegraphics[height=0.2\textheight]{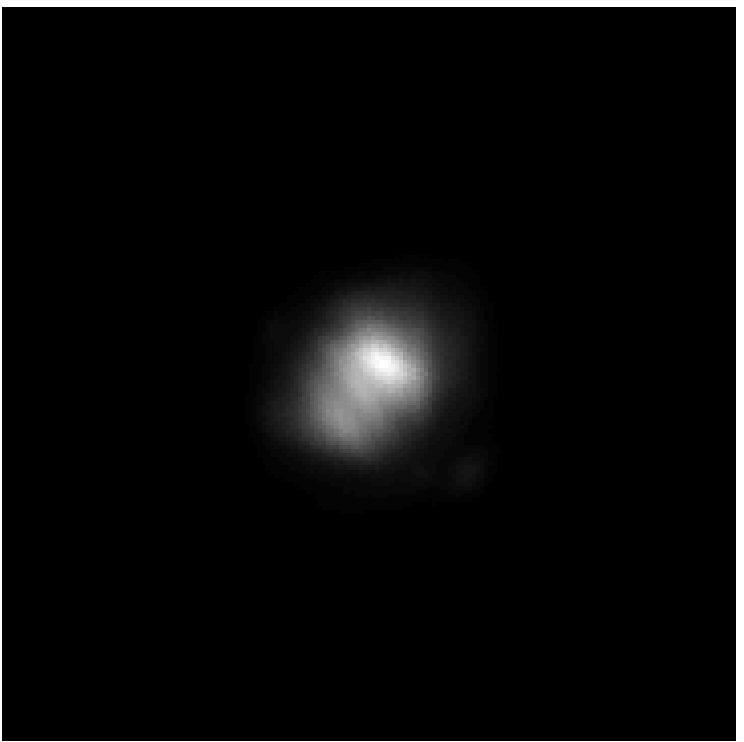}\\
d)Original $f(x,y)$ & e)BP $\widehat{f}(x,y)$ & f)MBP $\widehat{f}(x,y)$\\ 
\includegraphics[height=0.2\textheight]{imagethesis/Tomo/f005}&
\includegraphics[height=0.2\textheight]{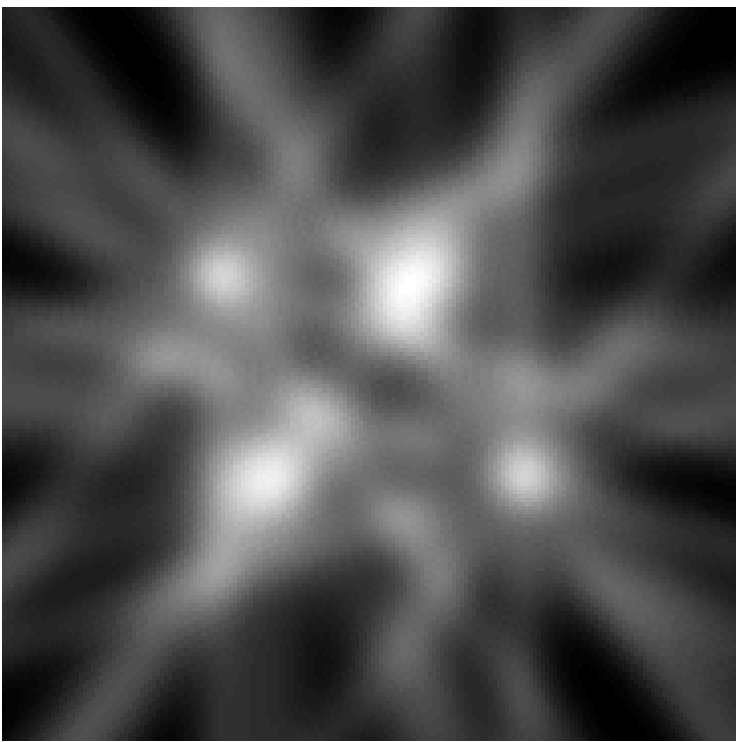}&
\includegraphics[height=0.2\textheight]{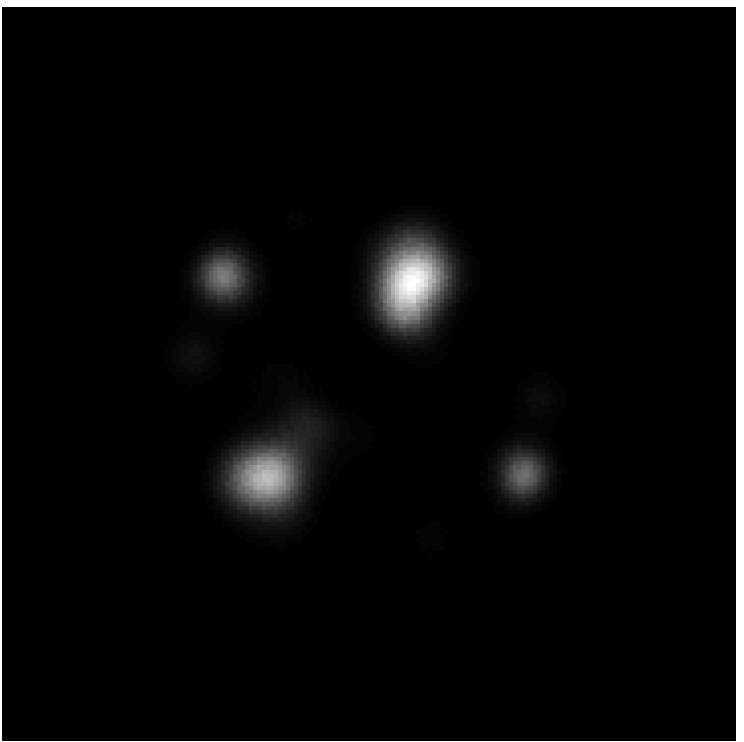}\\
g)Original $f(x,y)$ & h)BP $\widehat{f}(x,y)$ & i)MBP $\widehat{f}(x,y)$\\ 
\etabu
\ecc
\caption{BP and MBP on three synthetic examples: using 05  projections.}
\efig

From equation(\ref{GMBP}) we have generated $f(x,y)$, then we compute the two marginal functions $f_1$ and $f_2$. 
Our reconstruction used a test based on choice of copulas densities $c(x,y)$. 

There is a comparative result from only 5 projections in Figure \ref{comparative2} between BP and MBP. 

We clearly observe that, in this particular example of simulation using one Gaussian pdf\footnote{see Figure \ref{comparative2} a), b) and c)} and four Gaussians pdf's \footnote{see Figure \ref{comparative2} g), h) and i)} as original image to be reconstructed; the originate(s) Gaussian(s) are reconstructed but having small variance(s).%  Bridging the gap between Copula and Tomography
\chapter{Conclusion}
In this report, we have started an other mathematical approach from the theory of copula which could be used in tomography.
The main contribution of this report is to find a link between the notion of \emph{copulas} in statistics and X-ray CT. For this, first we have presented briefly the bivariate copulas and the image reconstruction problem in CT.
We have also implemented a Matlab code for copula and classical tomography method in order to simulate an image reconstruction and then present some preliminary result.

We could make a link between the two problems of 
\begin{enumerate}[i)]
 \item determining a joint bivariate pdf from its two marginals and
 \item the CT image reconstruction from only two horizontal and vertical projections, by emphasising that in both cases, we have the same inverse problem of determining a bivariate function (an image) from the line integrals.
\end{enumerate}

There are clearly many questions that we have left unanswered, among those questions we draw attention to:
\begin{enumerate}[1)]
\item the way to develop a strong statistical framework and methods including mixture of copulas and taking account of parameters in order to reconstruct with accuracy any complicated shape of objects,
\item how to choose and/or construct a new family of copula having a nice property for image reconstruction.
\end{enumerate}

Further research may be undertaken in order to deepen understanding the relation between copula and tomography for applications. Nevertheless, we hope we have managed to give the reader some useful insight into, and sparked their interest in this discussion situated in the border of statistical mathematics and the fascinating area of imaging sciences.
%  conclusion 

\appendix
\chapter{Tomography-copula package}
\label{app1}
Here is a short description of the package we have written using Matlab which allows users to simulate a tomographic image reconstruction via a wide range of copula family and mixture of Gaussians pdf's. We hope that 
this package (still under development) will help to enjoy the joy of copula in tomography.

\begin{figure}[!h!]
 \begin{center}
 {
 \includegraphics[height=0.4\textheight]{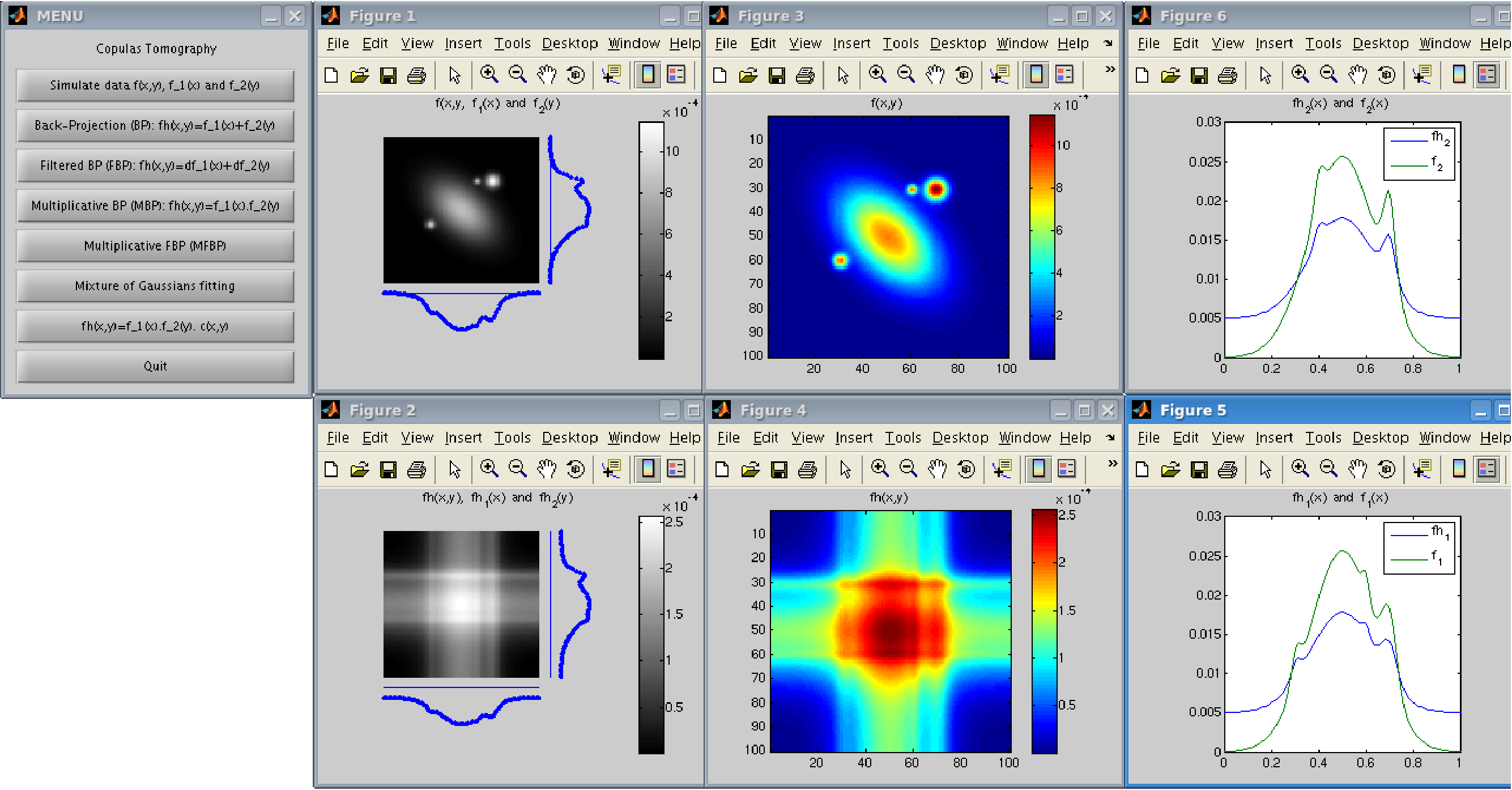}
 }
 \caption{<<Menu>> showing BP simulation: 4 Gaussians mixtures}
 \label{interface}
 \end{center}
\end{figure}

\section{How it works}
\subsection{Menu}
The main menu is titled : Copulas Tomography \\
It gives a user interface that allows to make selections and choices from the following preset lists:
\begin{enumerate}
 \item Simulate data $f(x,y),\, f_1(x) \, \mbox{and} \, f_2(y)$
 \item Back-Projection (BP): $f_h(x,y)=f_1(x)+f_2(y)$
 \item Filtered BP (FBP): $f_h(x,y)=df_1(x)+df_2(y)$
 \item Multiplicative BP (MBP): $f_h(x,y)=f_1(x).f_2(y)$
 \item Multiplicative FBP (MFBP): $f_h(x,y)=df_1(x).df_2(y)$
 \item Mixture of Gaussians fitting
 \item $f_h(x,y)=f_1(x).f_2(y). c(x,y)$
 \item Quit
\end{enumerate}

\subsection{Menu List}
Let us explain each part of the previous list:
\begin{enumerate}
\item \fbox{Simulate data $f(x,y)$, $f_1(x)$ and $f_2(y)$}\\

In fact, this list is a pop-up menu which contains a group of five choices in its own window.
Those choices are related to the kind of data to be simulated. 
Those data represent $f(x,y)$, the original image.  We offer to the user the choice between one Gaussian 
or two to nine mixture of Gaussians. After the selection of the data, clicking on <<quit>> will close 
the pop-up menu and allow the user to return to the main menu.

\item \fbox{Back-Projection (BP): $f_h(x,y)=f_1(x)+f_2(y)$}\\

This list offers a visualisation of the reconstruction using Back-Projection method.
$f_h(x,y)$ is the image reconstructed.

\item \fbox{Filtered BP (FBP): $f_h(x,y)=df_1(x)+df_2(y)$}\\

This list offers a visualisation of the reconstruction using Filtered Back-Projection method.

\item \fbox{Multiplicative BP (MBP): $f_h(x,y)=f_1(x).f_2(y)$}\\

This list offers a visualisation of the tomographic image reconstruction using Multiplicative Back-Projection method.

\item \fbox{Multiplicative FBP (MFBP) $f_h(x,y)=df_1(x).df_2(y)$}\\

This list offers a visualisation of the reconstruction using Multiplicative Filtered Back-Projection method.

\item \fbox{Mixture of Gaussians fitting}\\

Source code under development.

\item \fbox{ $f_h(x,y)=f_1(x).f_2(y). c(x,y)$}\\

Click on this list of the main menu yields to a large number of copula family $c(x,y)$ to reconstruct the original object. We offer the reconstruction through Archimedean families of copula, discrete copula and much more.
One could also add many other families of copulas.

\item \fbox{ Quit} : To exit from the main menu.

\end{enumerate}

\chapter{How do copulas look like ?}
\label{app2}
Here are some graphical representation of copulas, we have discussed in this report.
We remind in 2D case, that $F_1(x)$ and $F_2(y)$ are the marginal cumulative distribution functions (cdf's) related to  joint cdf $F(x,y)$ and the marginal probability density functions $f_1(x)$ and $f_2(y)$ are linked to their joint probability density function $f(x,y)$ via the horizontal and vertical line integrals.\\
We have also $c(x,y)$ for the copula pdf, to be distinguished from the copula cdf $C(x,y)$ (with capital letter $C$).
\bfig[h]
\bcc
\btabu{ccc}
\includegraphics[height=0.17\textheight]{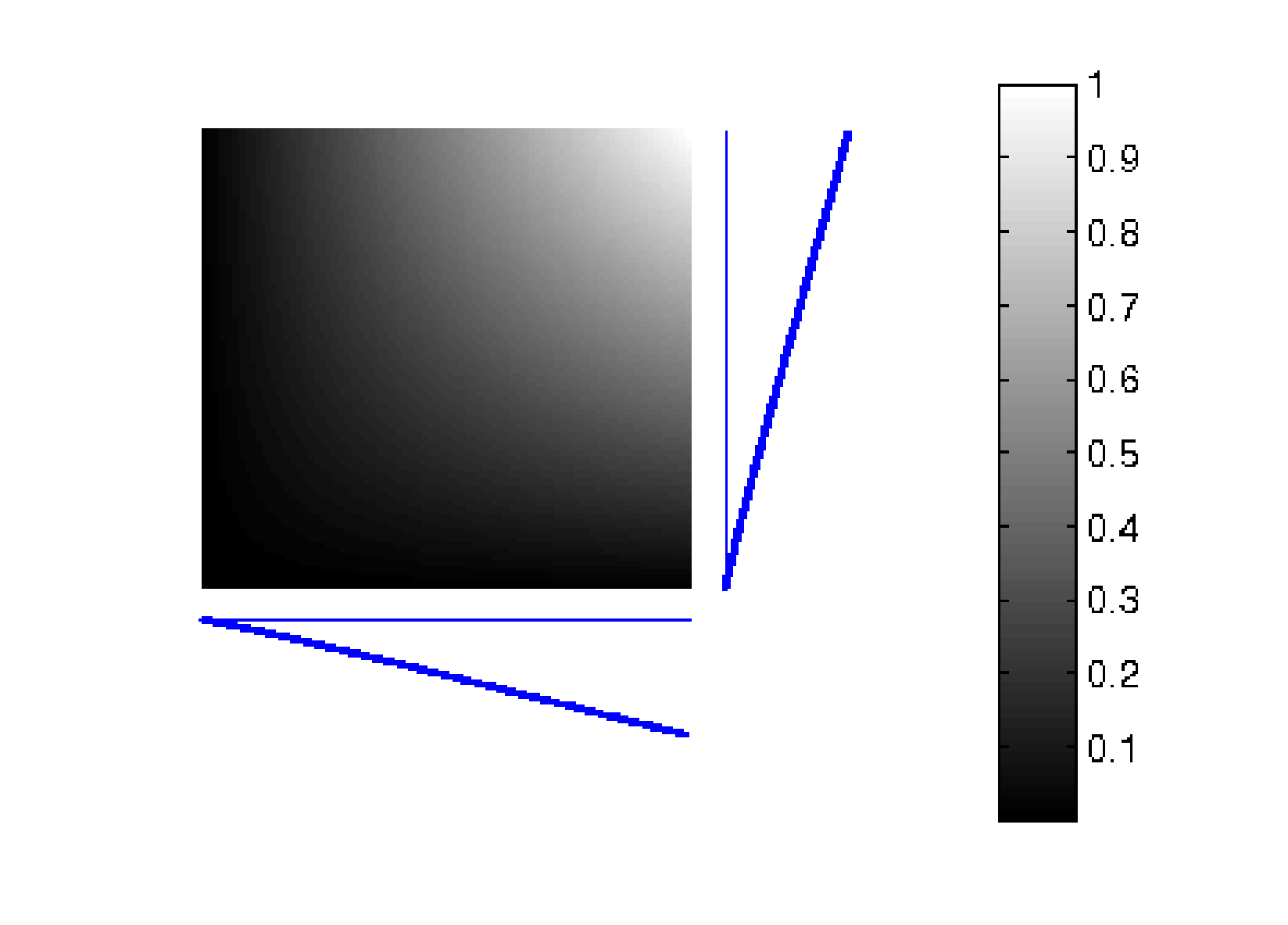}&
\includegraphics[height=0.17\textheight]{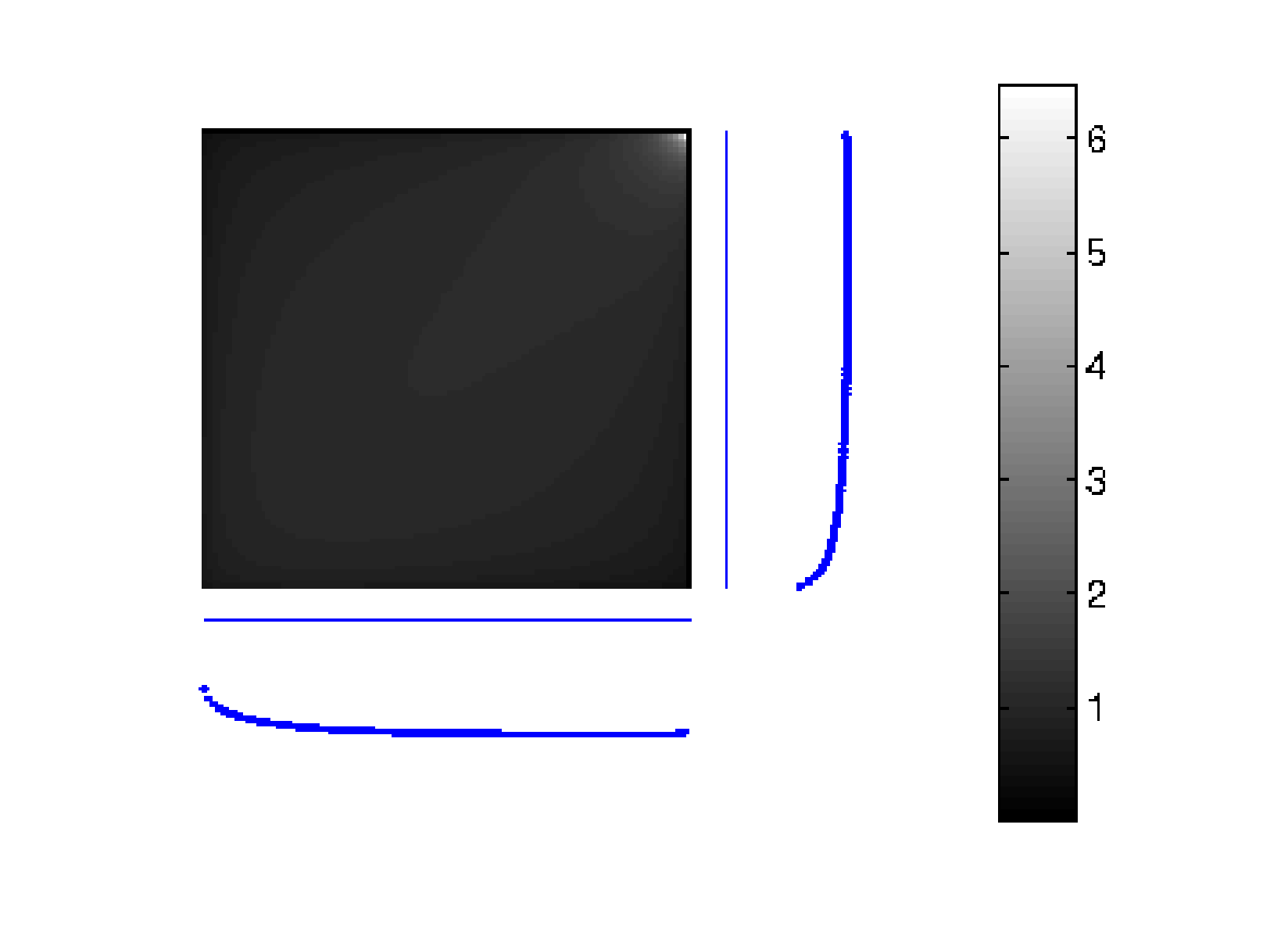}&
\includegraphics[height=0.17\textheight]{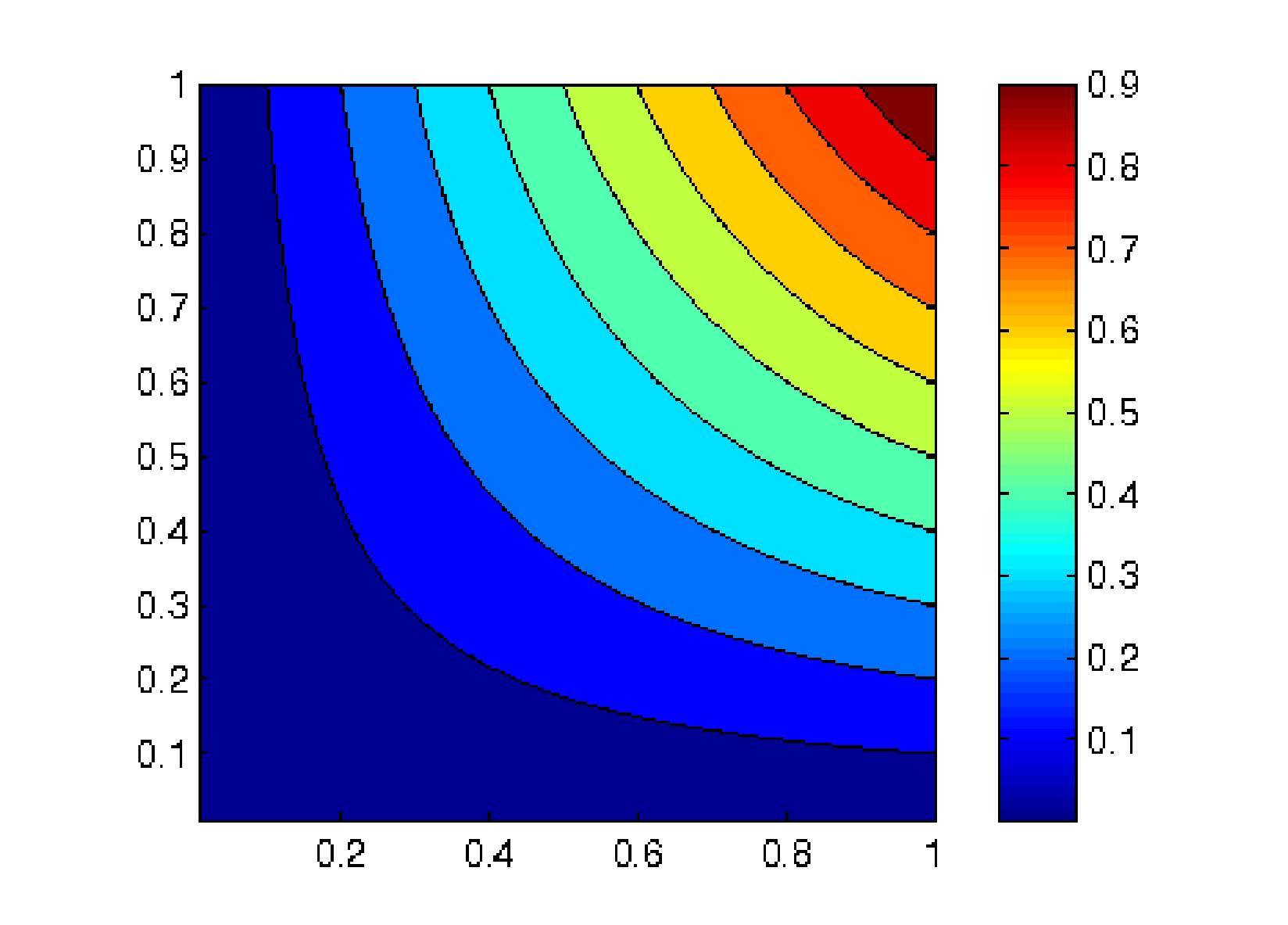}\\
 $F(x,y)$,$F_1(x)$,$F_2(y)$ & $f(x,y)$ with $f_1(x)$ and $f_2(y)$ & $C(x,y)$ contours plot\\ 
\includegraphics[height=0.17\textheight]{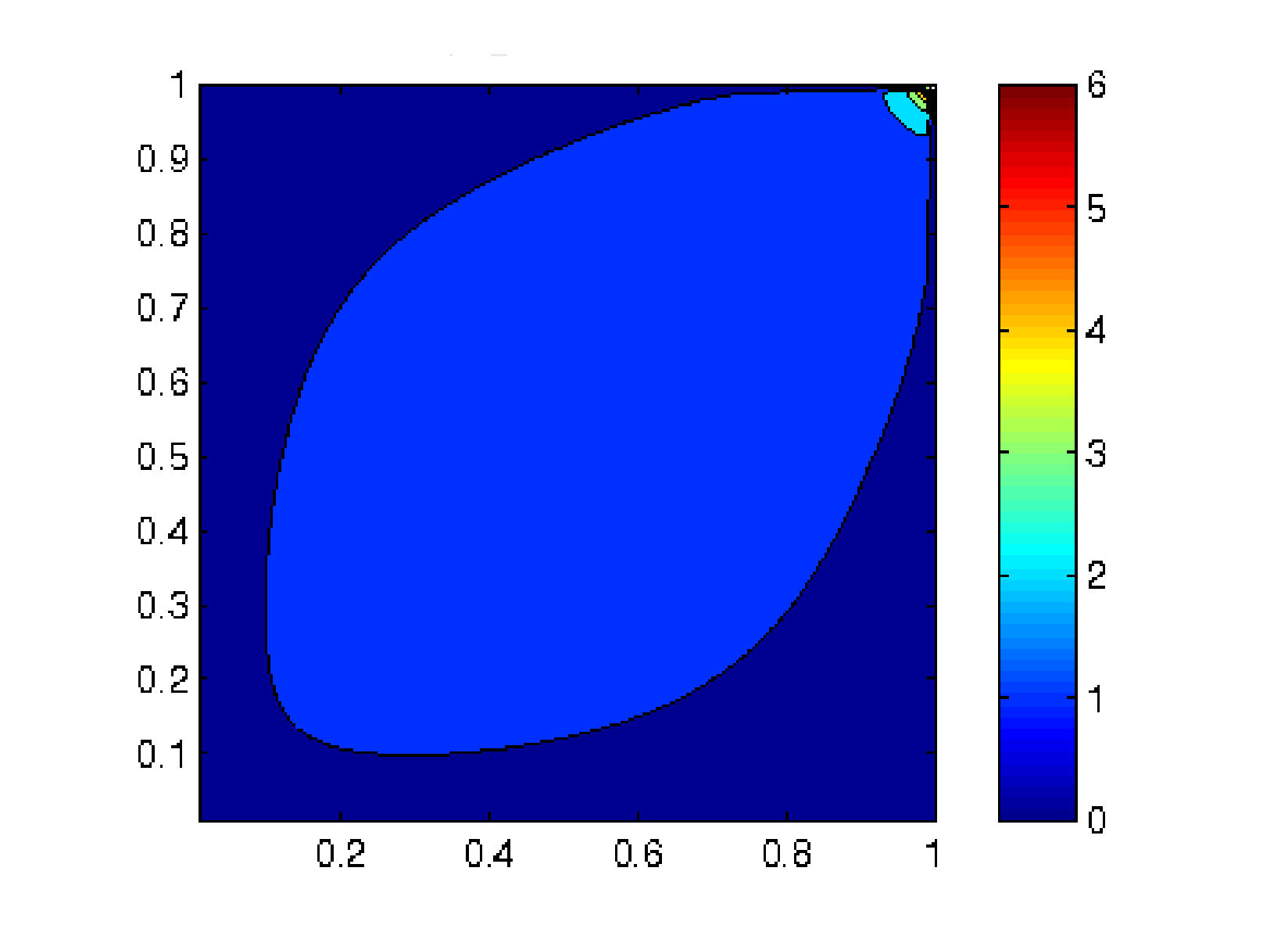}&
\includegraphics[height=0.17\textheight]{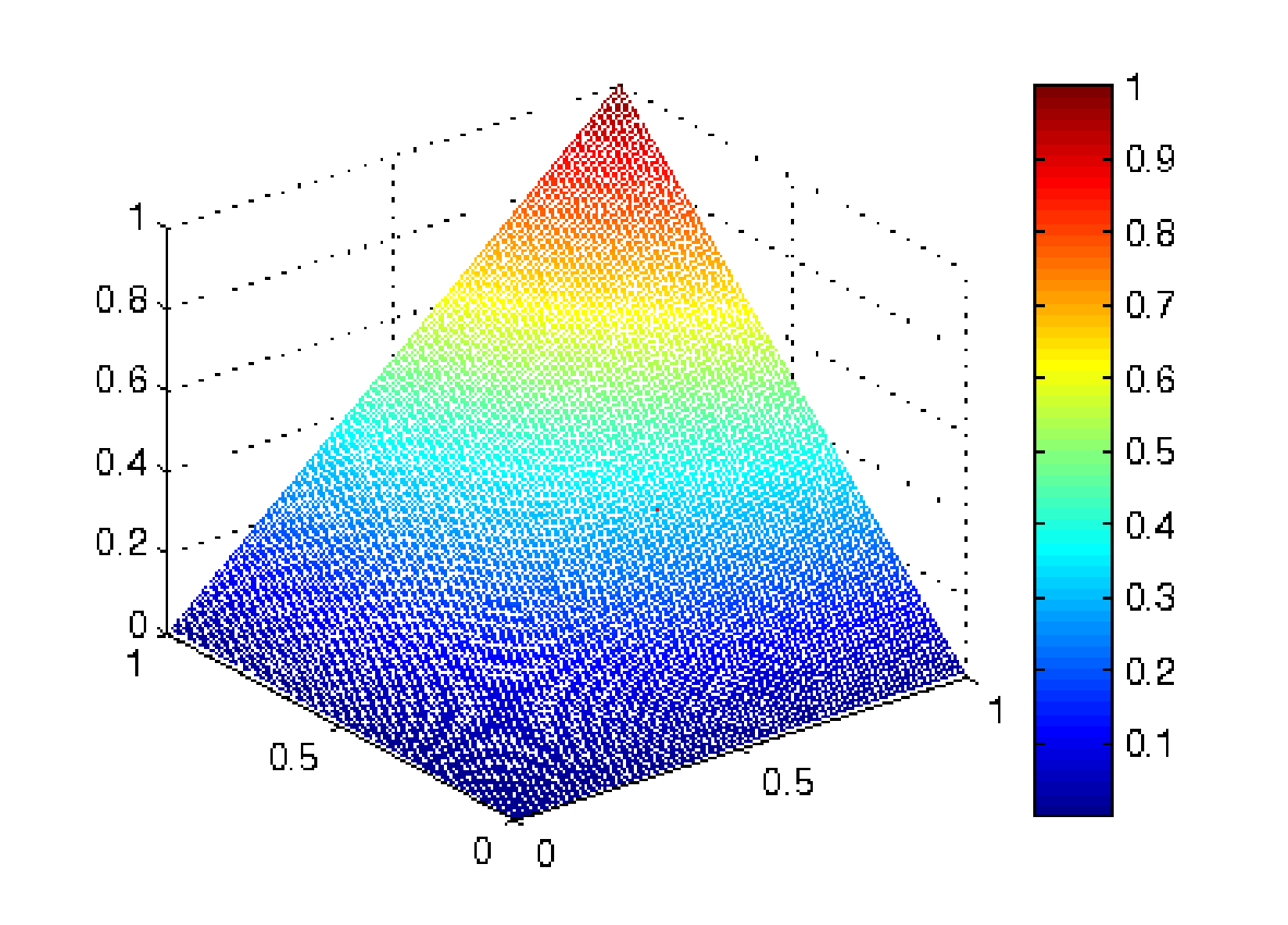}&
\includegraphics[height=0.17\textheight]{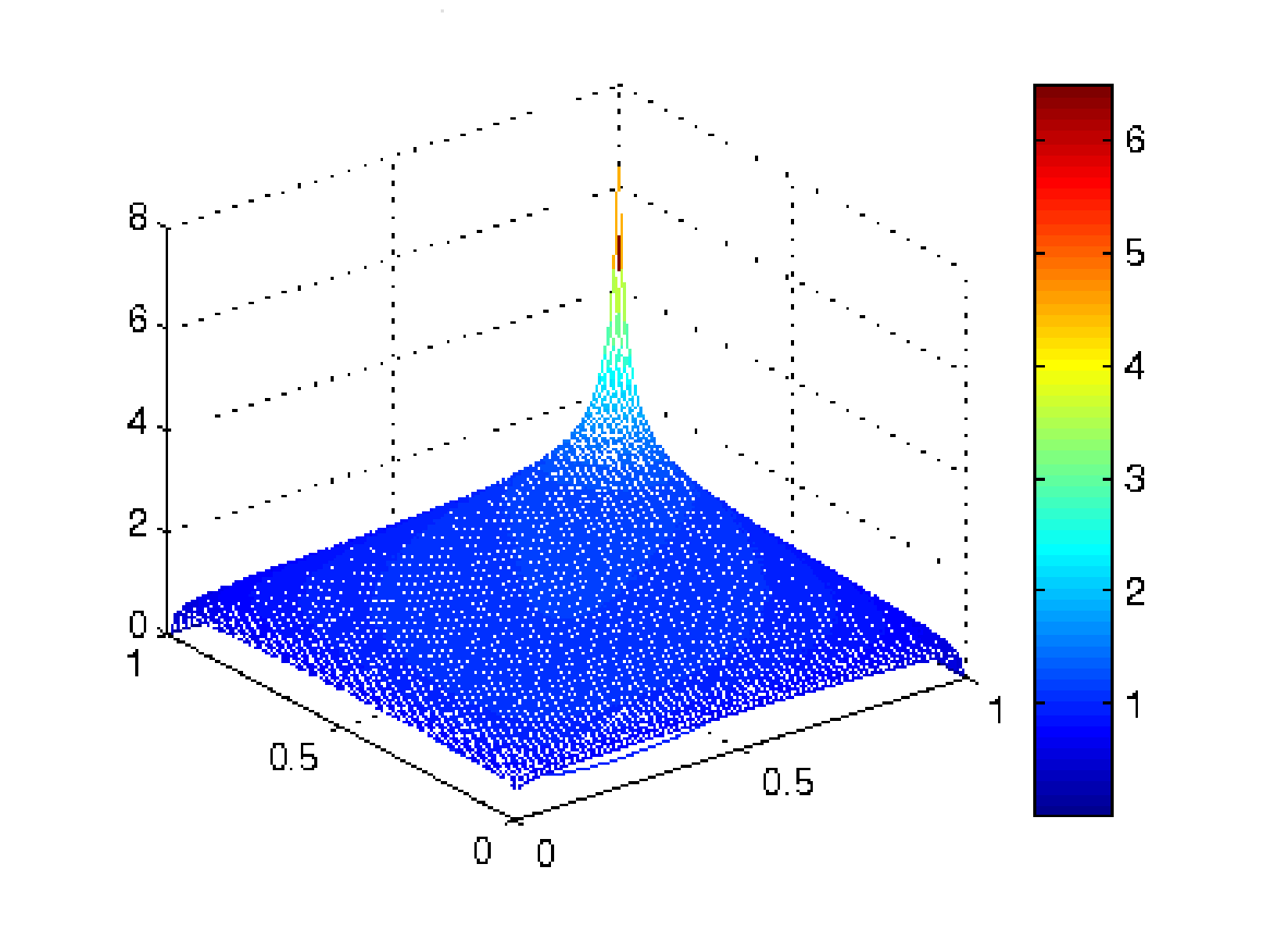}\\
$c(x,y)$ contours plot & $C(x,y)$ mesh plot & $c(x,y)$ mesh plot\\ 
\etabu
\ecc
\caption{Gumbel copula with $\alpha=1.1$.}
\efig

\bfig
\bcc
\btabu{ccc}
\includegraphics[height=0.17\textheight]{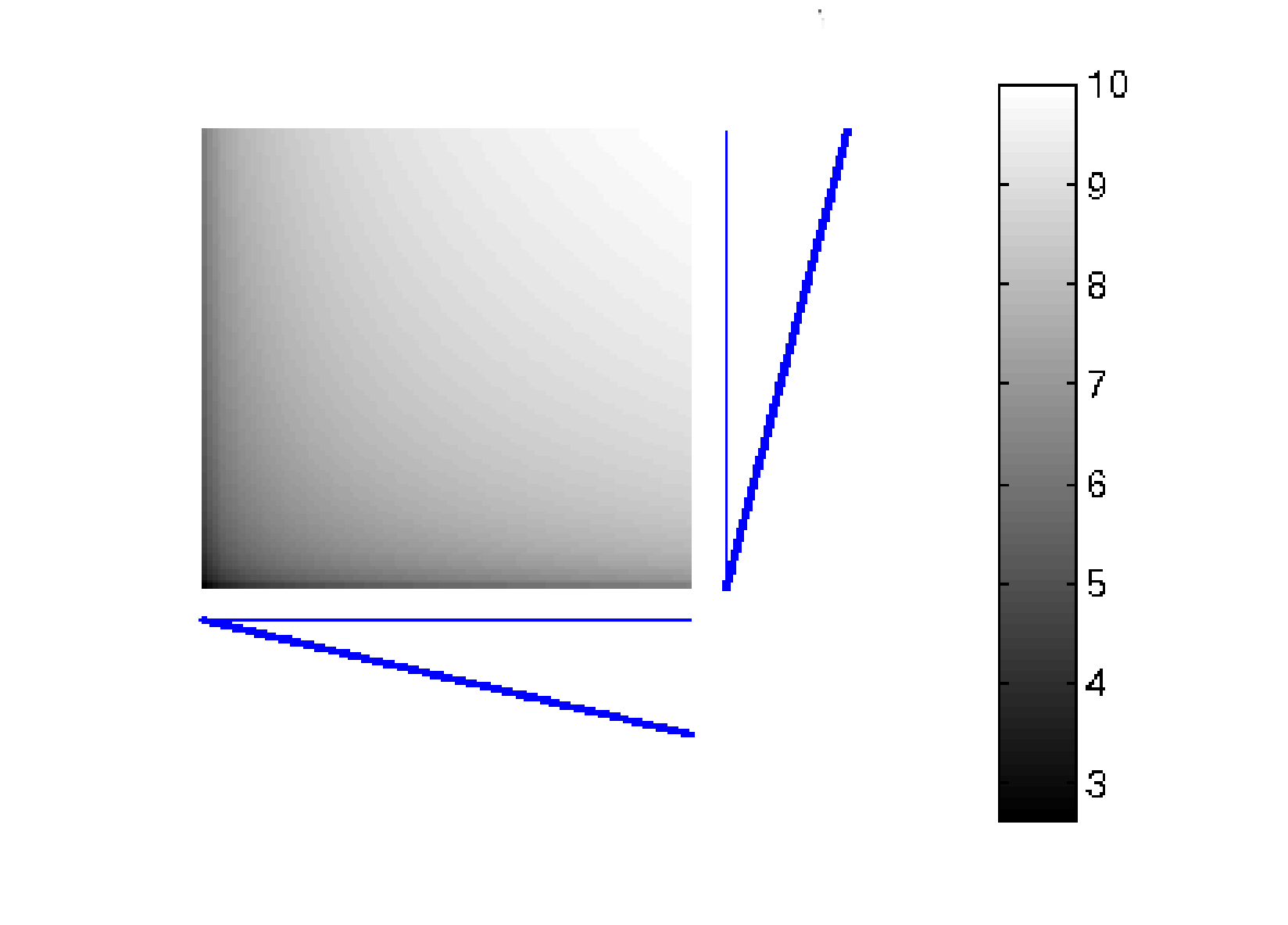}&
\includegraphics[height=0.17\textheight]{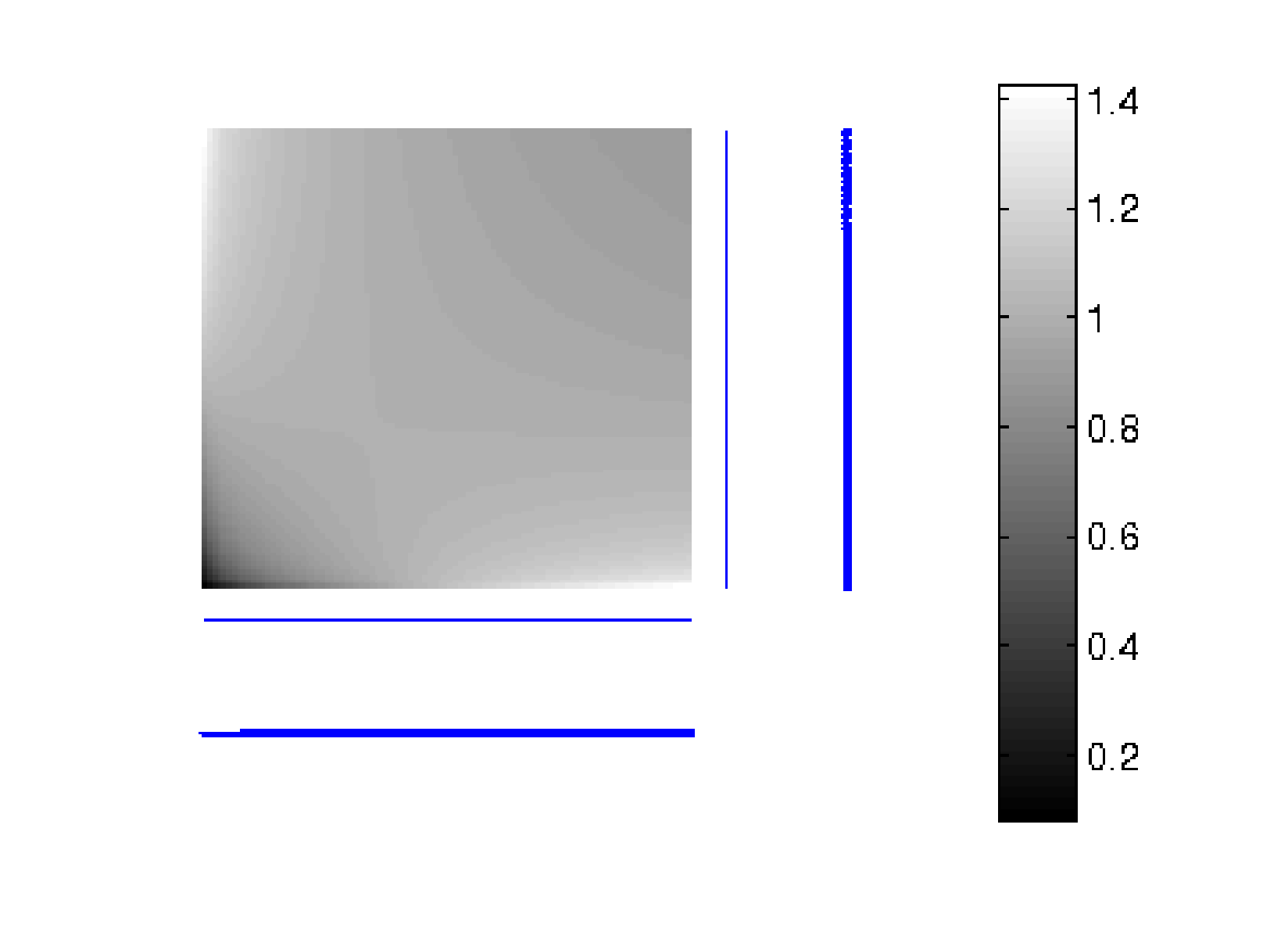}&
\includegraphics[height=0.17\textheight]{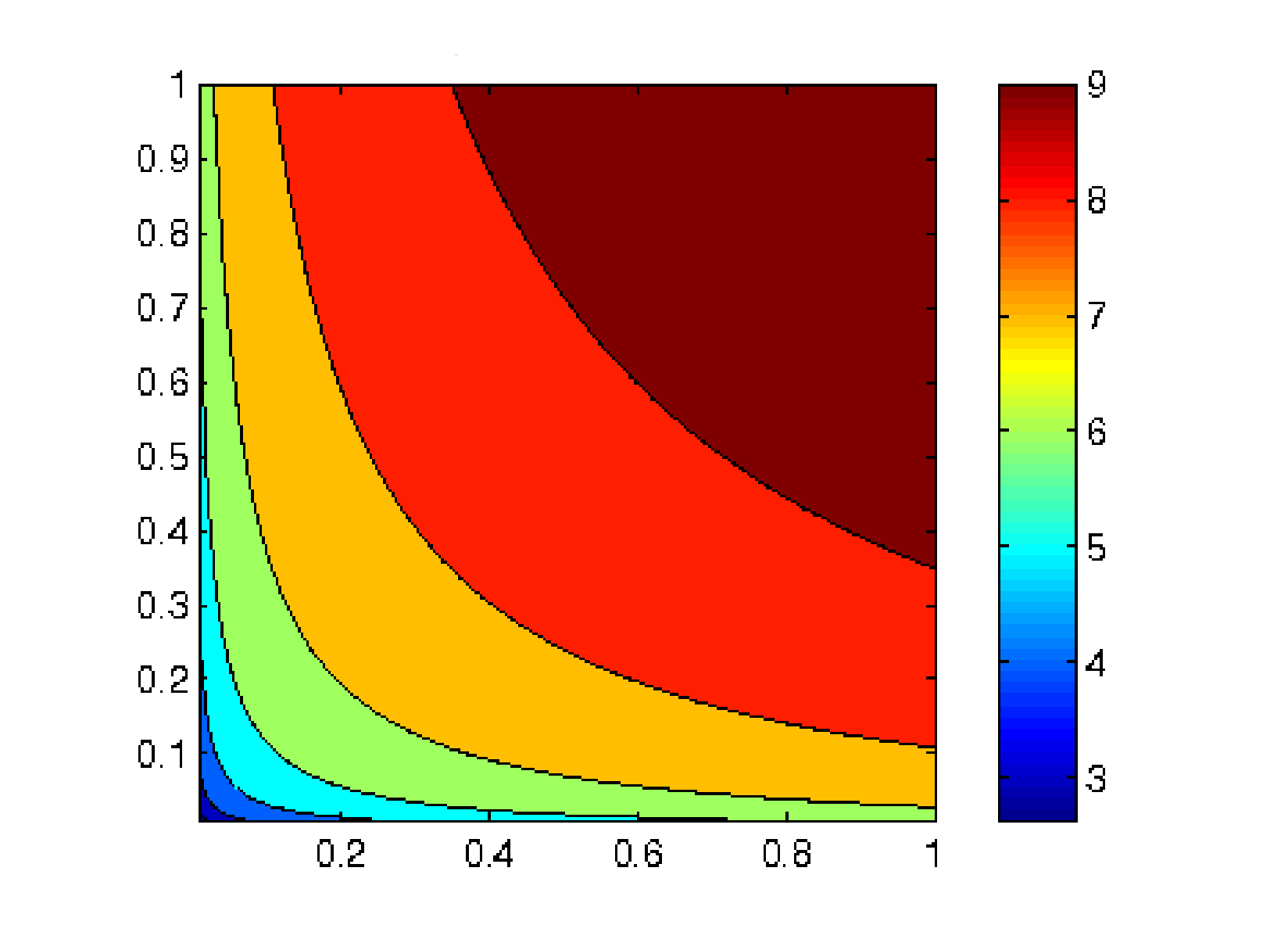}\\
$F(x,y)$,$F_1(x)$,$F_2(y)$ & $f(x,y)$ with $f_1(x)$ and $f_2(y)$ & $C(x,y)$ contours plot\\ 
\includegraphics[height=0.17\textheight]{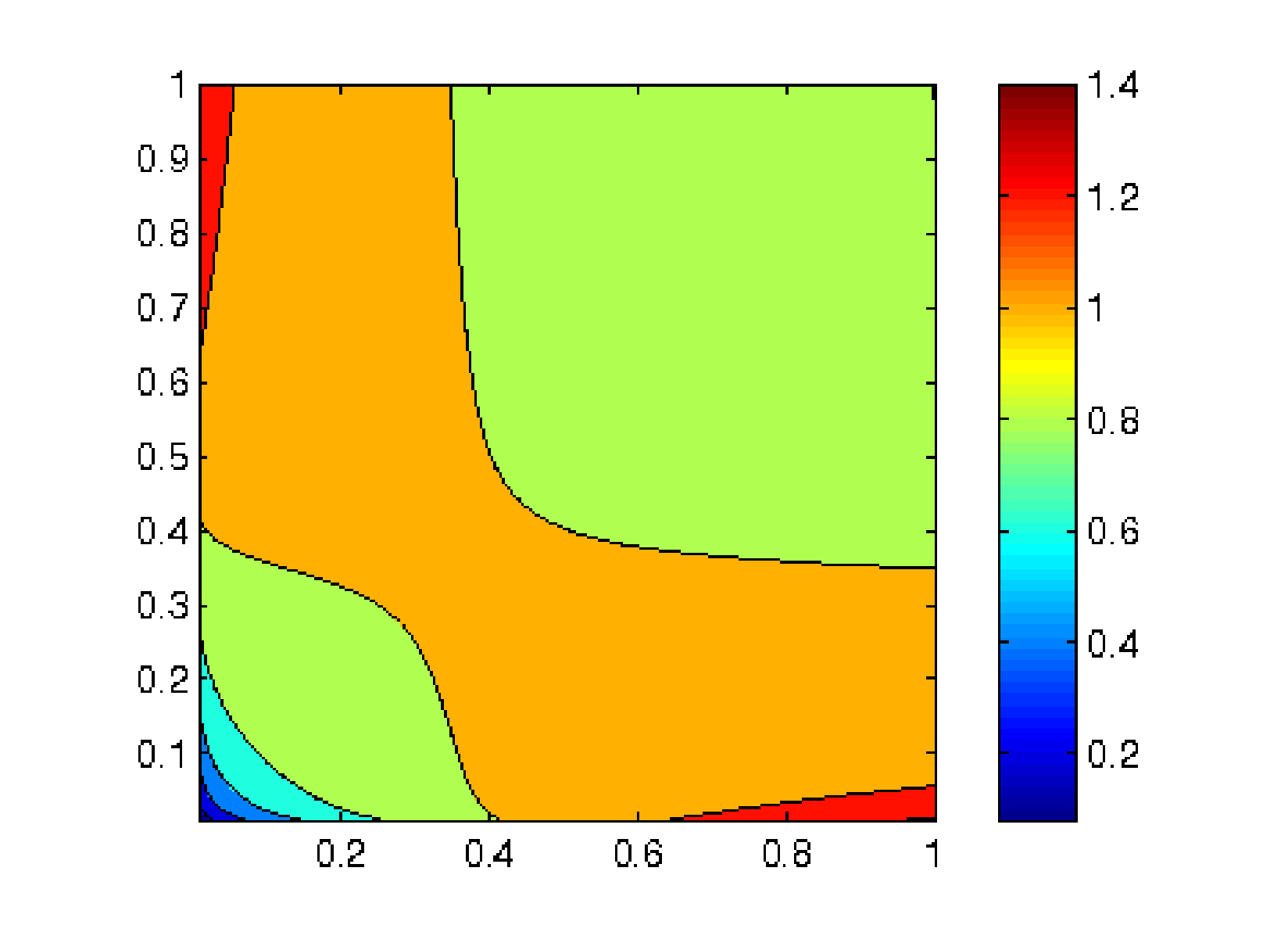}&
\includegraphics[height=0.17\textheight]{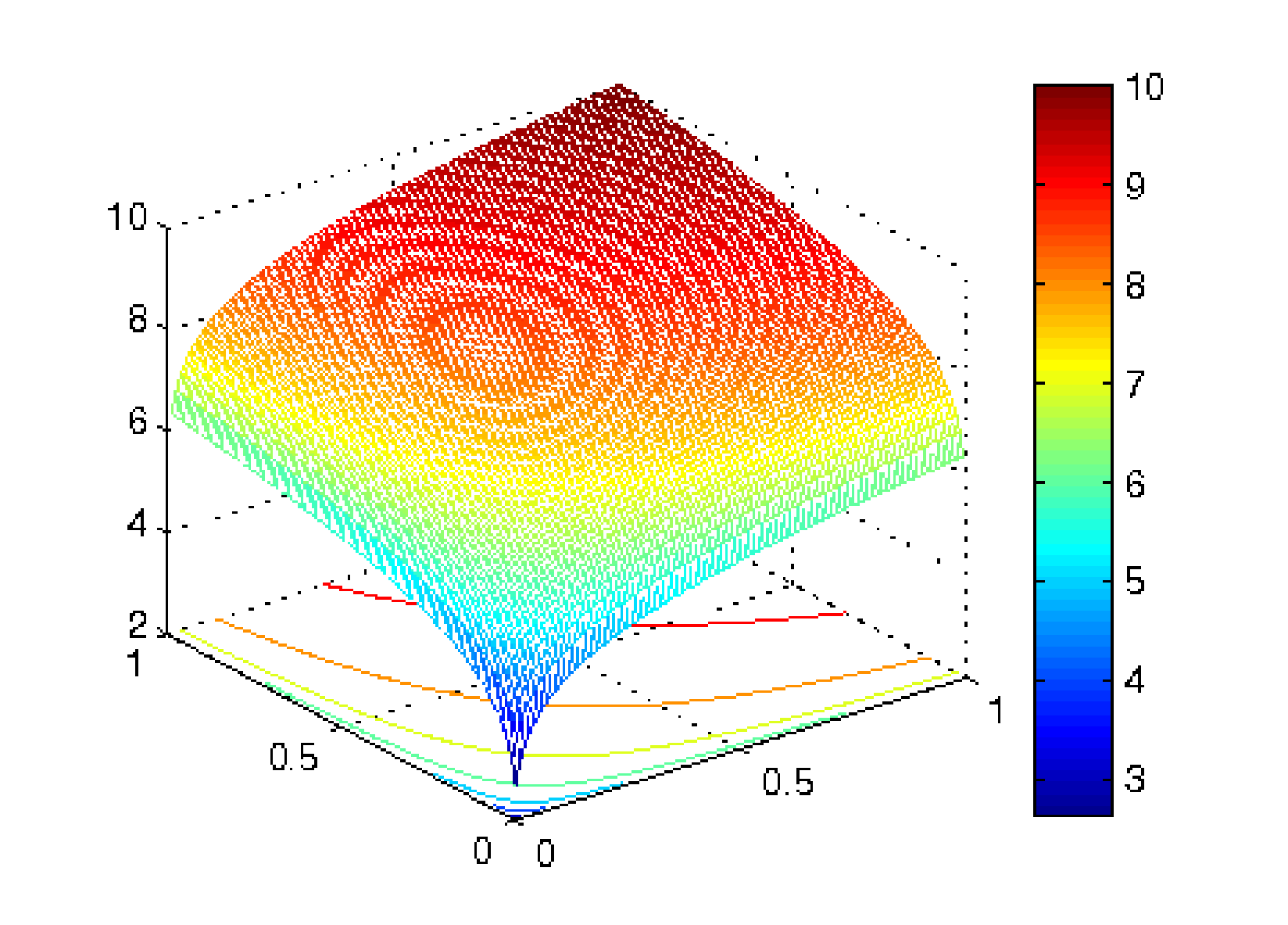}&
\includegraphics[height=0.17\textheight]{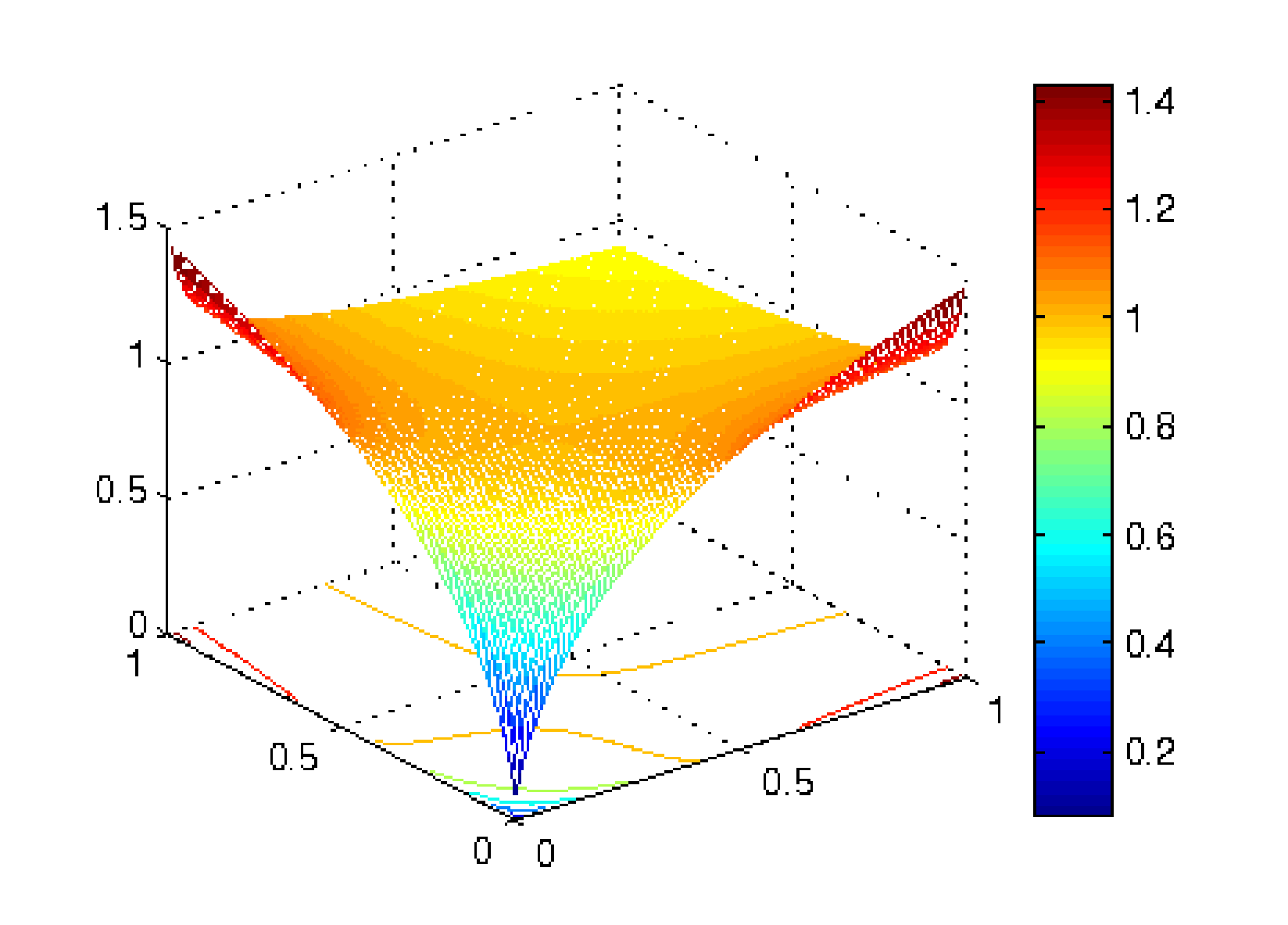}\\
$c(x,y)$ contours plot  & $C(x,y)$ mesh plot  & $c(x,y)$ mesh plot\\ 
\etabu
\ecc
\caption{Clayton copula with $\alpha=-0.1$.}
\efig

\bfig
\bcc
\btabu{ccc}
\includegraphics[height=0.17\textheight]{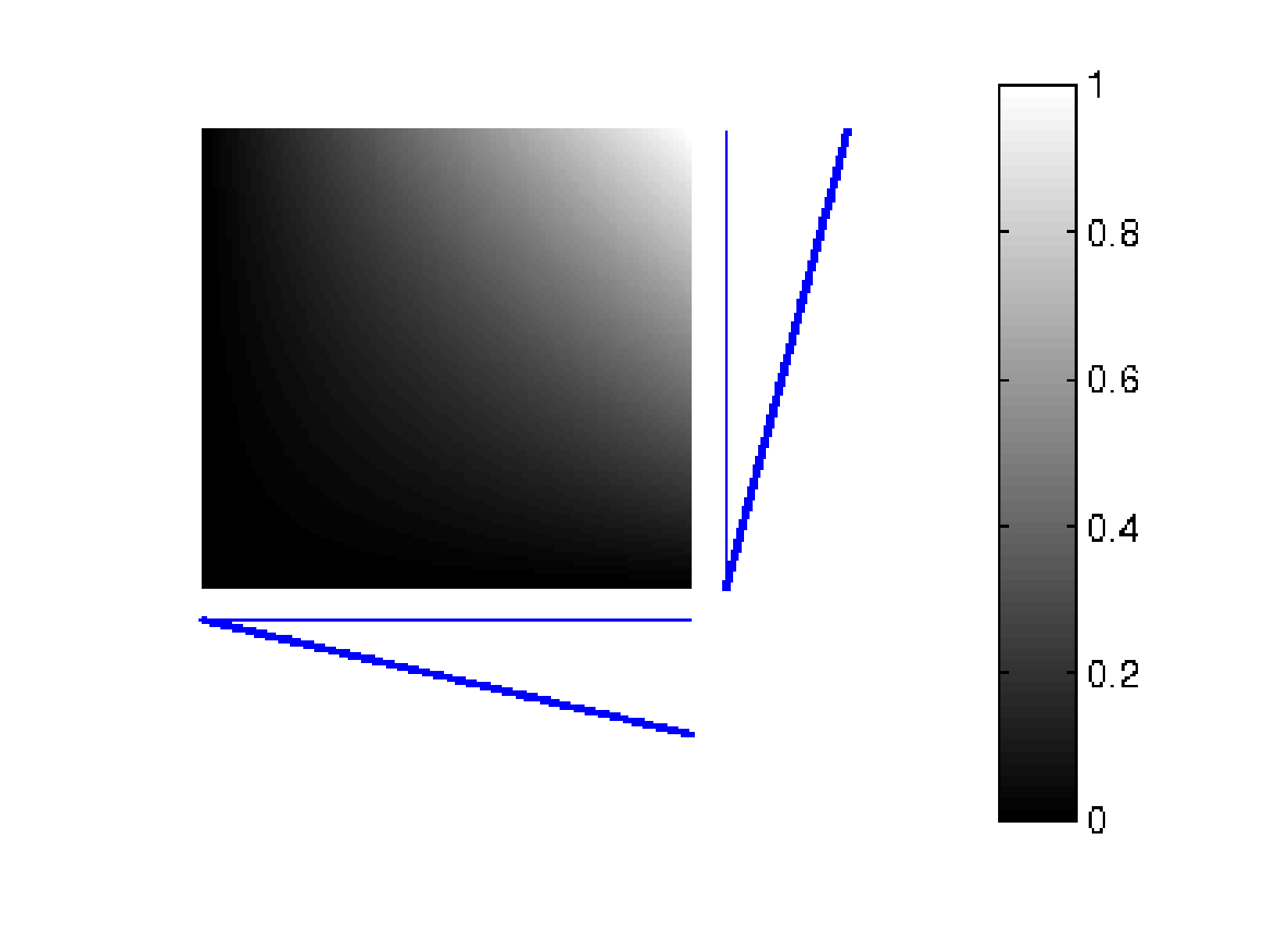}&
\includegraphics[height=0.17\textheight]{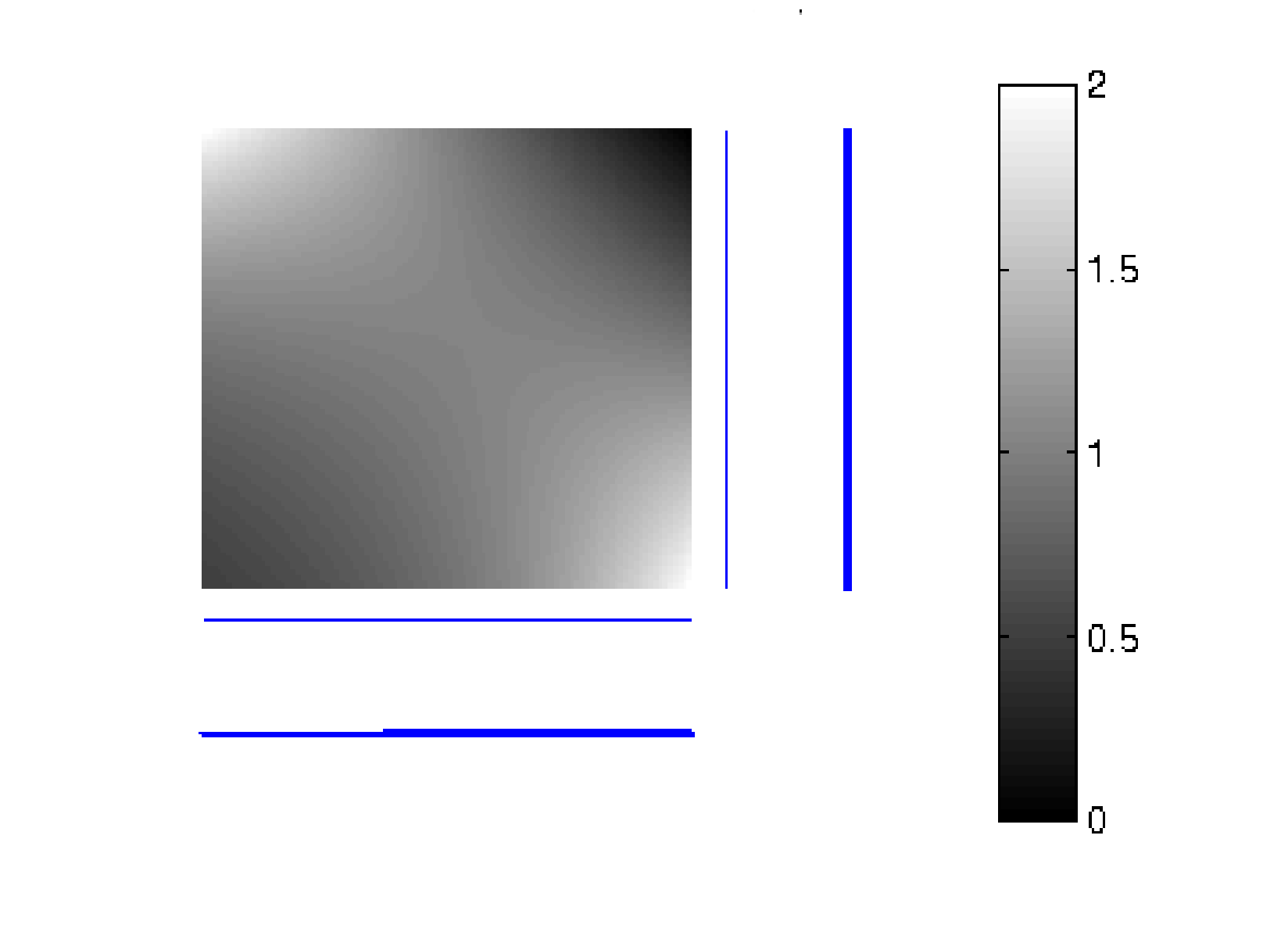}&
\includegraphics[height=0.17\textheight]{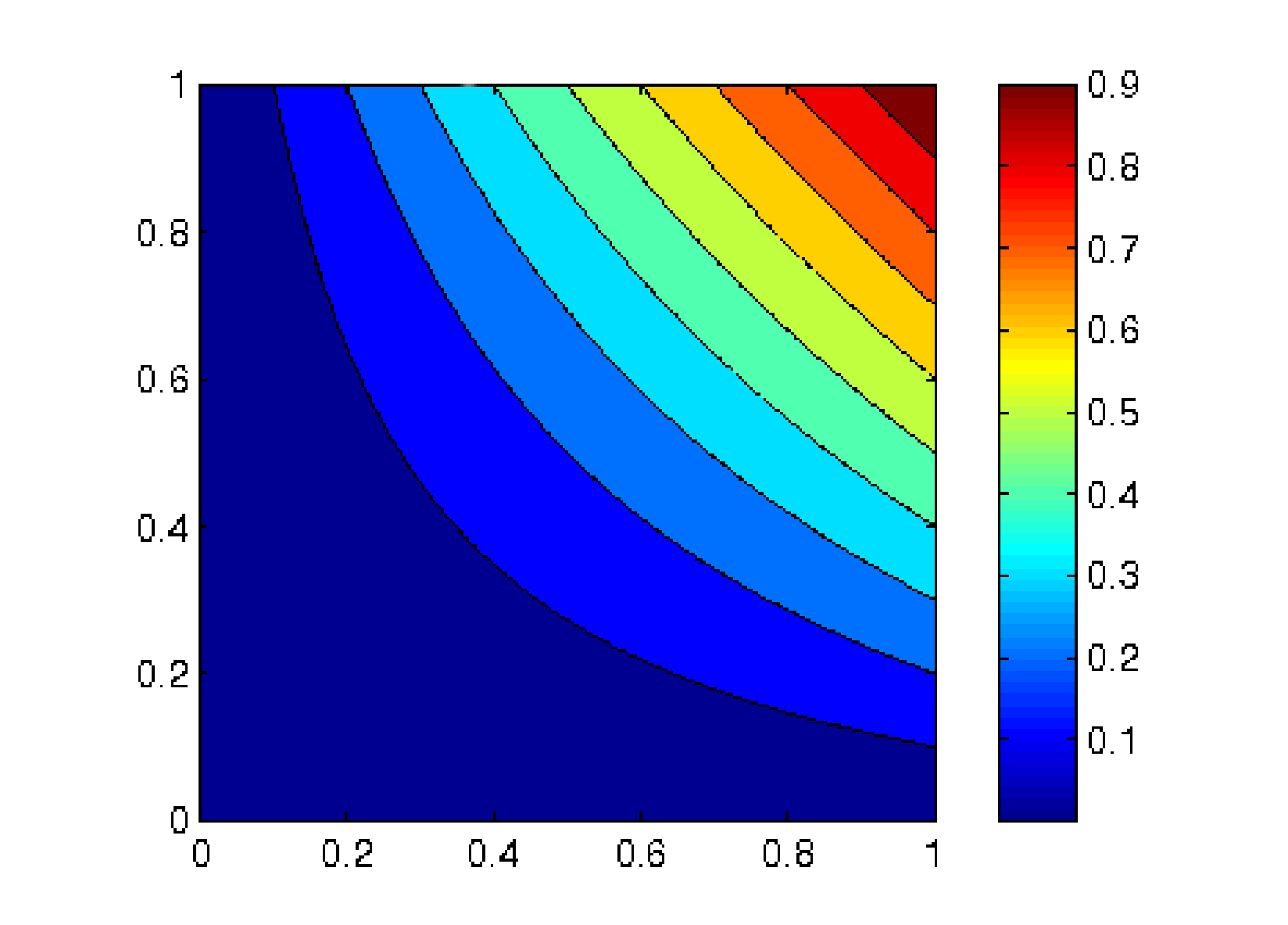}\\
$F(x,y)$ with $F_1(x)$ and $F_2(y)$ & $f(x,y)$ with $f_1(x)$ and $f_2(y)$ & $C(x,y)$ contours plot\\ 
\includegraphics[height=0.17\textheight]{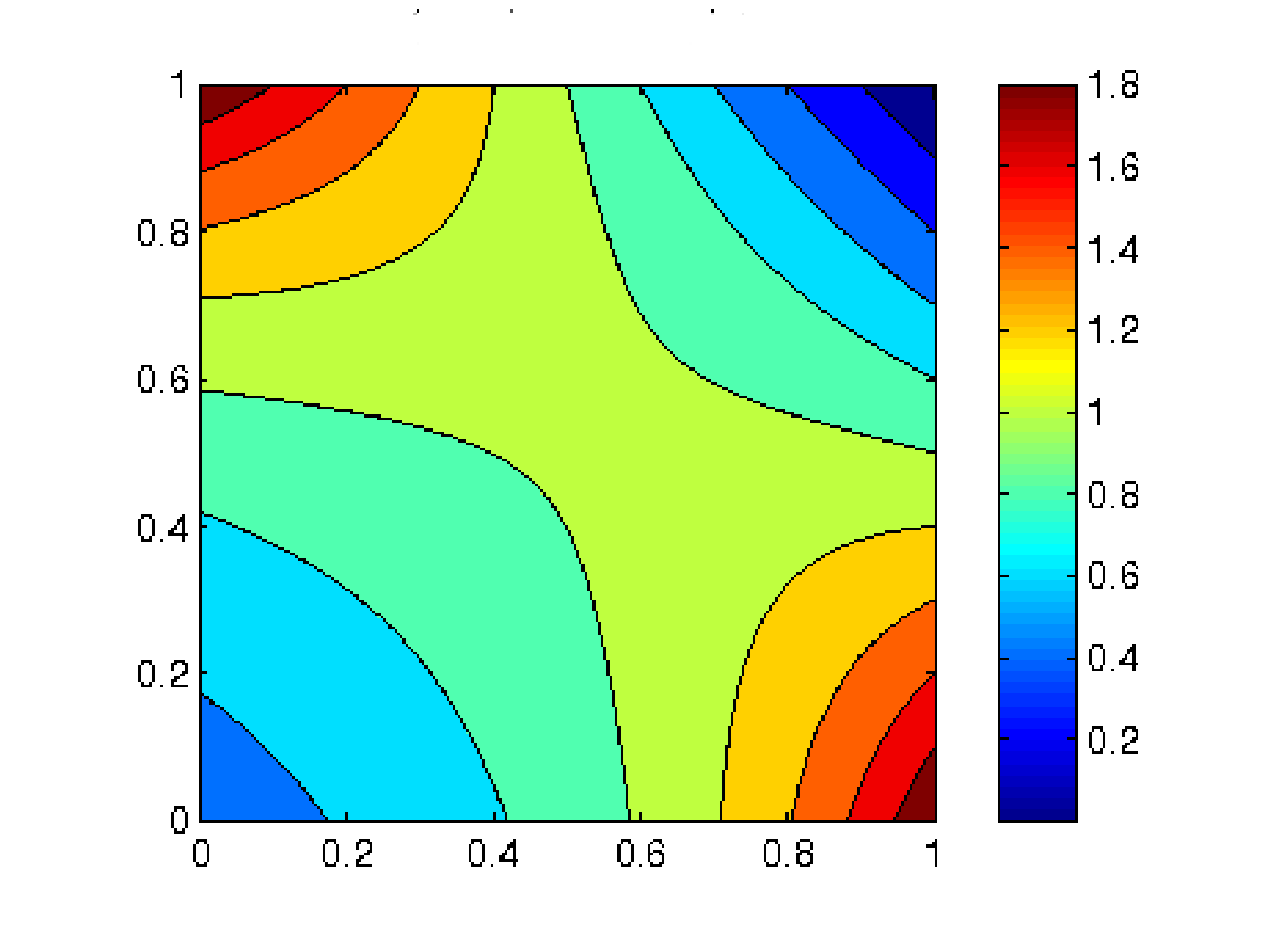}&
\includegraphics[height=0.17\textheight]{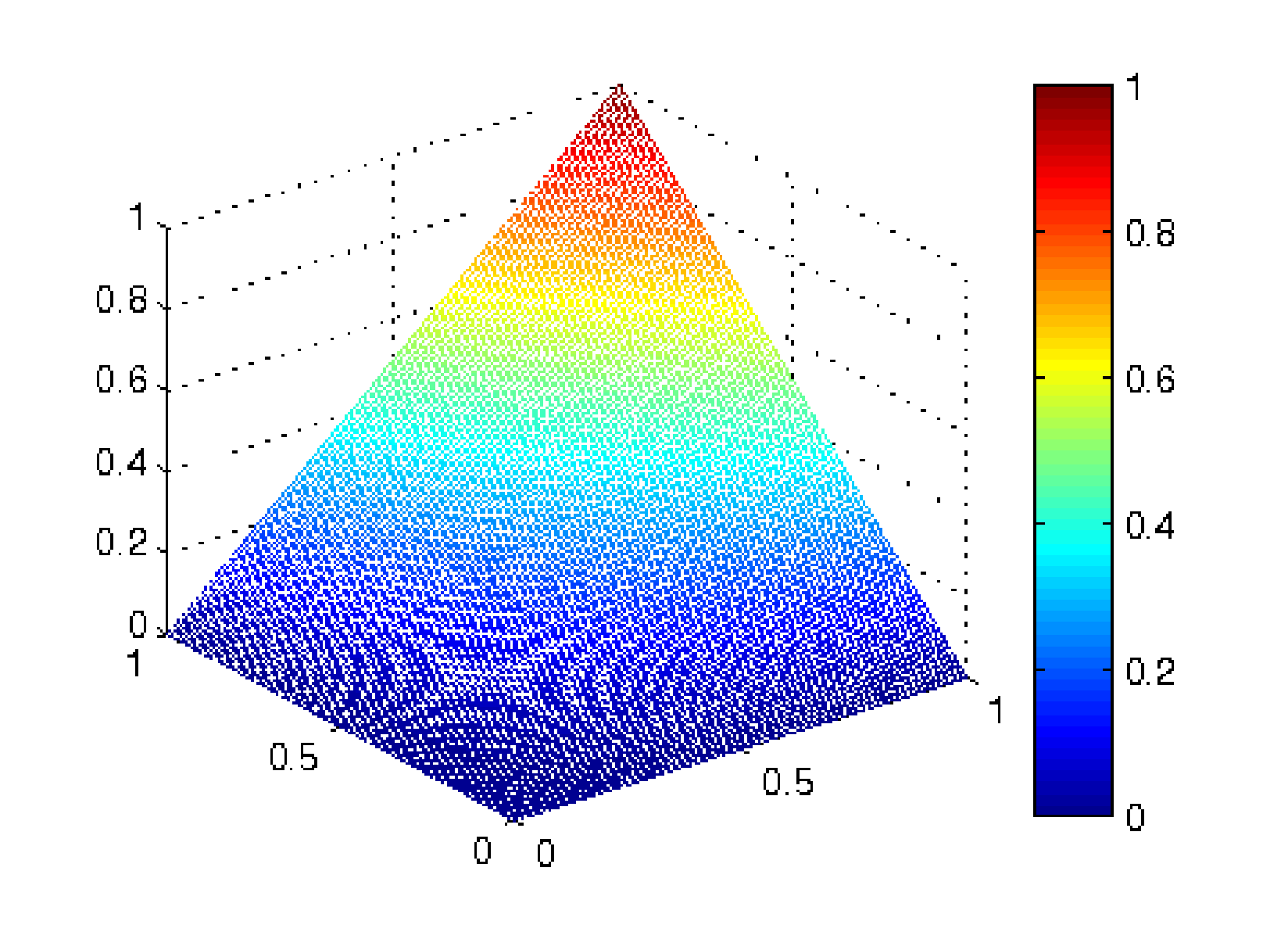}&
\includegraphics[height=0.17\textheight]{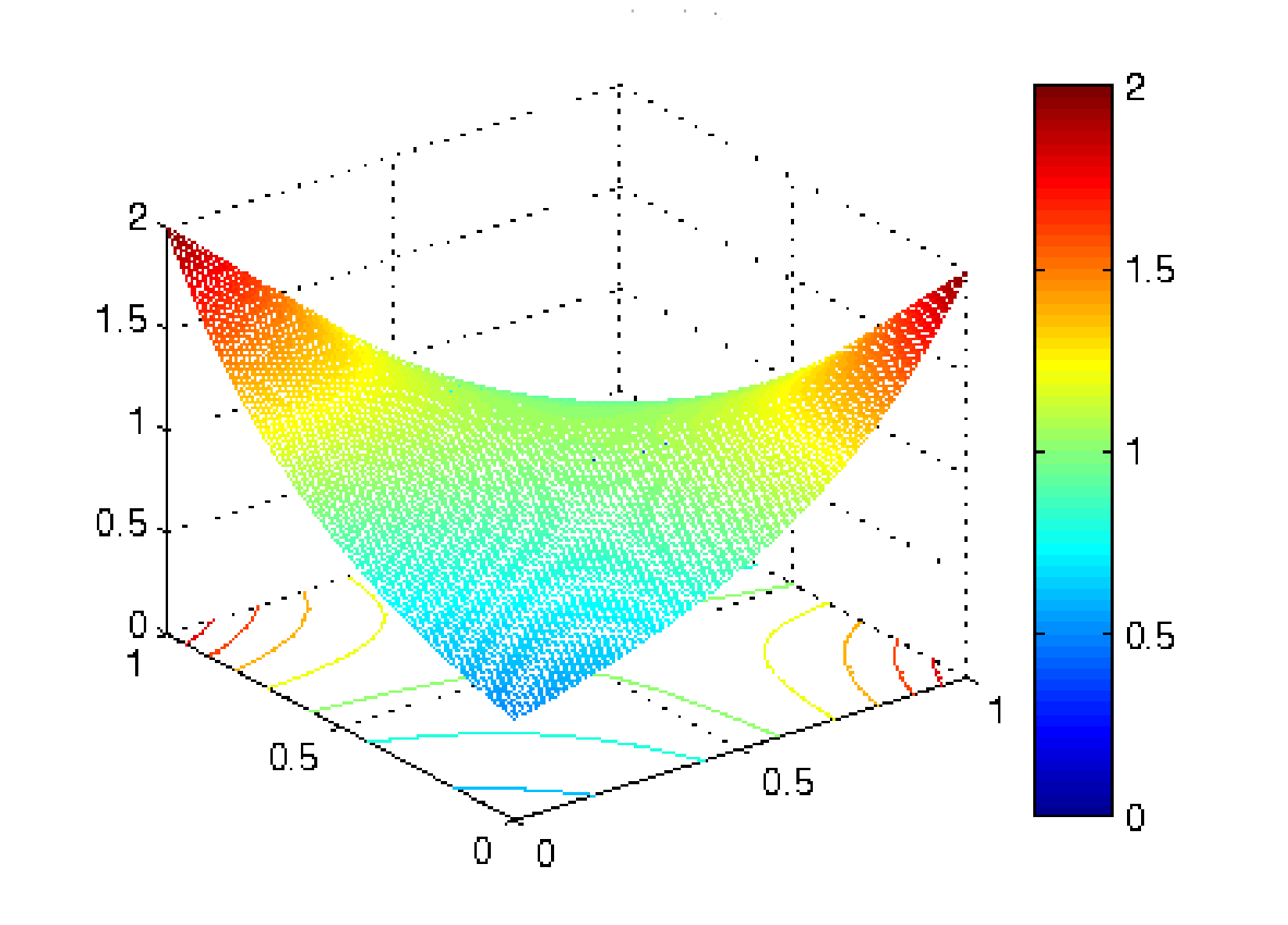}\\
$c(x,y)$ contours plot  & $C(x,y)$ mesh plot & $c(x,y)$ mesh plot\\ 
\etabu
\ecc
\caption{Ali-Mikhail-Haq copula with $\alpha=-1$.}
\efig

\bfig
\bcc
\btabu{ccc}
\includegraphics[height=0.17\textheight]{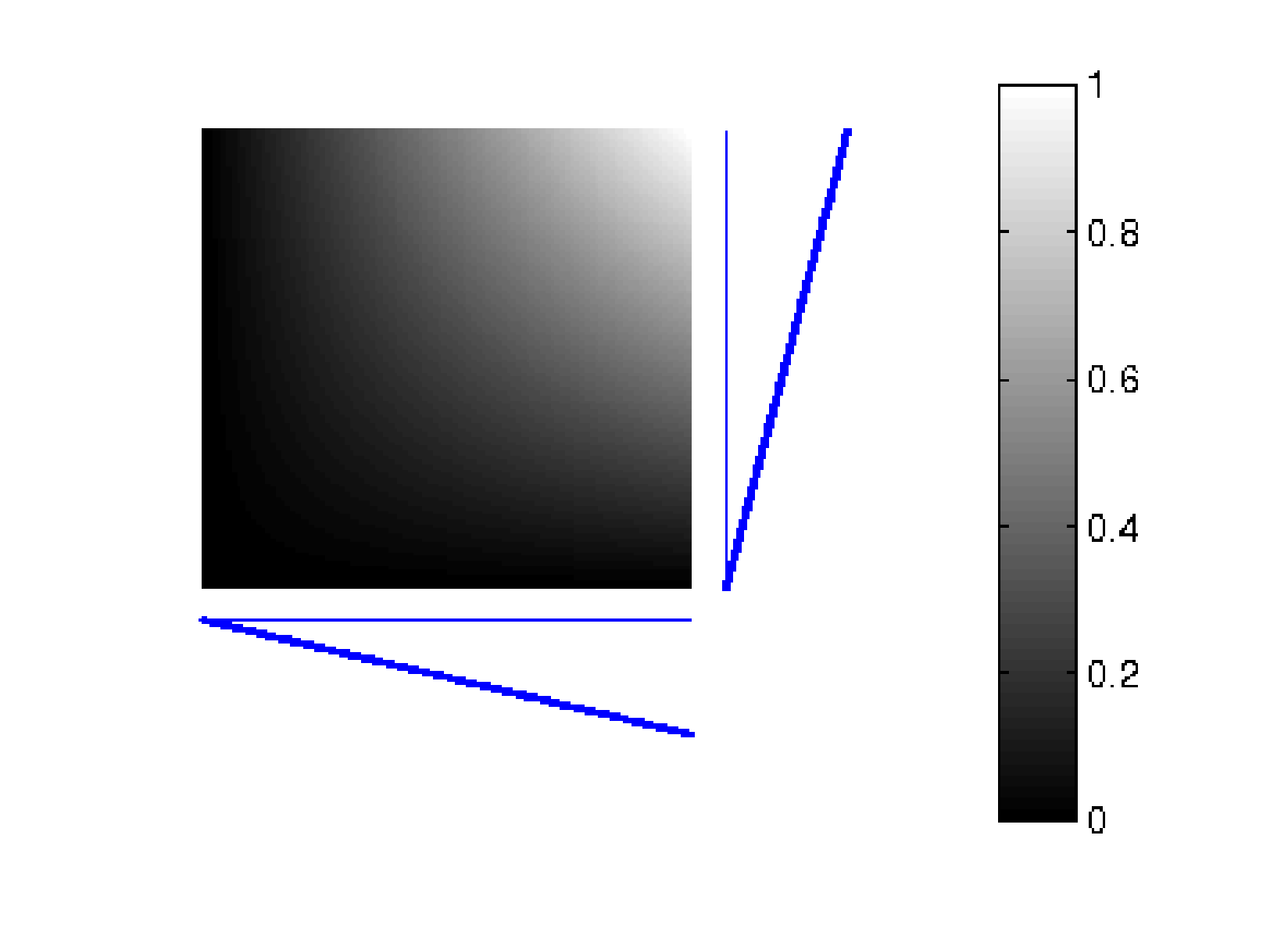}&
\includegraphics[height=0.17\textheight]{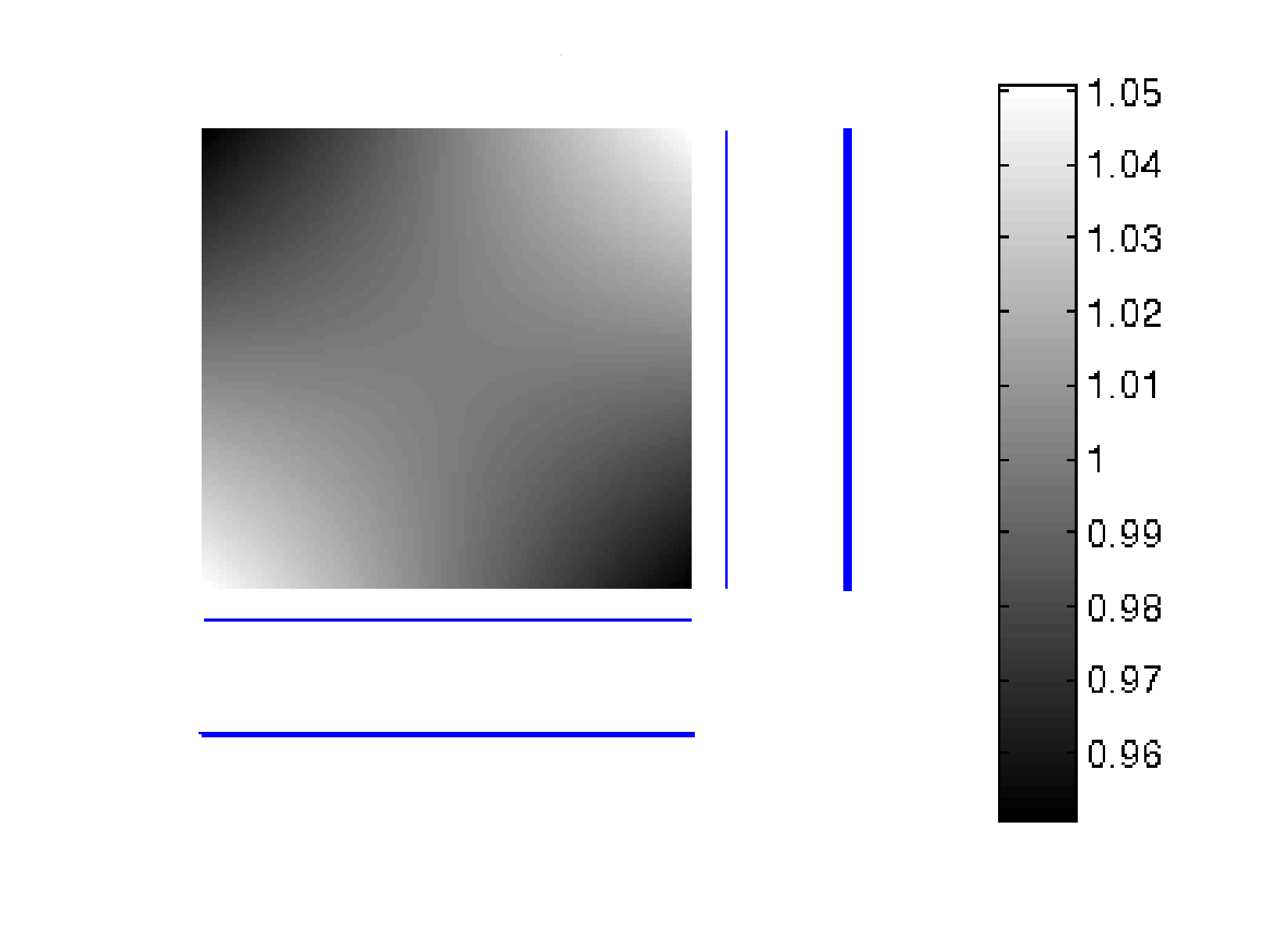}&
\includegraphics[height=0.17\textheight]{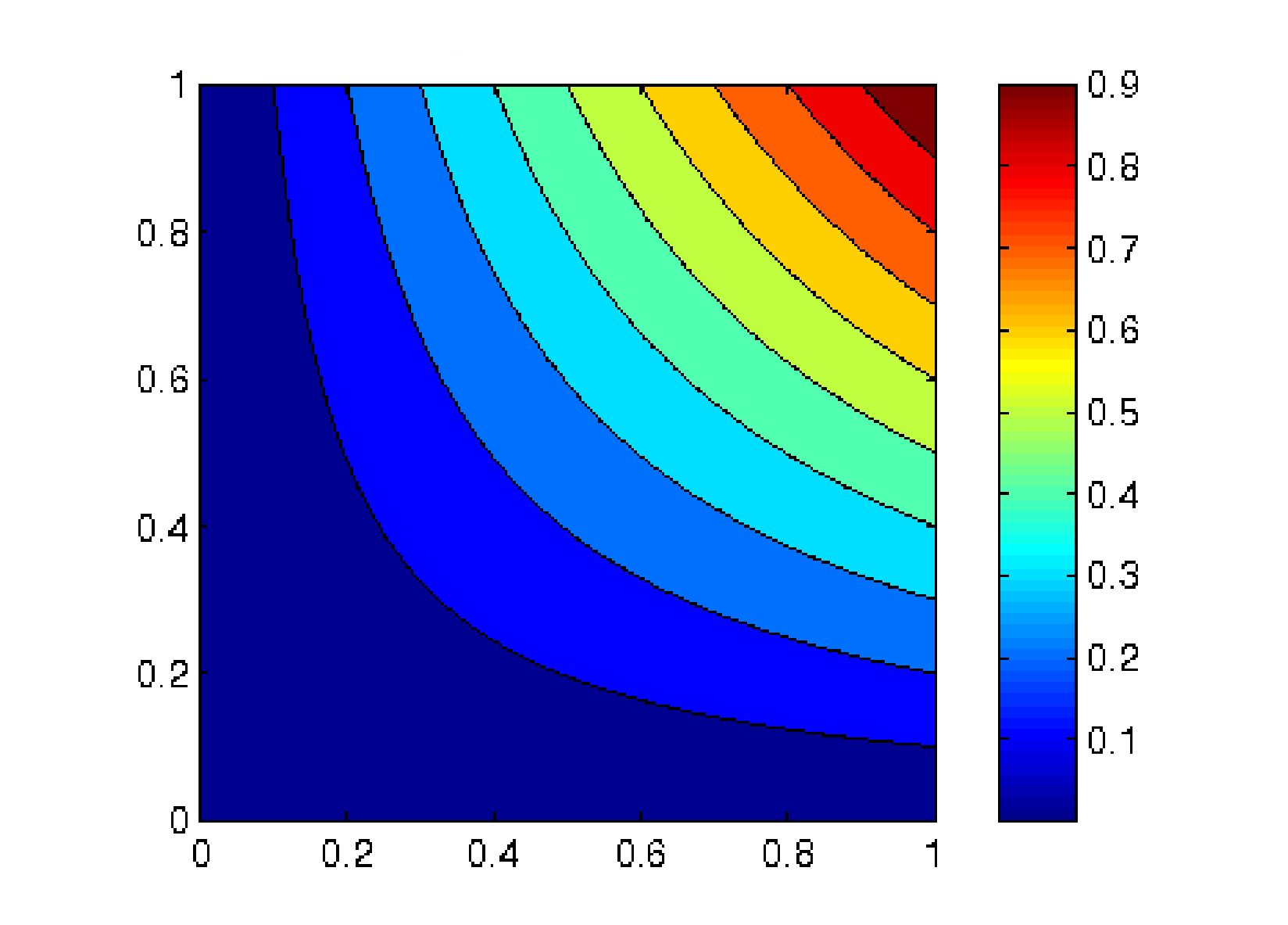}\\
$F(x,y)$,$F_1(x)$,$F_2(y)$& $f(x,y)$ with $f_1(x)$ and $f_2(y)$ & $C(x,y)$ contours plot\\ 
\includegraphics[height=0.17\textheight]{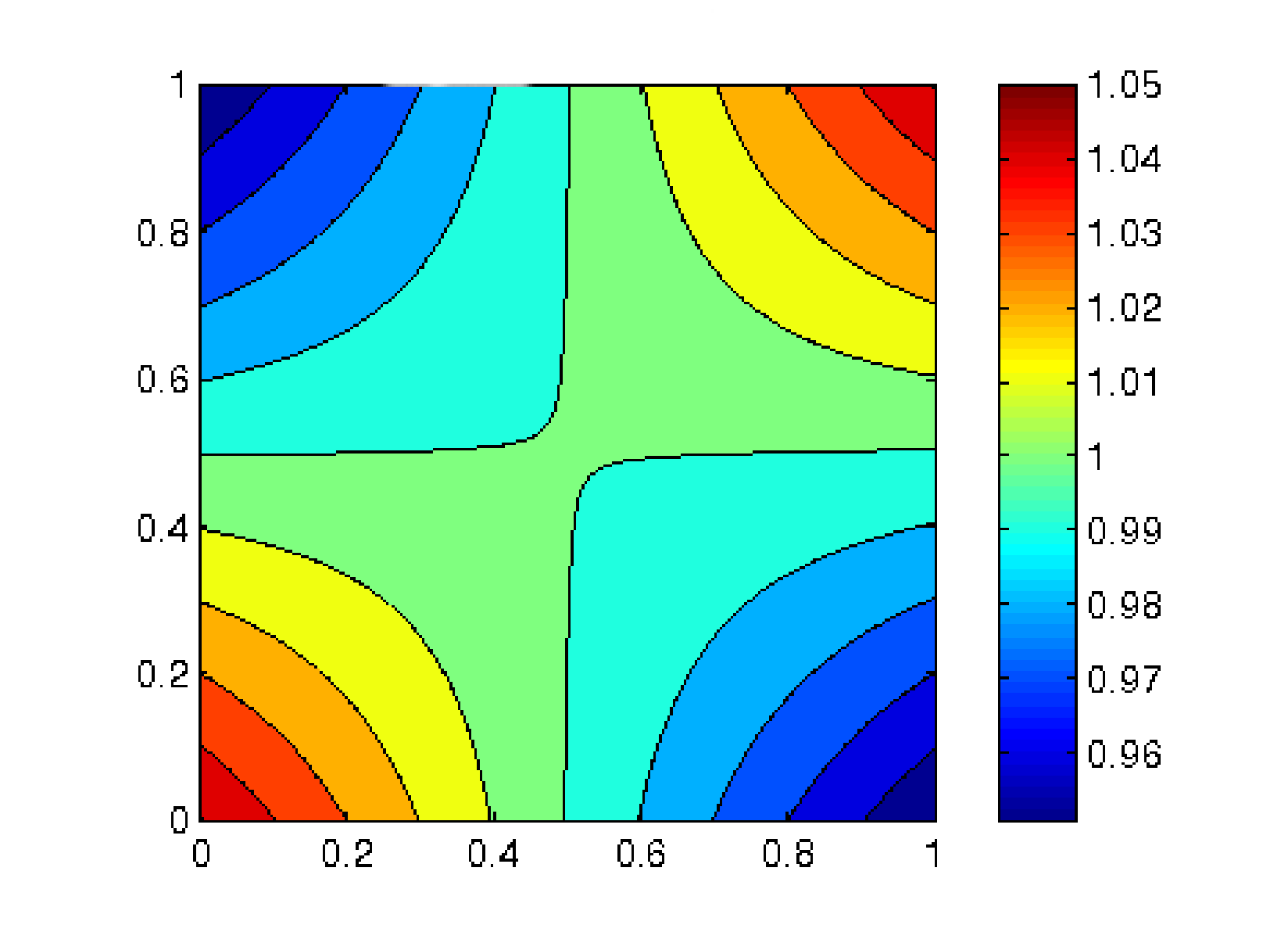}&
\includegraphics[height=0.17\textheight]{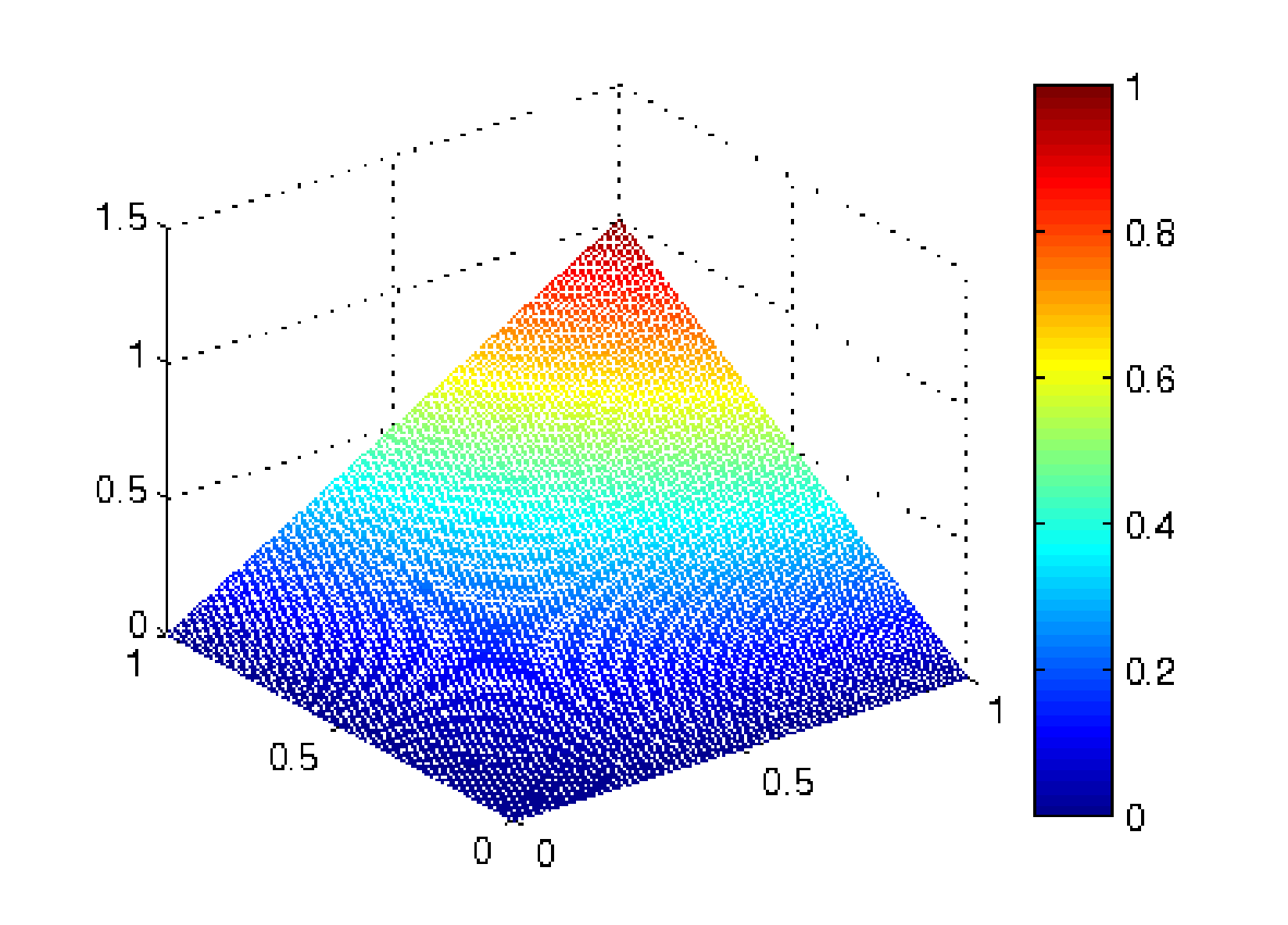}&
\includegraphics[height=0.17\textheight]{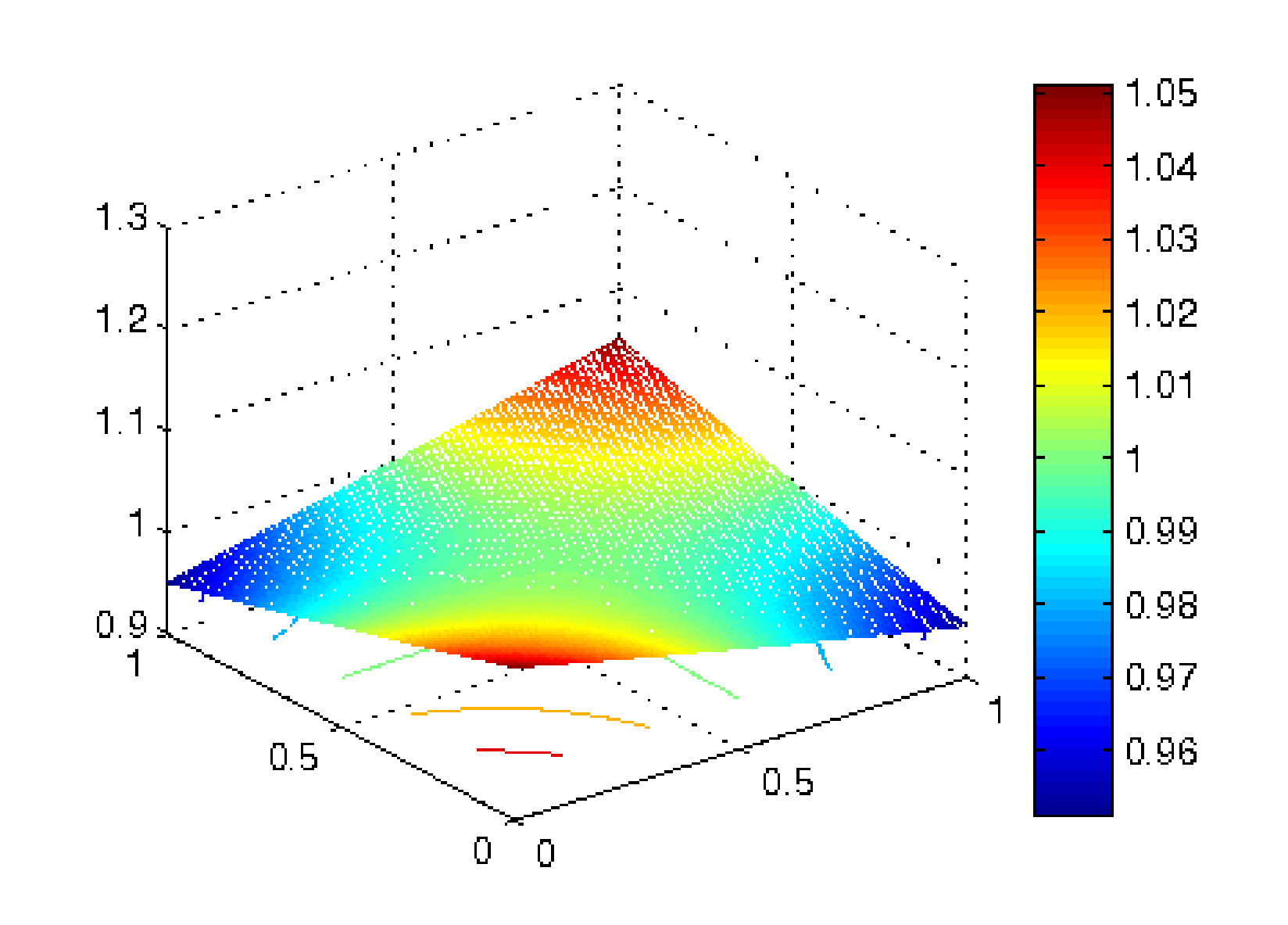}\\
$c(x,y)$ contours plot & $C(x,y)$ mesh plot & $c(x,y)$ mesh plot\\ 
\etabu
\ecc
\caption{Frank copula with $\alpha=0.1$.}
\efig

\bfig
\bcc
\btabu{ccc}
\includegraphics[height=0.17\textheight]{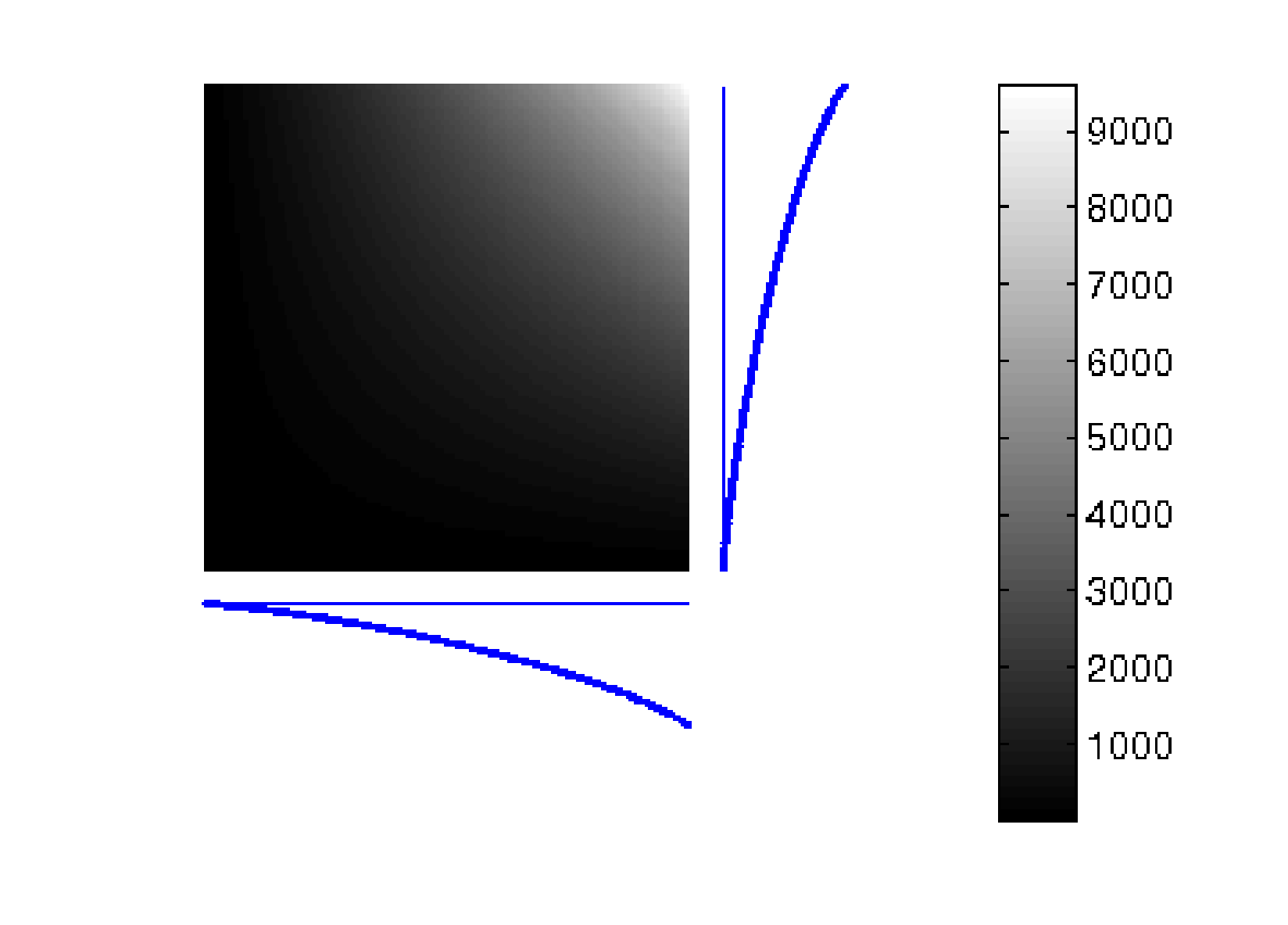}&
\includegraphics[height=0.17\textheight]{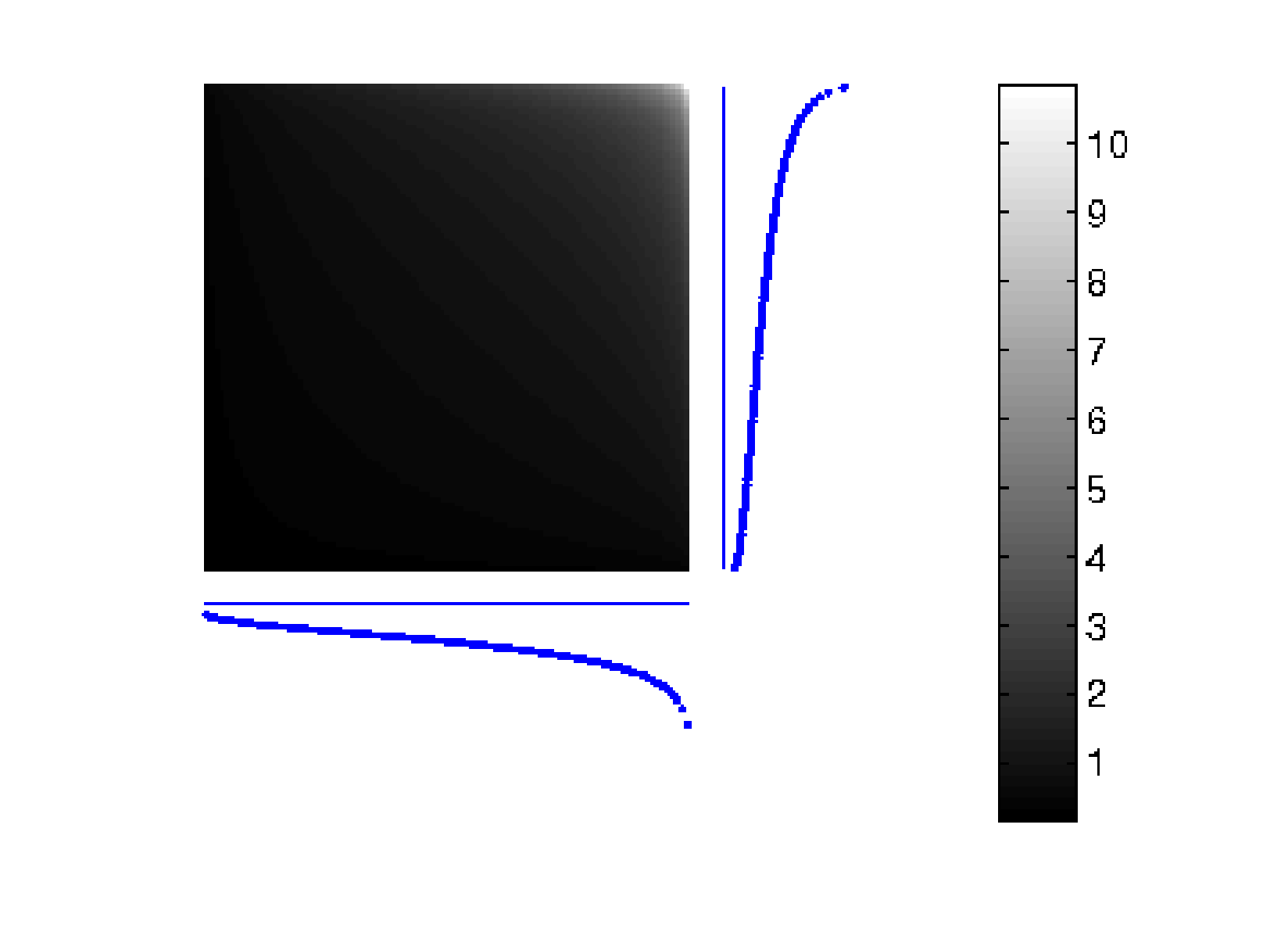}&
\includegraphics[height=0.17\textheight]{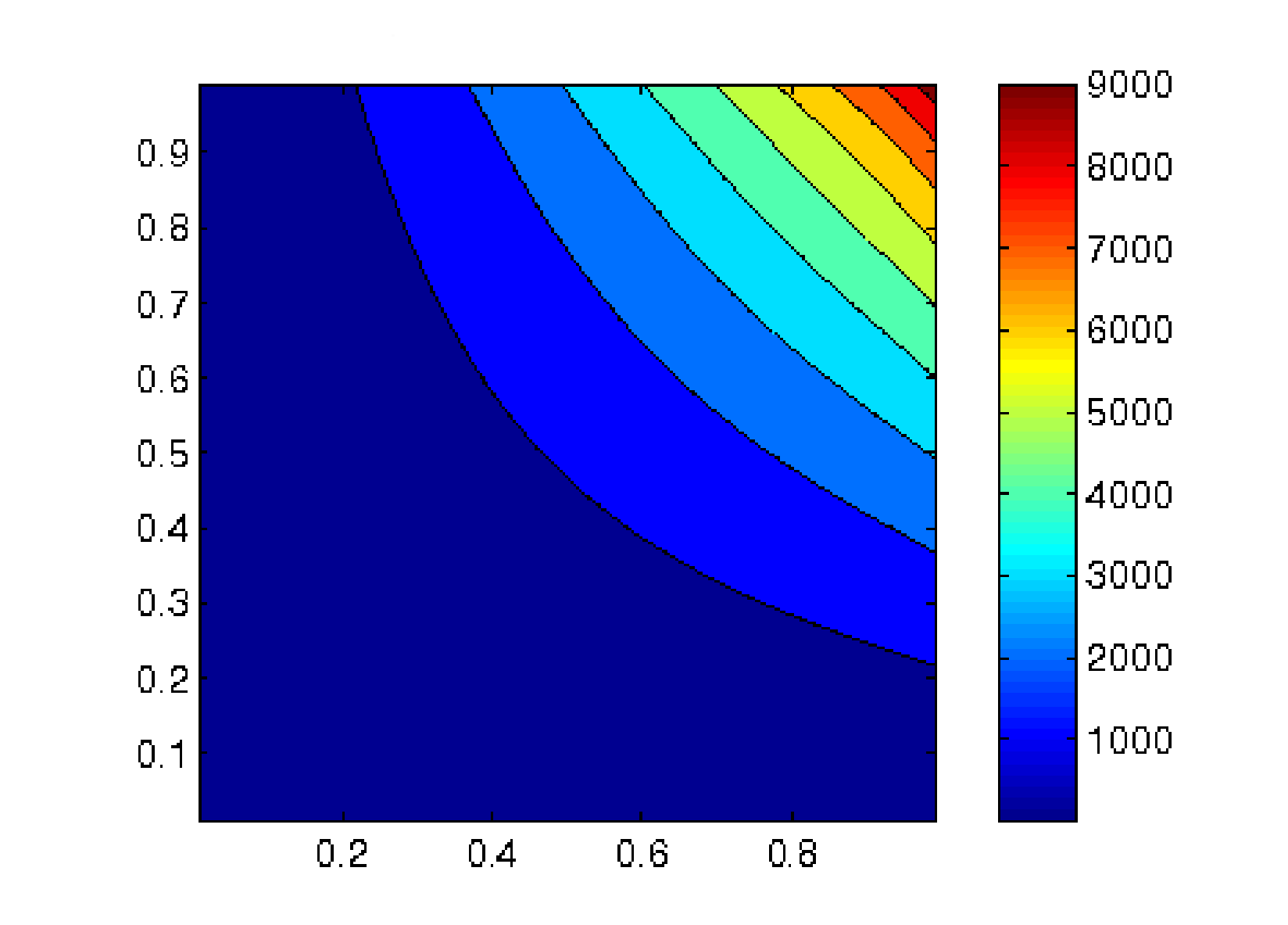}\\
$F(x,y)$,$F_1(x)$,$F_2(y)$ & $f(x,y)$ with $f_1(x)$ and $f_2(y)$ & $C(x,y)$ contours plot\\ 
\includegraphics[height=0.17\textheight]{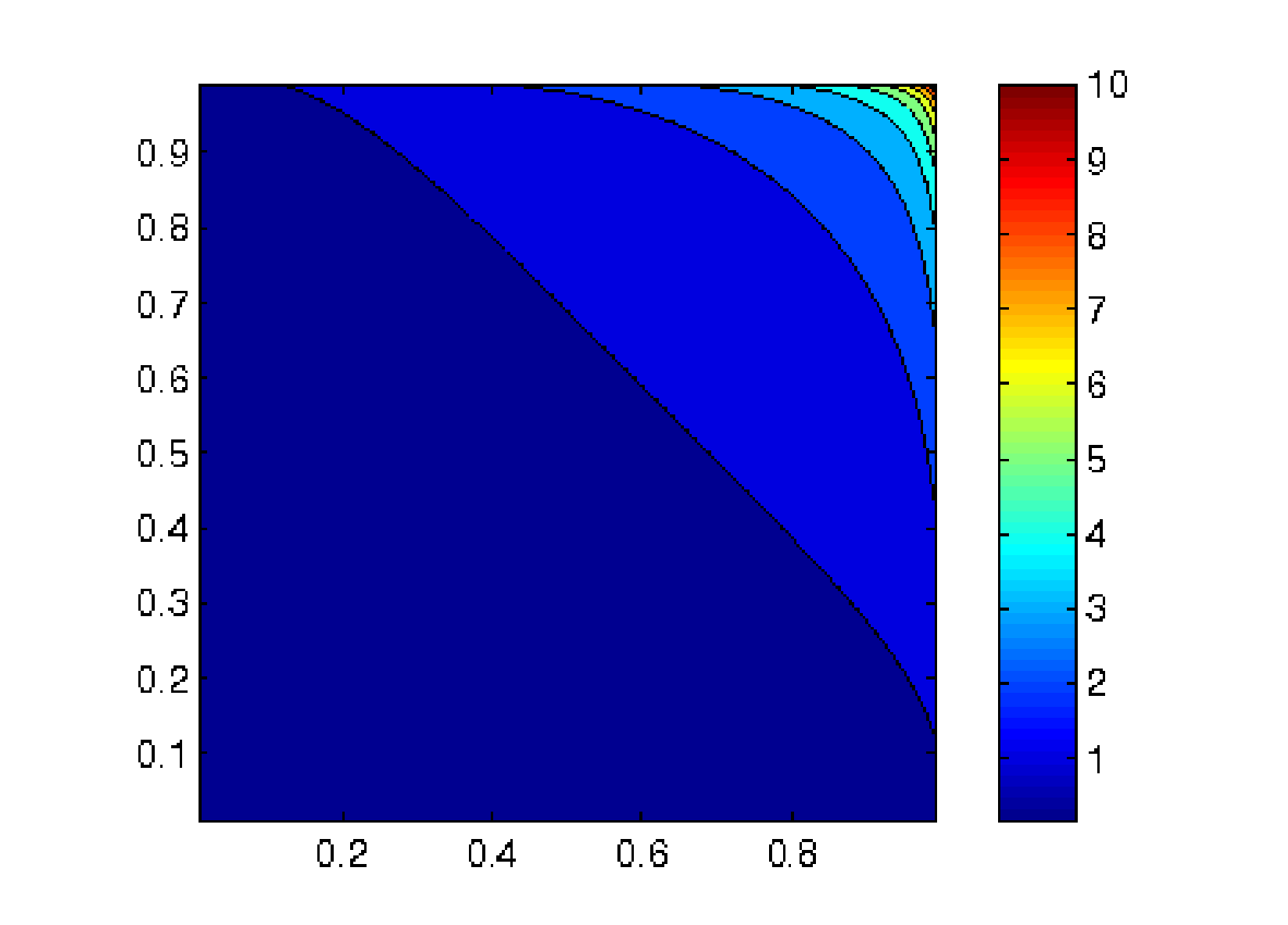}&
\includegraphics[height=0.17\textheight]{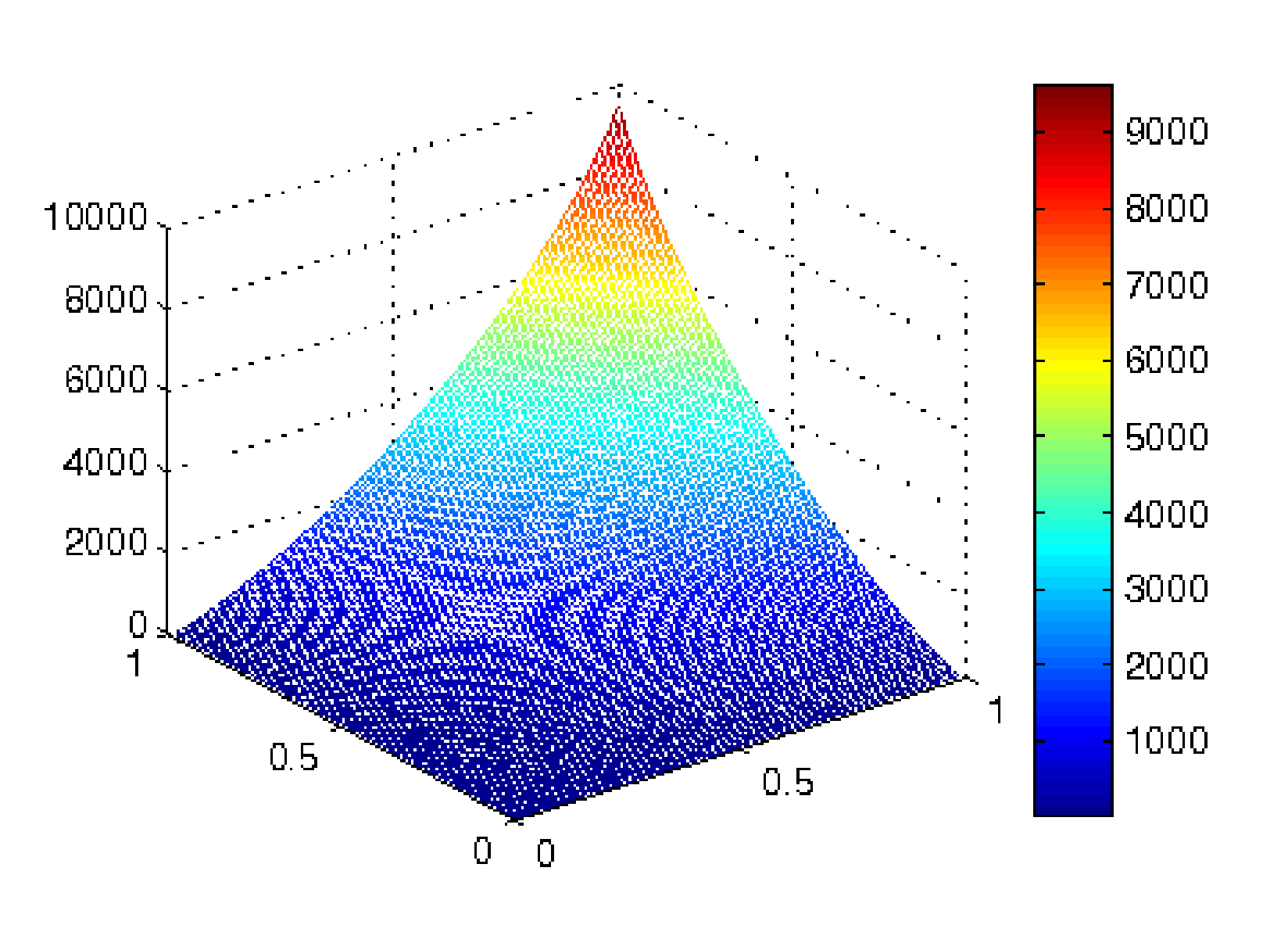}&
\includegraphics[height=0.17\textheight]{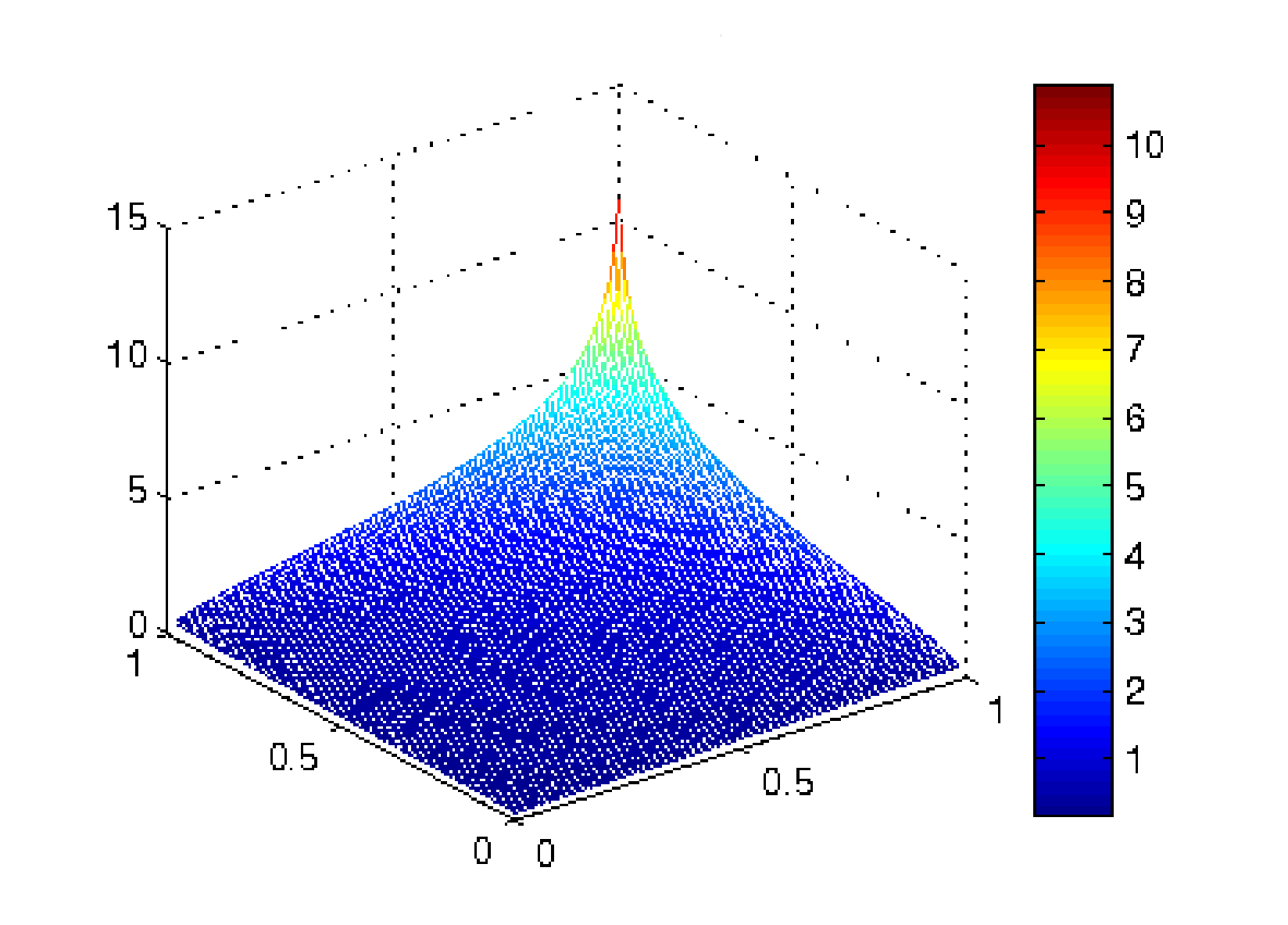}\\
$c(x,y)$ contours plot& $C(x,y)$ mesh plot & $c(x,y)$ mesh plot\\ 
\etabu
\ecc
\caption{Gaussian copula with $\rho=0.1$.}
\efig

\bfig
\bcc
\btabu{ccc}
\includegraphics[height=0.17\textheight]{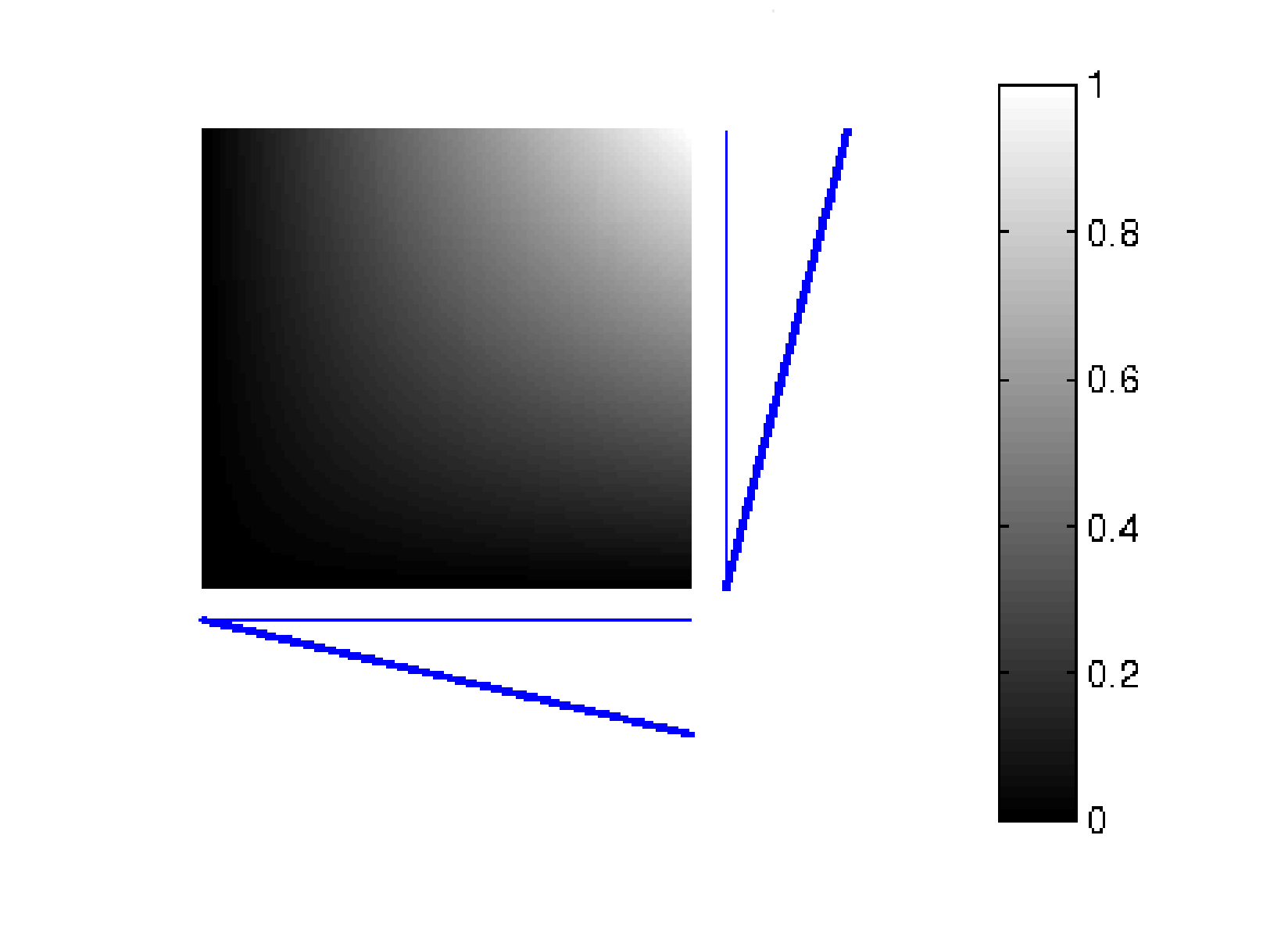}&
\includegraphics[height=0.17\textheight]{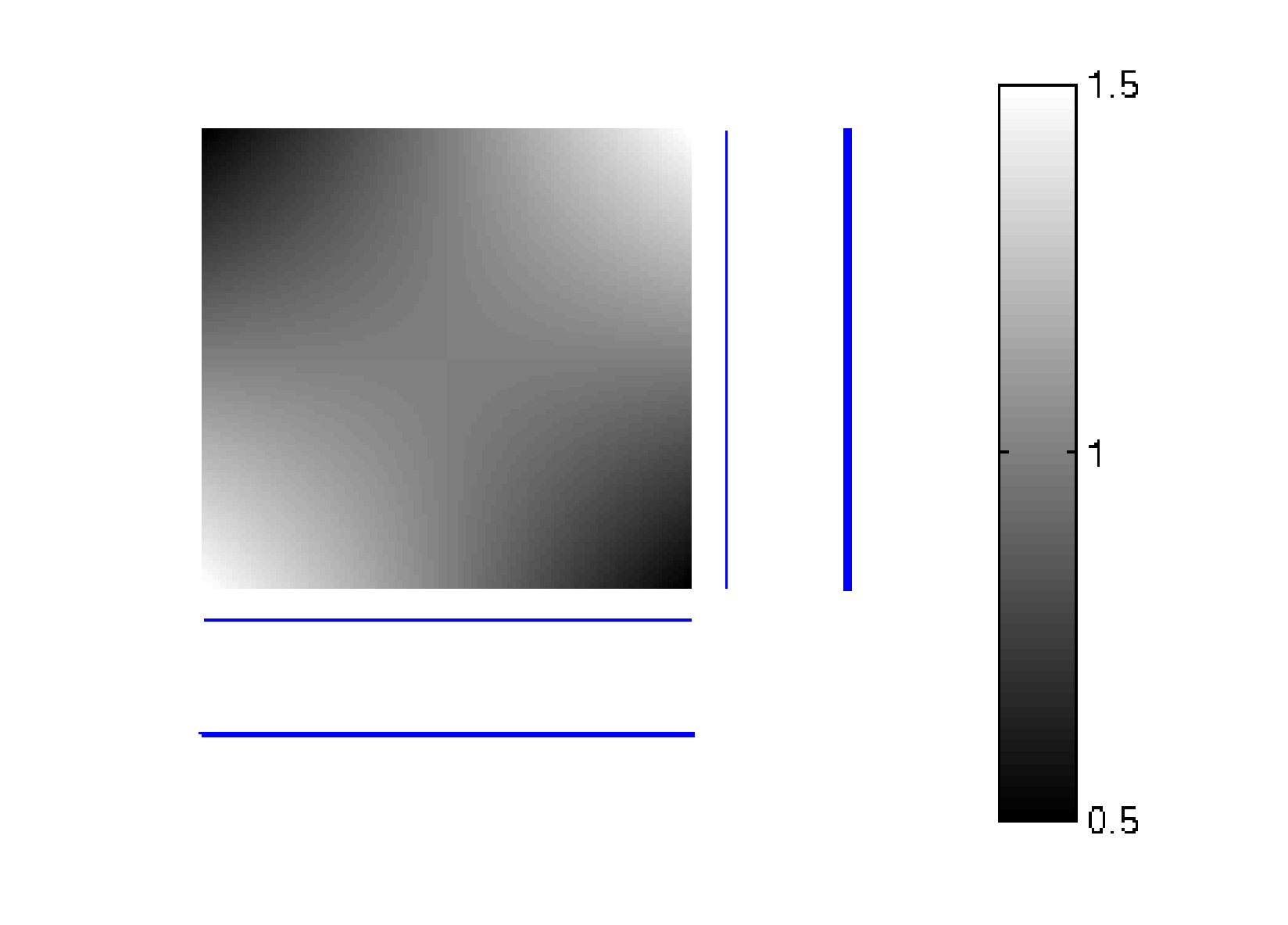}&
\includegraphics[height=0.17\textheight]{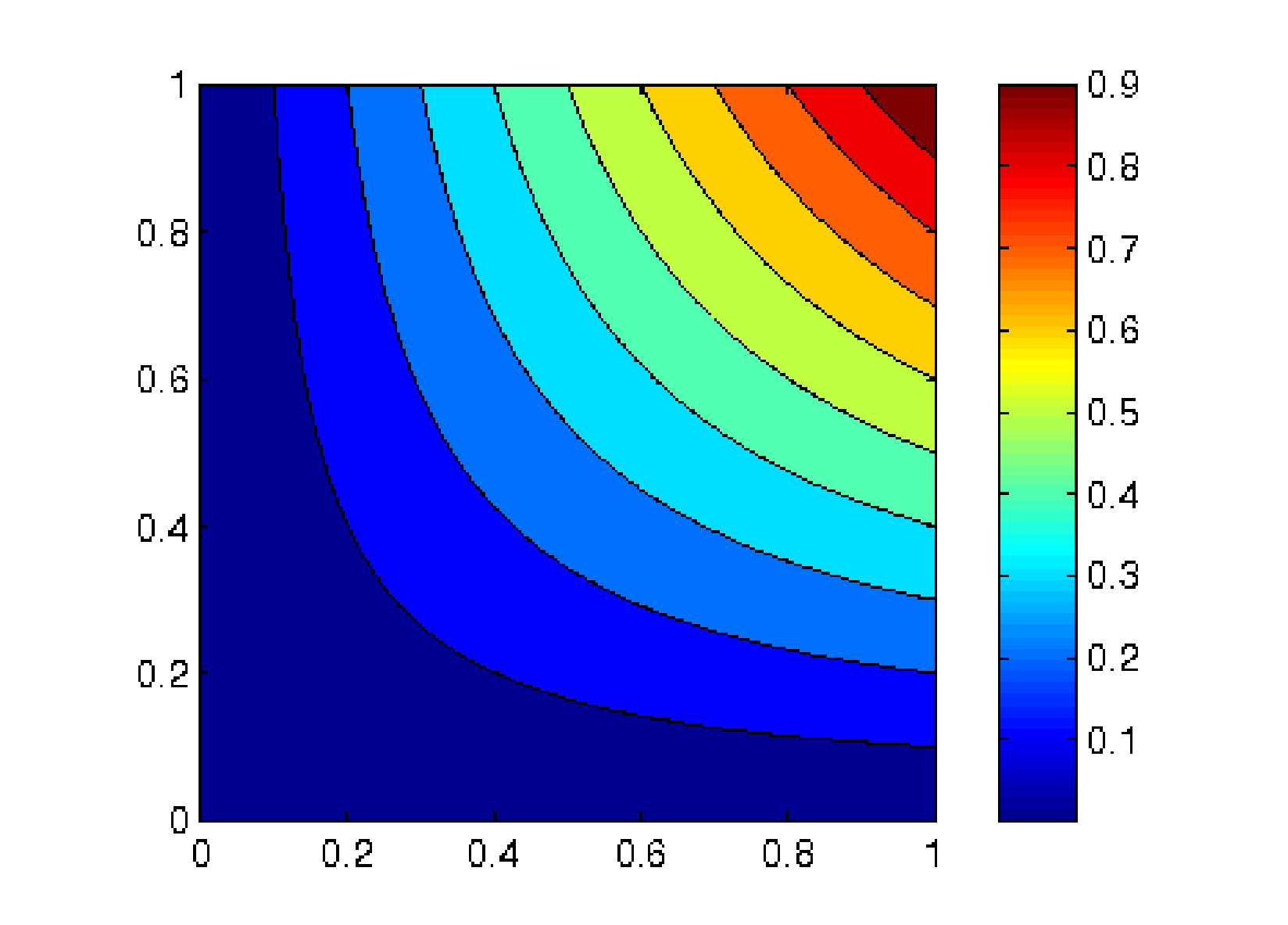}\\
$F(x,y)$,$F_1(x)$,$F_2(y)$ & $f(x,y)$ with $f_1(x)$ and $f_2(y)$ & $C(x,y)$ contours plot\\ 
\includegraphics[height=0.17\textheight]{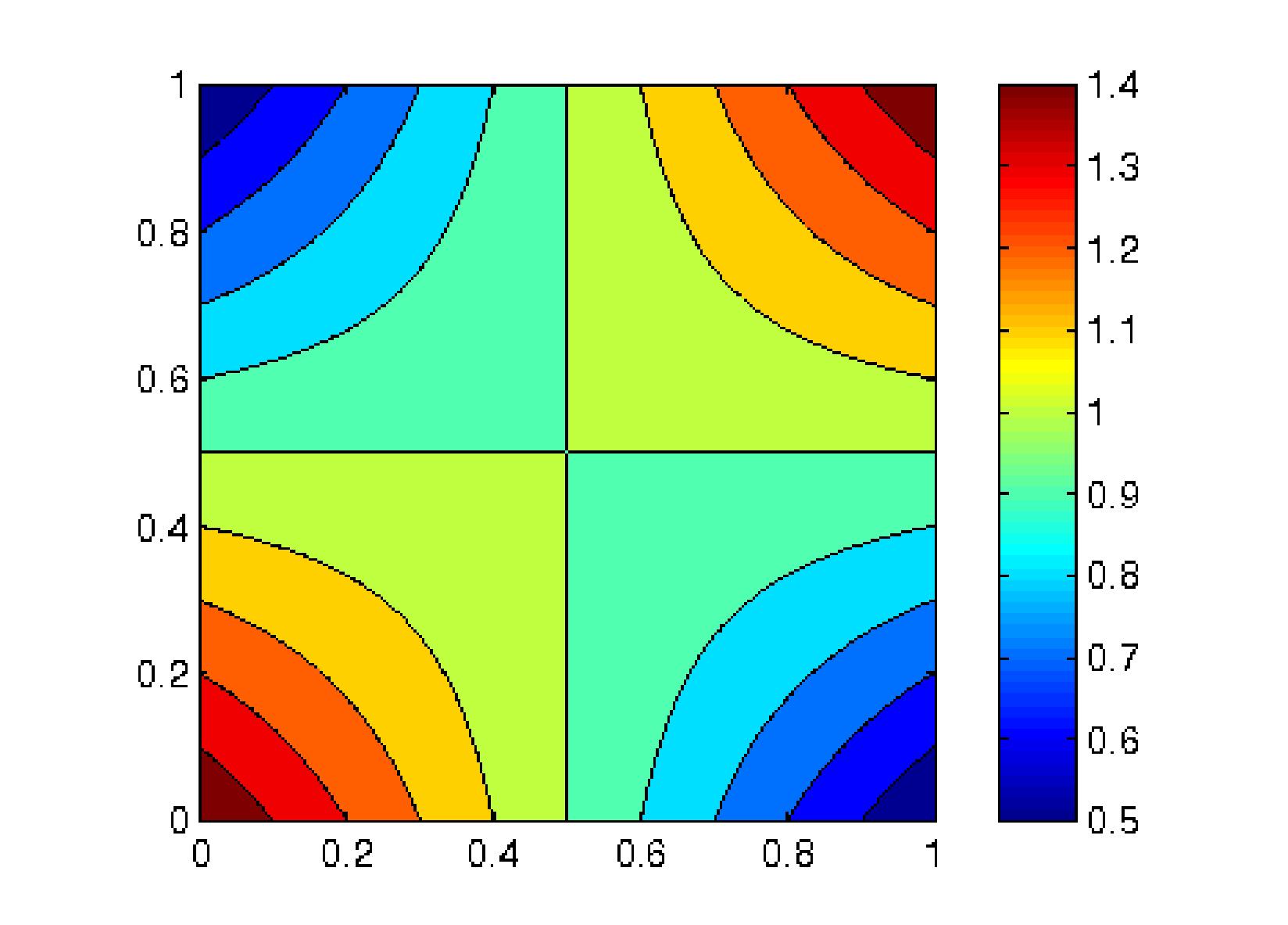}&
\includegraphics[height=0.17\textheight]{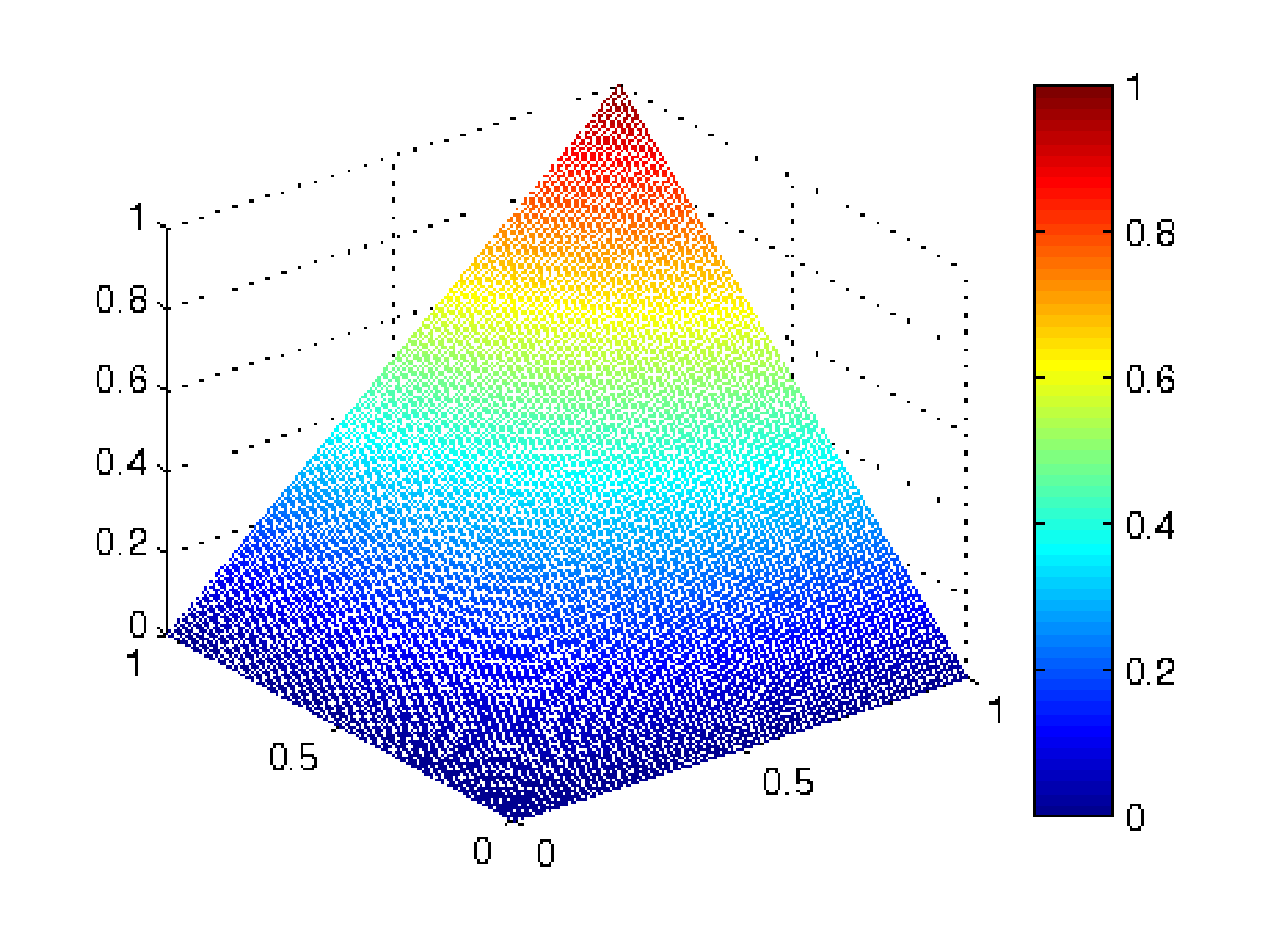}&
\includegraphics[height=0.17\textheight]{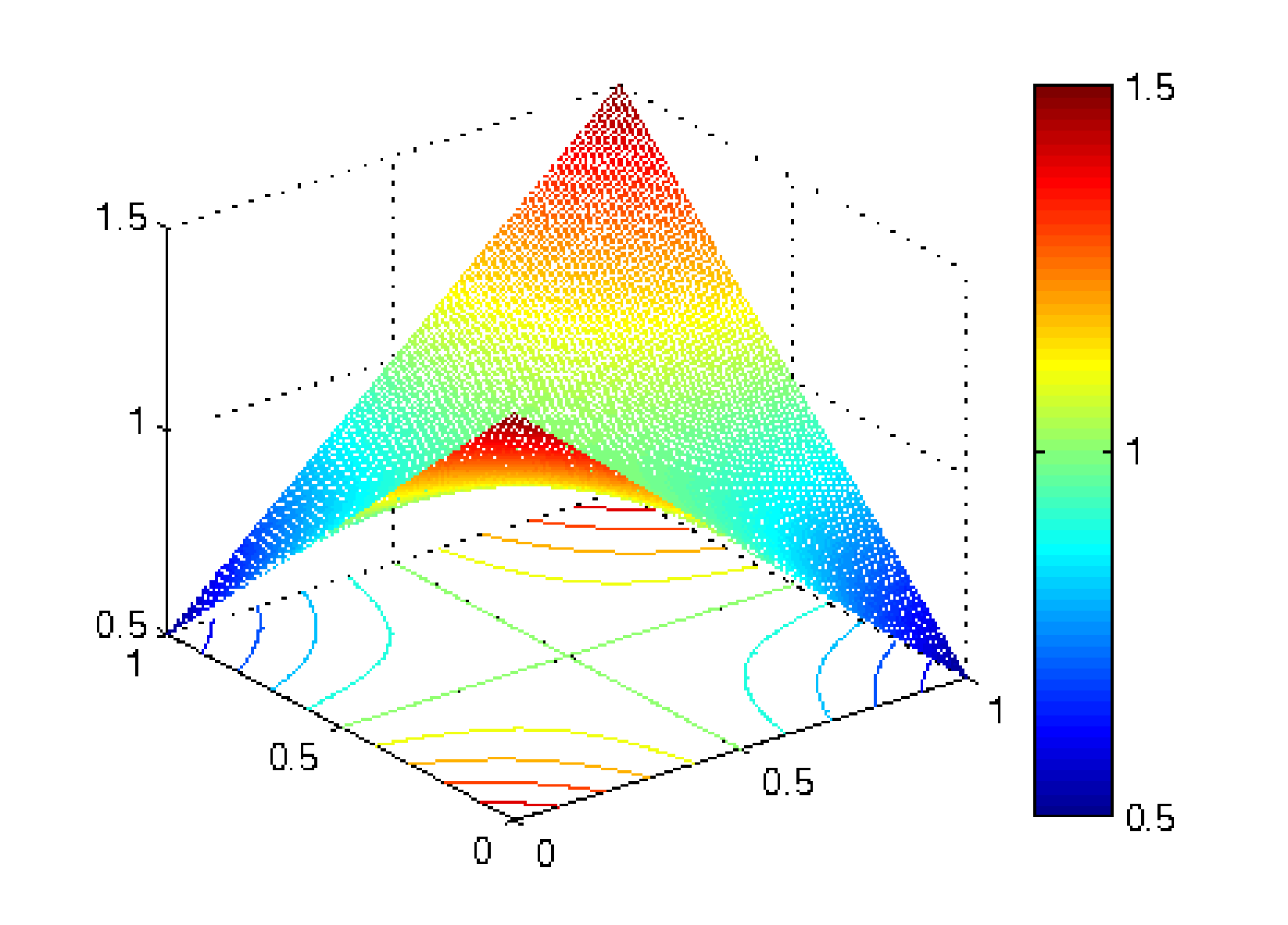}\\
$c(x,y)$ contours plot & $C(x,y)$ mesh plot & $c(x,y)$ mesh plot\\ 
\etabu
\ecc
\caption{Farlie-Gumbel-Morgenstern family with $\alpha=0.5$}
\efig

\bfig
\bcc
\btabu{ccc}
\includegraphics[height=0.17\textheight]{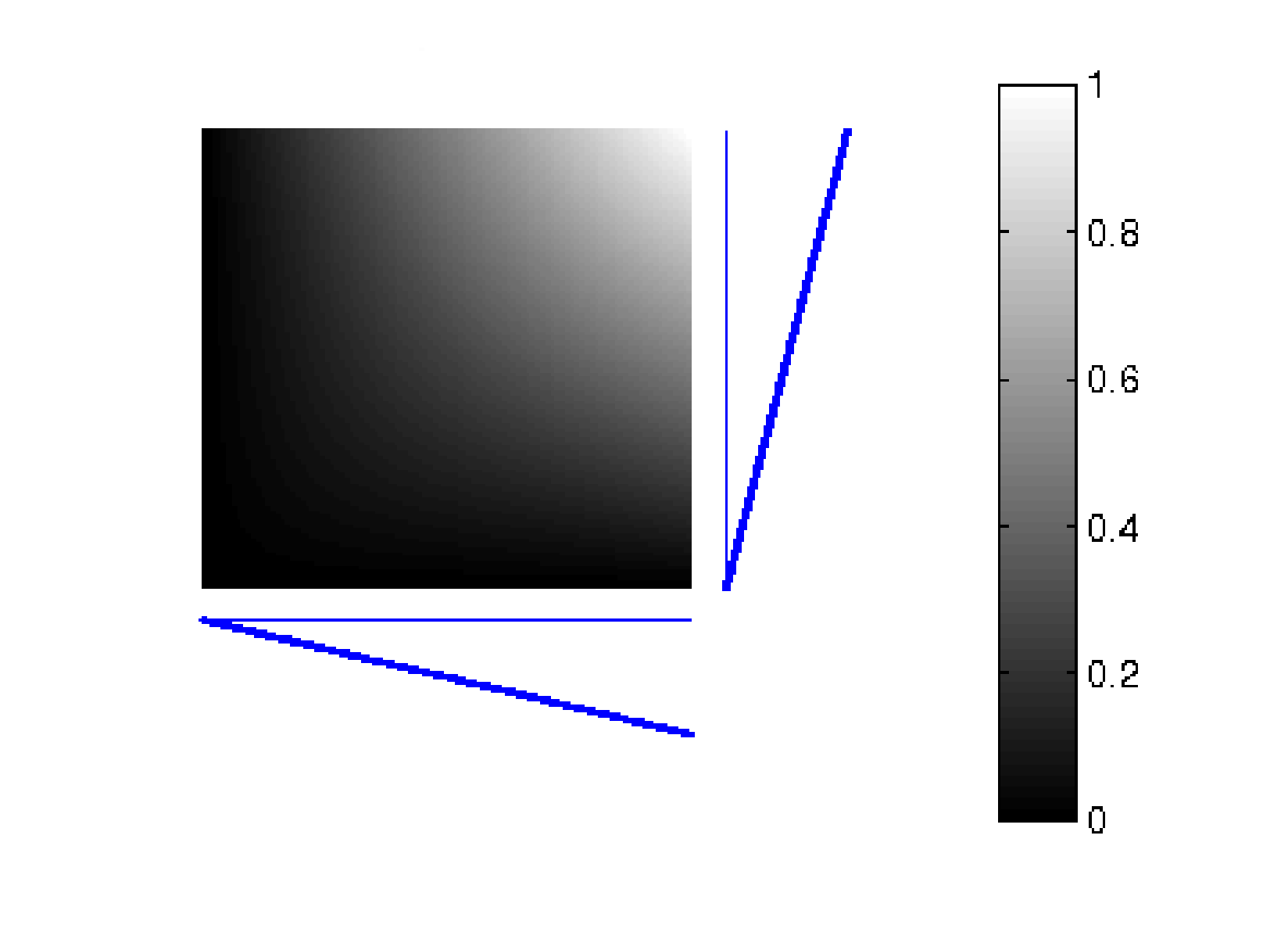}&
\includegraphics[height=0.17\textheight]{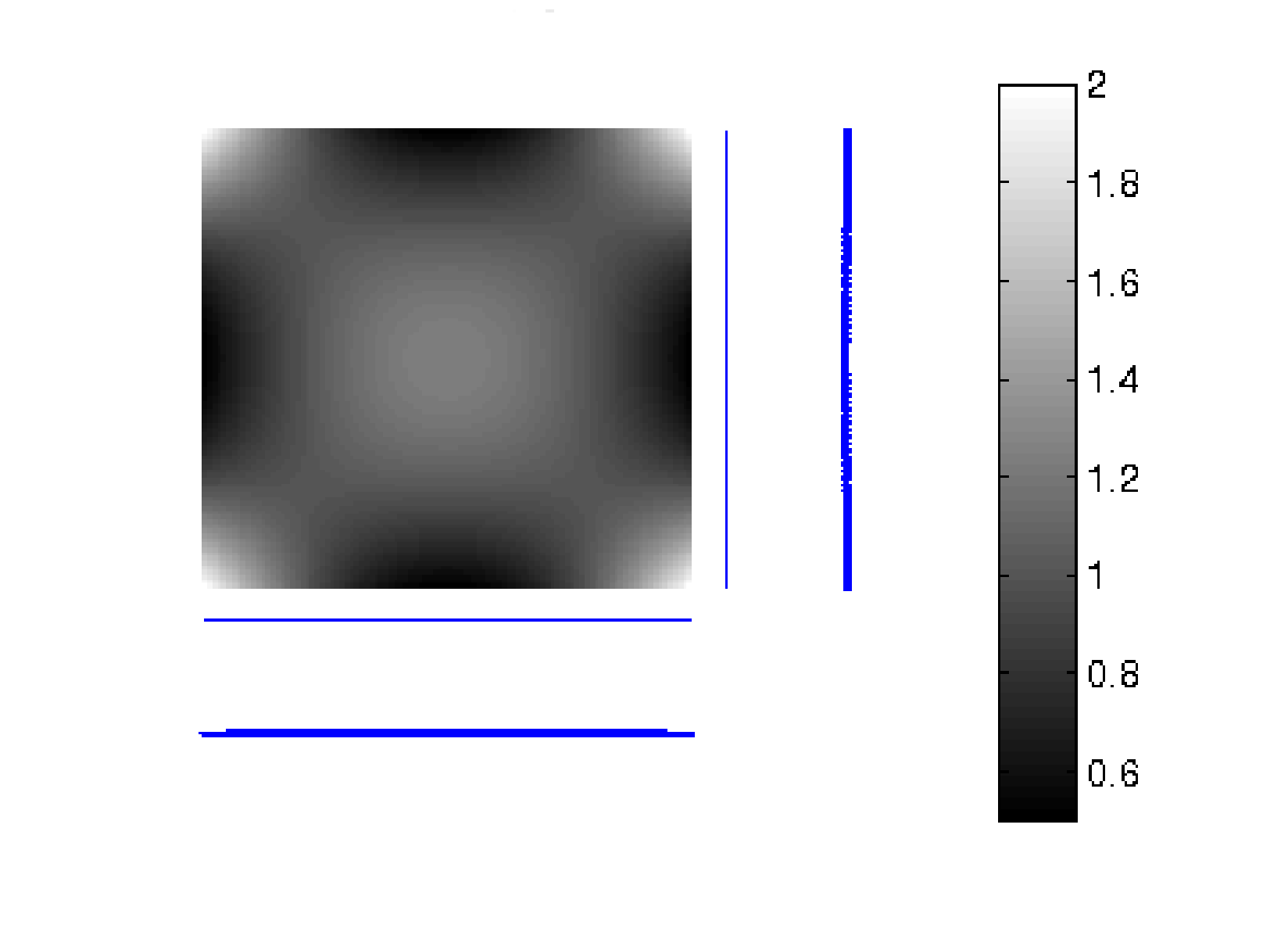}&
\includegraphics[height=0.17\textheight]{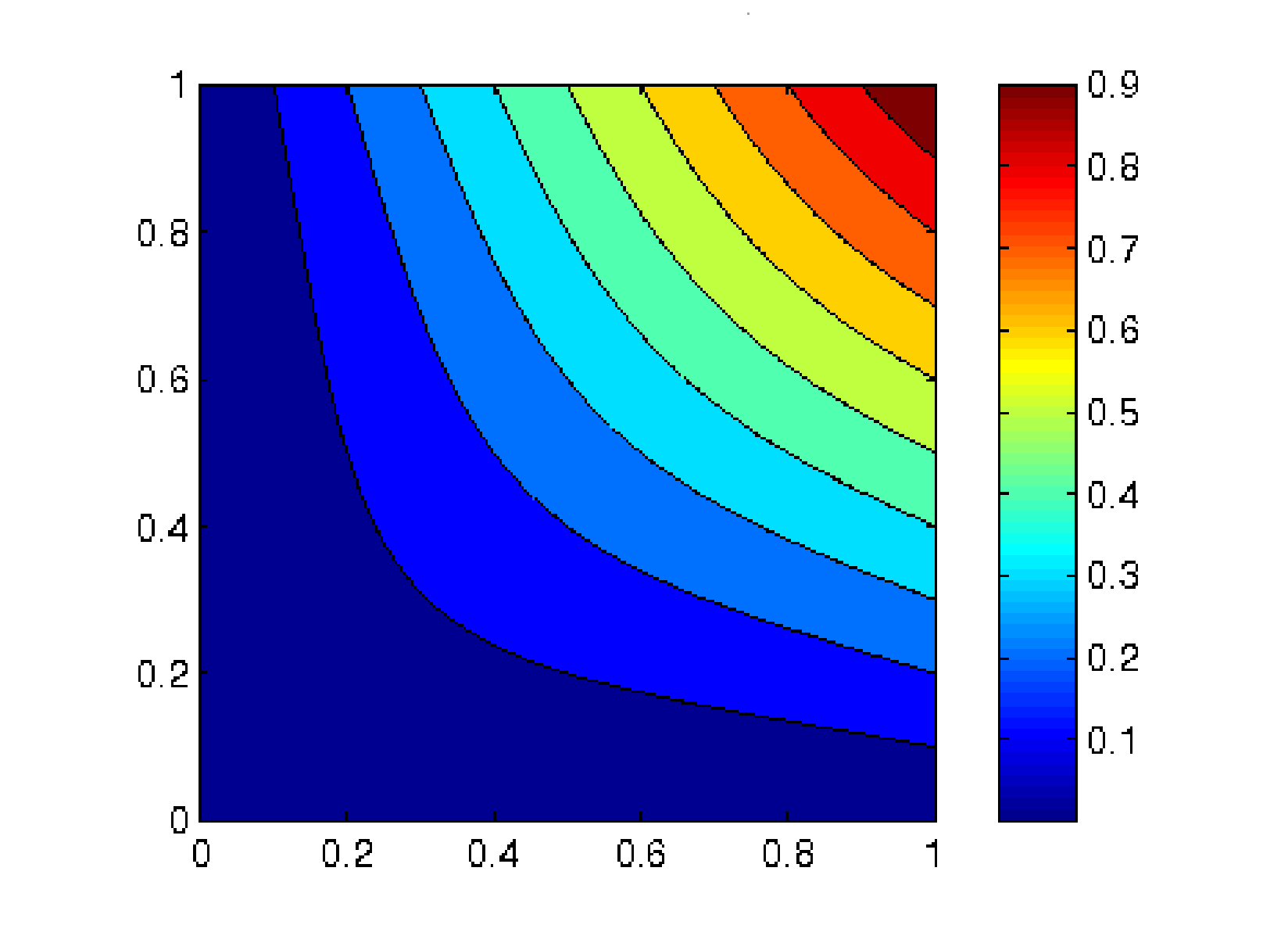}\\
$F(x,y)$,$F_1(x)$,$F_2(y)$ & $f(x,y)$ with $f_1(x)$ and $f_2(y)$ & $C(x,y)$ contours plot\\ 
\includegraphics[height=0.17\textheight]{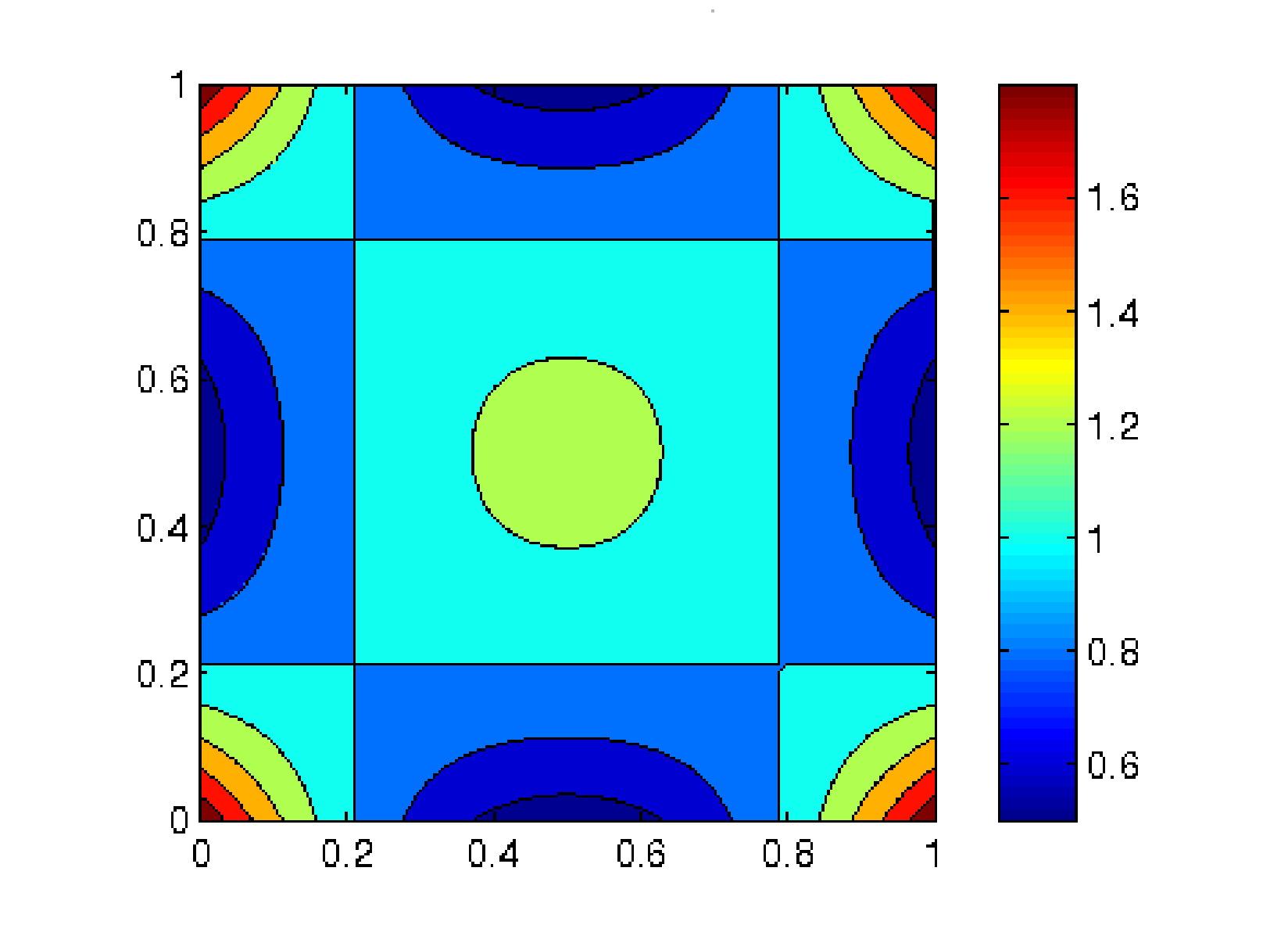}&
\includegraphics[height=0.17\textheight]{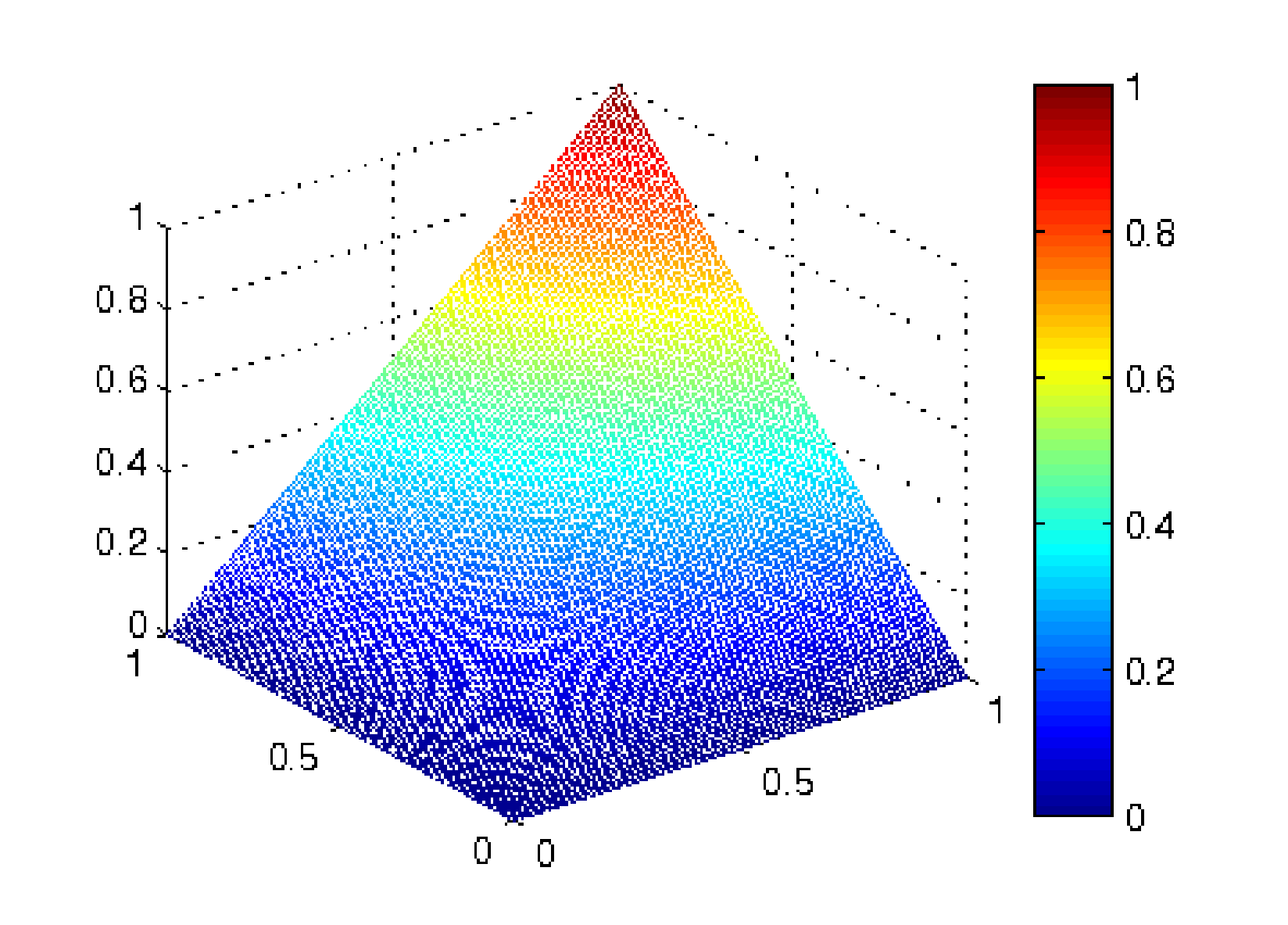}&
\includegraphics[height=0.17\textheight]{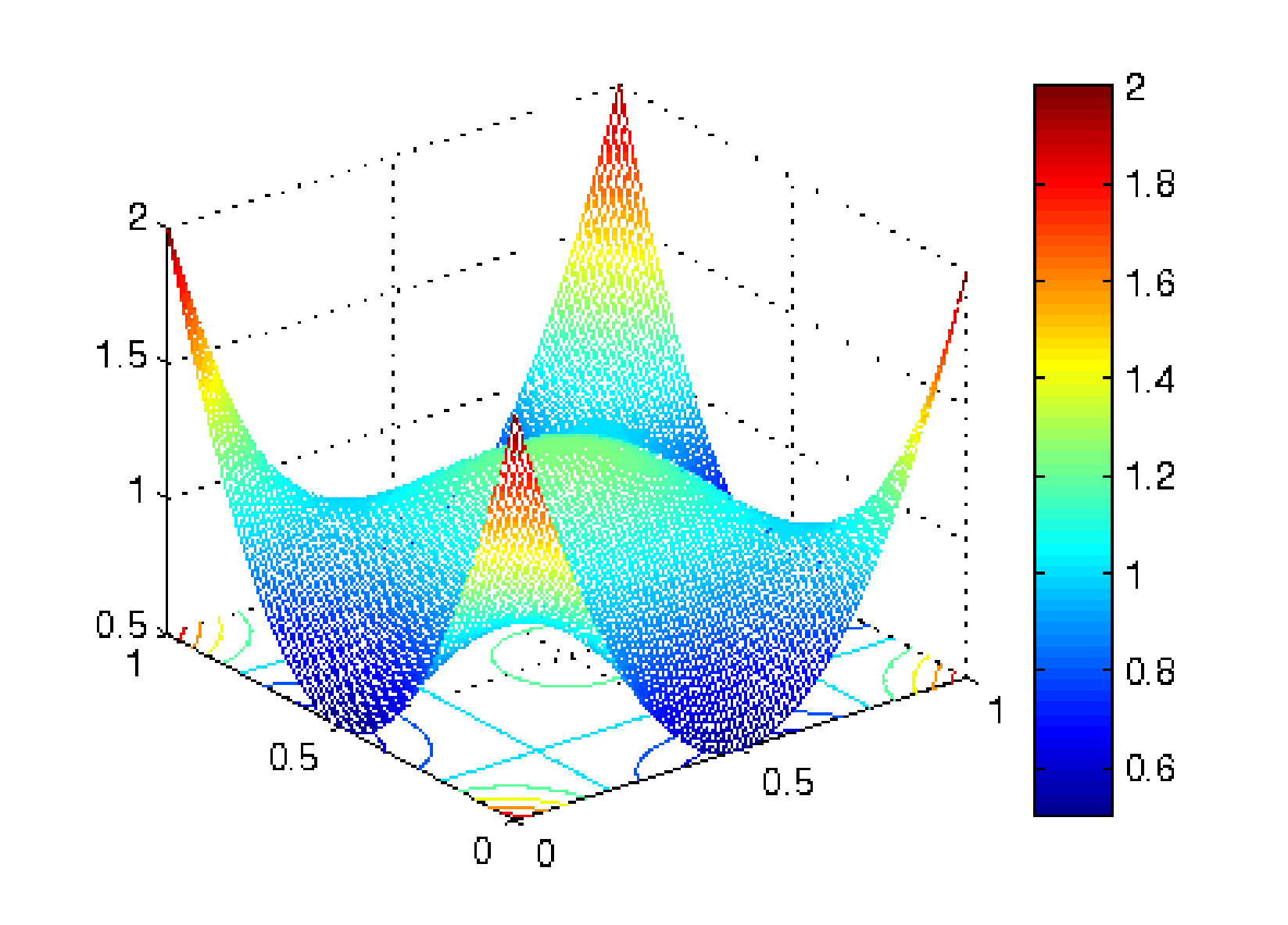}\\
$c(x,y)$ contours plot & $C(x,y)$ mesh plot & $c(x,y)$ mesh plot\\ 
\etabu
\ecc
\caption{Copula with cubic section with $\alpha= 1$ and $\beta=0$.}
\efig

%\endappendix
%-----------------------------------------------------------------------------
\nocite{*}
\bibliographystyle{plain}
\bibliography{masterthesis}
\addcontentsline{toc}{chapter}{Bibliography}
%-----------------------------------------------------------------------------
\end{document}